\documentclass[fleqn,usenatbib]{mnras}

\usepackage{newtxtext,newtxmath}

\usepackage[T1]{fontenc}

\DeclareRobustCommand{\VAN}[3]{#2}
\let\VANthebibliography\thebibliography
\def\thebibliography{\DeclareRobustCommand{\VAN}[3]{##3}\VANthebibliography}

\usepackage{graphicx}	
\usepackage{amsmath}	
\usepackage{xspace}
\usepackage{comment}
\usepackage{gensymb}
\usepackage[export]{adjustbox}
\usepackage{physics}
\usepackage{bm}

\usepackage{siunitx}

\usepackage{threeparttable} 
\newlength{\abovecaptionskip}%
\newcommand{\bahamas}{BAHAMAS}
\newcommand{\flamingo}{FLAMINGO}

\newcommand{\healpix}{\textsc{HEALPix}\xspace}

\newcommand{\vr}{\textsc{VELOCIraptor}\xspace}

\newcommand{\beq} {\begin{equation}}
\newcommand{\eeq} {\end{equation}}
\newcommand{\bal} {\begin{aligned}}
\newcommand{\eal} {\end{aligned}}

\title[kSZ effect in FLAMINGO]{FLAMINGO: combining kinetic SZ effect and galaxy-galaxy lensing measurements to gauge the impact of feedback on large-scale structure}
\author[I. G. McCarthy et al.]{Ian G. McCarthy,$^{1}$\thanks{E-mail: i.g.mccarthy@ljmu.ac.uk}
Alexandra Amon,$^{2}$
Joop Schaye,$^{3}$
Emmanuel Schaan,$^{4,5}$
Raul E. Angulo,$^{6,7}$ \newauthor
Jaime Salcido,$^{1}$
Matthieu Schaller,$^{8,3}$
Leah Bigwood,$^{9,10}$
Willem Elbers,$^{11}$
Roi Kugel,$^{3}$
John C. Helly,$^{11}$ \newauthor
Victor J. Forouhar Moreno,$^{3}$
Carlos S. Frenk,$^{11}$ 
Robert J. McGibbon,$^{3}$
Lurdes Ondaro-Mallea,$^{6,12}$ \newauthor
Marcel P. van Daalen$^{3}$
\\
$^{1}$Astrophysics Research Institute, Liverpool John Moores University, Liverpool, L3 5RF, UK\\
$^{2}$Department of Astrophysical Sciences, Princeton University, Peyton Hall, Princeton, NJ 08544, USA\\
$^{3}$Leiden Observatory, Leiden University, PO Box 9513, 2300 RA Leiden, the Netherlands\\
$^{4}$Kavli Institute for Particle Astrophysics and Cosmology, 382 Via Pueblo Mall Stanford, CA 94305-4060, USA\\
$^{5}$SLAC National Accelerator Laboratory 2575 Sand Hill Road Menlo Park, California 94025, USA\\
$^{6}$Donostia International Physics Center (DIPC), Paseo Manuel de Lardizabal, 4, D-20018 Donostia-San Sebastian, Guipuzkoa, Spain\\
$^{7}$IKERBASQUE, Basque Foundation for Science, D-48013 Bilbao, Spain\\
$^{8}$Lorentz Institute for Theoretical Physics, Leiden University, PO box 9506, 2300 RA Leiden, the Netherlands\\
$^{9}$ Institute of Astronomy, University of Cambridge, Madingley Road, Cambridge CB3 0HA, UK\\
$^{10}$ Kavli Institute for Cosmology Cambridge, Madingley Road, Cambridge, CB3 OHA\\
$^{11}$Institute for Computational Cosmology, Department of Physics, University of Durham, South Road, Durham, DH1 3LE, UK\\
$^{12}$Department of Theoretical Physics, University of the Basque Country UPV/EHU, Bilbao, E-48080, Spain
}

\date{Accepted XXX. Received YYY; in original form ZZZ}

\pubyear{2024}

\begin{document}
\label{firstpage}
\pagerange{\pageref{firstpage}--\pageref{lastpage}}
\maketitle

\begin{abstract}
Energetic feedback processes associated with accreting supermassive black holes can expel gas from massive haloes and significantly alter various measures of clustering on $\sim$Mpc scales, potentially biasing the values of cosmological parameters inferred from analyses of large-scale structure (LSS) if not modelled accurately.  Here we use the state-of-the-art \flamingo\ suite of cosmological hydrodynamical simulations to gauge the impact of feedback on large-scale structure by comparing to \textit{Planck} + ACT stacking measurements of the kinetic Sunyaev-Zel'dovich (kSZ) effect of SDSS BOSS galaxies.  We make careful like-with-like comparisons to the observations, aided by high precision KiDS and DES galaxy-galaxy lensing measurements of the BOSS galaxies to inform the selection of the simulated galaxies.  In qualitative agreement with several recent studies using dark matter only simulations corrected for baryonic effects, we find that the kSZ effect measurements prefer stronger feedback than predicted by simulations which have been calibrated to reproduce the gas fractions of low redshift X-ray-selected groups and clusters.  We find that the increased feedback can help to reduce the so-called $S_8$ tension between the observed and CMB-predicted clustering on small scales as probed by cosmic shear (although at the expense of agreement with the X-ray group measurements).  However, the increased feedback is only marginally effective at reducing the reported offsets between the predicted and observed clustering as probed by the thermal SZ (tSZ) effect power spectrum and tSZ effect--weak lensing cross-spectrum, both of which are sensitive to higher halo masses than cosmic shear. 
\end{abstract}

\begin{keywords}
large-scale structure of Universe -- cosmology: theory -- methods: numerical -- galaxies: clusters: general -- galaxies: formation
\end{keywords}



\section{Introduction}
\label{sec:intro}

Measurements of how matter is spatially clustered in the Universe can be used to place strong constraints on cosmological models, including allowing one to test theories of gravity as well as the natures of dark matter and dark energy.  The standard model of cosmology, the so-called $\Lambda$CDM model, is based on General Relativity and assumes dark matter is composed of relatively ``cold'' and weakly interacting particles and that dark energy takes the form of a cosmological constant. This relatively simple model, which only has six adjustable parameters, describes a wealth of large-scale cosmological data remarkably well.  Nevertheless, there are some notable anomalies which may be hinting at deviations from the predictions of $\Lambda$CDM \citep{Peebles2024}.  One of these is the so-called `$S_8$ tension', where $S_8$ is defined as $\sigma_8 \sqrt{\Omega_\textrm{m}/0.3}$, where $\Omega_\textrm{m}$ represents the present-day matter density, and $\sigma_8$ is the linearly evolved variance of the current matter density field filtered on an 8 Mpc/$h$ scale. Several low-redshift observations of the large-scale structure (LSS), including measurements of total matter clustering via cosmic shear (e.g., \citealt{Heymans2021,Abbott2022,Amon2023}), yield a best-fitting value of $S_8$ that is smaller than, and in mild ($\approx1.5$-$3\sigma$) tension with, the predictions of the standard model based on parameter values specified by the primary CMB and BAO (e.g. \citealt{Planck2020cosmopars}). It is notable that this tension has persisted in some comparisons for nearly a decade, starting with the first \textit{Planck} data release, and encompasses several independent probes that consistently show discrepancies of similar significance and in the same direction (see \citealt{McCarthy2018}).

Various possible solutions have been put forward to potentially reconcile the low-redshift LSS observations with the primary CMB and BAO data. These include unidentified or mischaracterised systematic uncertainties in the LSS observations, or possibly even in the primary CMB measurements. On the theoretical side, LSS tests of cosmology often probe deep into the non-linear regime, and it has been proposed that new physics on those scales could explain the tension (e.g., \citealt{Amon2022,Preston2023}), such as new dark sector models \citep[e.g.][]{Rogers2023,Elbers2025}, or a mis-understanding of galaxy formation effects \citep[e.g.][]{McCarthy2018}.  Modelling non-linear scales necessitates cosmological simulations (or models calibrated on such simulations) to predict the clustering of matter on small scales and at late times. Additionally, as matter collapses to form self-gravitating halos, densities increase to the point where radiative cooling of the gas becomes efficient, leading to further collapse and galaxy formation. This process is accompanied by various energetic feedback mechanisms related to star formation and the accretion of matter onto supermassive black holes.

Modern cosmological hydrodynamical simulations, particularly of volumes that are large enough for clustering studies, cannot resolve all of the relevant scales in order to include such feedback processes in an ab initio way (see the discussion in \citealt{Schaye2015}).  These processes must be included using subgrid prescriptions and it has been demonstrated that certain predictions of the simulations are sensitive to the details of the feedback implementations.  One of these is the fraction of baryons that are retained by (and how they are radially distributed around) massive galaxy groups and clusters (e.g., \citealt{LeBrun2014,Planelles2014,Velliscig2014,McCarthy2017,Henden2018}), which dominate the total matter clustering signal \citep{VanDaalen2015,Mead2020}.  AGN feedback is energetically capable of ejecting large quantities of baryons from haloes and this reduces the amplitude of the clustering signal on small scales (e.g., \citealt{VanDaalen2011,Mummery2017,Springel2018,Chisari2019,VanDaalen2020}).  If such effects are not accounted for, or are included inaccurately, this can lead to an incorrect prediction for the matter clustering signal and potentially bias the recovered cosmological parameters when fitting to the observed clustering (e.g., \citealt{Semboloni2011,Semboloni2013,Schneider2020,Debackere2020,Castro2021}).

In several previous studies, we demonstrated that there is a strong quantitative link between the suppression of the matter power spectrum and the baryon fractions of groups and clusters \citep{Semboloni2013,VanDaalen2020,Salcido2023}.  External observations of the hot gas, which dominates the baryon budget of groups and clusters, can be used to help to evalute the impact of baryon physics on the matter clustering in cosmological analyses.  Traditionally, X-ray observations have provided the highest quality measurements of the state of the hot gas.  When combined with measurements of the total mass, either via the X-ray observations themselves (under the assumption of hydrostatic equilibrium) or from other probes such as weak lensing measurements, the hot gas mass fractions can be inferred (e.g., \citealt{Sun2009,Lovisari2015,Eckert2016,Akino2022}).  Recent cosmological hydrodynamical simulation campaigns, including \bahamas~\citep{McCarthy2017, McCarthy2018}, FABLE \citep{Henden2018,Henden2020}, and \flamingo~\citep{Schaye2023,Kugel2023}, have used X-ray-based measurements of the hot gas fractions to help calibrate the efficiencies of feedback in the simulations, although variations in the feedback efficiencies about the fiducial calibrated models were also considered in \bahamas~and \flamingo. 
Comparisons of these simulations to LSS observables, including the auto- and cross-power spectra of cosmic shear, the thermal Sunyaev-Zel'dovich effect, and CMB lensing, suggest that baryon feedback is incapable of resolving the $S_8$ tension \citep{McCarthy2018,McCarthy2023}.  This is consistent with other recent studies that have used different approaches, such as the baryonification formalism, together
with X-ray measurements as constraints on the baryon physics (e.g., \citealt{Grandis2024}).

While X-ray measurements remain a valuable source of information on the hot gas properties of groups/clusters, particularly with new insights coming in from eROSITA data (e.g., \citealt{Bulbul2024,Popesso2024}), observations of the thermal and kinetic Sunyaev-Zel'dovich (tSZ and kSZ) effects around groups and clusters are yielding increasingly high signal-to-noise measurements when stacking analyses are employed.  The SZ effects have some advantages over the X-ray measurements.  First, the amplitudes of the SZ effects are independent of redshift, in principle allowing one to measure the hot gas over a wide range of cosmic times [by contrast the X-ray surface brightness fades as $(1+z)^4$ and measurements of galaxy groups in particular are generally confined to relatively low redshifts, $z \la 0.3$].  In addition, the tSZ effect yields the total thermal energy density of the gas when integrated over the surface area of the cluster while the kSZ effect yields the gas mass (or gas momentum), both of which are more directly linked to the impact of baryon physics on the matter clustering than the X-ray emission.  Note that high angular resolution data is required, so that the groups/clusters under study can be spatially resolved, allowing for measurements of their gaseous properties on the scales where feedback is important (e.g., \citealt{LeBrun2015,Yang2022}).

\textit{Planck} data has been used to produce full-sky maps of the tSZ effect, enabling the stacking of many sources and detections over a wide range of masses (e.g., \citealt{Planck2013,Greco2015}).  However, the typical 10 arcmin angular resolution of \textit{Planck} prevents resolved measurements (radial profiles) for all but most massive and nearby sources.  The Atacama Cosmology Telescope (ACT) and the South Pole Telescope (SPT) offer almost an order of magnitude better angular resolution than \textit{Planck} and higher sensitivity measurements, although with a smaller sky coverage and a limited number of frequencies, which can be key to removing foreground and background contaminants (e.g., dust in the Galaxy, clustered radio sources, the cosmic infrared background, or CIB).  It is now becoming common place to combine \textit{Planck} data with ACT or SPT data to study groups and clusters (e.g., \citealt{Aghanim2019,Melin2021,Bleem2022}), taking advantage of \textit{Planck}'s multi-frequency measurements and sensitivity to large scales together with ACT and SPT's higher angular resolutions and sensitivities. 

\citet{Schaan2021} have used a combination of \textit{Planck} and ACT data to measure the stacked kSZ profiles of SDSS BOSS galaxies \citep{Ahn2014}.  The kSZ effect is proportional to the line of sight radial (peculiar) velocity of galaxies and, statistically speaking, we are as likely to find a galaxy moving towards us as away from us (if we subtract our motion relative to the CMB), so simply stacking the CMB temperature maps of galaxies without regard for the direction of motion would imply that the kSZ effect will cancel out.  To address this, \citet{Schaan2021} used the 3D clustering signal of BOSS galaxies to reconstruct the implied linear velocity field, allowing the galaxies to be weighted according to their predicted velocities in the stack.  Strong detections were made for both the CMASS and LOWZ samples from the BOSS survey.  The derived radial profiles show that the hot gas around the BOSS galaxies is very extended in comparison to the expectation for the dark matter distribution from dissipationless simulations, implying that non-gravitational processes (particularly feedback) have significantly altered the gas distribution.  Using a simple halo model formalism to model the kSZ effect profiles suggests a significant impact on the matter clustering on small scales \citep{Amodeo2021} with possible implications for the $S_8$ tension \citep{Amon2023}.

Recently, \citet{Schneider2022} and \citet{Bigwood2024} have used the baryonification formalism (e.g., \citealt{Schneider2015,Schneider2019,Arico2021emulator,Arico2023}) to jointly model the \citet{Schaan2021} kSZ effect measurements together with measurements of the cosmic shear correlation functions.  The baryonification formalism employs a parametric model to describe the radial distributions of the matter components (gas, stars, dark matter) of haloes which it uses to radially perturb the mass distribution of dissipationless (`dark matter only') simulations.  The parameters of the model can either be determined by fitting to external data sets (e.g., X-ray and SZ effect measurements) or they can be left free in the fit to cosmological observables such as the cosmic shear correlation functions.  \citet{Schneider2022} and \citet{Bigwood2024} have shown that the inclusion of the kSZ effect measurements in the cosmic shear analysis better constrains the parameters which characterise the impact of baryons.  Furthermore, \citet{Bigwood2024} demonstrated that the joint analysis of cosmic shear and kSZ effect data prefers a stronger impact of baryons compared to that implied by models which are fitted to resolved X-ray observations (see also \citealt{Salcido2024}).  

A major source of uncertainty in the modelling of the \citet{Schaan2021} kSZ effect observations is the choice of halo mass of the sample.  The amplitude of the kSZ effect scales proportionally with the gas mass, and therefore (approximately) proportionally with the total halo mass.  Given that measurements of the mean halo mass of the BOSS CMASS galaxies vary by nearly an order of magnitude between previous studies (see the discussion in \citealt{Bigwood2024}), marginalising over this uncertainty significantly weakens the constraints on the feedback model parameters and in turn weakens the cosmological constraints.  Note also that the LOWZ sample has no reliable mean halo mass measurement to date, which is the main reason why the stacked kSZ effect measurement from this sample has not yet been utilised as a constraint on the baryon modelling. In addition, previous attempts to model the signal have implicitly assumed that the BOSS CMASS sample is composed entirely of central galaxies with no mis-centring of the galaxies with respect to the total gravitating mass or hot gas distributions.  A main reason for adopting these assumptions, whose impact is presently difficult to assess, is that there is currently no straightforward way to select a realistic mock BOSS-like sample in the context of the baryonification formalism.

In this study, we address these issues using self-consistent full cosmological hydrodynamical simulations.  Specifically, we use full-sky lightcones constructed using the FLAMINGO hydro simulations, which systematically vary the important parameters (and parametrisations) controlling the impact of baryon feedback.  We also explore the cosmology dependence of the signal.  We select galaxy populations from the simulations that are constructed to carefully match the observed galaxy-galaxy lensing profiles of the BOSS CMASS and LOWZ samples, as recently measured by \citet{Amon2023} using high-quality Kilo Degree Survey (KiDS 1000) and Dark Energy Survey Year Three (DES Y3) data.  We will show that the galaxy-galaxy lensing measurements place very stringent constraints (better than $10\%$ at $2\sigma$) on the mean halo masses of these samples.  Using the lensing-selected galaxy populations, we extract the simulation kSZ effect profiles following the methodology of \citet{Schaan2021}, stacking the profiles at the locations of the galaxies in the lightcone-based maps in a way that is faithful to that done for the real observations.  We quantify the impact of satellite contamination on the derived galaxy-galaxy lensing and kSZ effect profiles, showing it to be non-negligible for both.  Finally, in agreement with several recent studies, we will show that the kSZ effect measurements imply a stronger impact of feedback relative to that adopted in the fiducial FLAMINGO model (which was calibrated on X-ray observations of low-redshift groups) and we discuss the implications of this finding for the $S_8$ tension.  

The paper is organised as follows.  In Section~\ref{sec:sims_and_obs}, we describe the \flamingo\ simulation suite and the observations employed in this study.  In Section~\ref{sec:analysis}, we describe our methodology, including the selection of galaxies in the simulations and the derivation of lensing and kSZ effect profiles from lightcone-based maps.  In Section \ref{sec:results} we present our main results, including an examination of the feedback and cosmological dependencies of the derived lensing and kSZ effect profiles, in comparison with the \citet{Amon2023} and \citet{Schaan2021} measurements.  In Section~\ref{sec:discuss}, we discuss the implications of our findings for feedback modelling and the $S_8$ tension. In Section \ref{sec:conclusions} we summarise our main findings and conclude.

\section{Simulation and observational datasets}
\label{sec:sims_and_obs}

\subsection{\flamingo~simulations}

We provide here a summary of the \flamingo\ simulations, referring the reader to \citet{Schaye2023} and \citet{Kugel2023} for in depth presentations.

The \flamingo\ suite consists of 16 hydrodynamical simulations presented in \citet{Schaye2023}, two decaying dark matter variants in \citet{Elbers2025}, plus two new hydrodynamical simulations introduced here (see below), and 12 gravity-only simulations.  The suite has variations in resolution, box size, subgrid modelling, and cosmology.  We mostly use the intermediate resolution
($m_{\mathrm{gas}} = 1.09\times10^9$ M$_\odot$) in box sizes of (1 Gpc)$^3$. These simulations use $2 \times 1800^3$ gas and dark matter particles and $1000^3$ neutrino particles, and most adopt cosmological parameters corresponding to the maximum likelihood DES Y3 `3×2pt + All Ext.' $\Lambda$CDM cosmology \citep{Abbott2022}, which we refer to as `D3A'. These values assume a spatially flat universe and are based on the combination of constraints from DES~Y3 `$3\times2$-point' correlation functions: cosmic shear, galaxy clustering, and galaxy-galaxy lensing, with constraints from external data from BAO, redshift-space distortions, SN Type Ia, and \textit{Planck} observations of the CMB (including CMB lensing), Big Bang nucleosynthesis, and local measurements of the Hubble constant (see \citealt{Abbott2022} for details).  We also consider two alternative cosmologies: a run with the \textit{Planck} 2018 maximum likelihood cosmology (`Planck'; \citealt{Planck2020cosmopars}) and the `lensing cosmology' from \citet{Amon2023} (`LS8'). The latter model has a lower amplitude of the power spectrum, $S_8 = 0.766$, compared with 0.815 and 0.833 for D3A and \textit{Planck}, respectively.  We assess the resolution dependence of our results by comparing with a higher resolution run, labelled L1\_m8.  This run adopts the fiducial D3A cosmology in a (1 Gpc)$^3$ volume but uses $2 \times 3600^3$ gas and dark matter particles and $2000^3$ neutrino particles.  The mass resolution is therefore a factor of 8 higher for this run (i.e., $m_{\mathrm{gas}} = 1.34\times10^8$ M$_\odot$) compared to the fiducial resolution runs.

The simulations were run with the cosmological smoothed particle hydrodynamics and gravity code SWIFT \citep{Schaller2024a} using the SPHENIX SPH scheme \citep{Borrow2022sphenix}. The initial conditions are obtained from a modified version of monofonIC \citep{Hahn2021,Elbers2022a}, and neutrinos are implemented with the $\delta f$ method \citep{Elbers2021}.  The modelling of subgrid physics (star formation, stellar evolution, radiative cooling, and sources of feedback) is described in \citet{Schaye2023} and references therein.

The subgrid physics was calibrated by requiring that the simulations should match the $z=0$ galaxy stellar mass function and the gas fractions in low-$z$ groups and clusters using machine learning-based emulators \citep{Kugel2023}. The emulators are not only used to design simulations that reproduce these observations, but also to create models in which the galaxy stellar mass function and/or cluster gas fractions are shifted to higher/lower values. This allows us to specify model variations in terms of the number of $\sigma$ by which they deviate from the calibration data. Of particular interest for this work are the variations in the group/cluster gas fractions and AGN model, which are denoted as fgas\_$\pm N\sigma$ and Jet\_fgas\_$\pm N\sigma$. For these models $N\sigma$ denotes by how many observed standard deviations the gas fractions have been shifted up or down with respect to the fiducial model. The Jet models make use of kinetic jets for the AGN feedback instead of the thermal model used for all other runs. These Jet models are calibrated to match the same data as the corresponding thermal AGN feedback models.

We introduce two new runs, both in 1 Gpc boxes at intermediate resolution.  The first is a run denoted `no cooling' which sets the net radiative cooling + radiative heating rate to zero for gas where the net rate would have been negative (i.e., net cooling).  Consequently there is no cooling and also no star formation or feedback present in this simulation.  While obviously unrealistic, comparisons to this run are helpful for quantifying the impact of feedback in the other \flamingo~runs.  The second new run, denoted `LS8\_fgas-$8\sigma$', is a strong feedback model in the LS8 lensing cosmology.  This run provides an opportunity to explore the degeneracy between feedback and cosmology, via comparison to the fgas-$8\sigma$ run in the fiducial D3A cosmology.

\subsubsection{Lightcones: \healpix\ maps and galaxy/halo catalogs}

A description of the on-the-fly lightcone implementation in \flamingo\ can be found in the appendix of \citet{Schaye2023}.  Here we give a brief summary, focusing on the details most relevant for the present study.

We work primarily with lightcone-based maps (as opposed to the particle lightcone output).  To produce the maps, the observer's past lightcone is split into a set of concentric spherical shells in comoving distance. For each shell one full sky \healpix\ \citep{Gorski2005} map for each quantity is created. Whenever a particle is found to have crossed the lightcone, we determine which shell it lies in at the time of crossing and accumulate the particle's contributions to the \healpix\ maps for that shell.  The shell radii are specified in terms of redshift. From redshifts $z=0$ to 3 we use shells of thickness $\Delta z=0.05$, with a larger $\Delta z$ at higher redshifts. 

The \healpix map resolution is set to $N_\text{side} = 16384$, which gives a maximum pixel radius of 13.46 arcseconds and $12*16384^2$ pixels in each full sky map. We note that the number of pixels exceeds the size of a signed 32-bit integer ($2^{31}$), which would prevent us from smoothing the kSZ effect maps with the ACT beam (necessary for a like-with-like comparison) using the \textsc{Healpy} smoothing function (sphtfunc.smoothing), as the function can currently only handle a maximum $N_\text{side}$ of 8192.  We therefore downsample the kSZ effect maps to this resolution using the \textsc{Healpy} function pixelfunc.ud\_grade, preserving the mean of the map in the downsampling operation.
We describe the production of galaxy-galaxy lensing and kSZ effect maps in Sections \ref{sec:delta_sigma_profs} and \ref{sec:ksz_profs}, respectively.

Computational limitations prevent us from running a structure finder on-the-fly during the simulation.  Instead, to produce galaxy/halo catalogs on the lightcone, structure finding\footnote{We have checked that our results and conclusions are not strongly dependent on the choice of substructure finder, by comparing our fiducial results with those derived from employing an independent halo catalog derived using the \vr~package \citep{Elahi2019vr}.} is done in post-processing on the snapshot particle data using a modified version of the \textsc{HBT-HERONS} algorithm \citep{Han2012,Han2018,ForouharMoreno2025}.  Note that at $z \le 3$ snapshots are also written out with a frequency of $\Delta z=0.05$, which was chosen to minimise any issues arising from the evolution of galaxy/halo properties between the snapshots and the \healpix\ maps.  We read in the snapshot subhalo catalog corresponding to a given snapshot and then read in a spherical shell from the black hole (BH) particle lightcone which spans the redshift range that is half way to the previous snapshot to halfway to the next snapshot.  Every time a BH particle in the lightcone shell appears as the most bound BH in a subhalo in the snapshot (identified by matching their unique particle IDs), we place the subhalo at the BH particle's position\footnote{Note that subhaloes can potentially contain multiple black holes.  To address this issue, we identify which of the BHs in a subhalo exist at both the next and previous snapshots and we pick the most bound BH particle from that subset to use as a tracer. In cases where the most bound BH survives, we then use the same BH as before. If the most bound BH disappears because it merges with a more massive BH before the next snapshot, we pick the next most bound BH in the same subhalo (which is not about to get swallowed) and use that as the tracer.} in the halo lightcone. We repeat this process for every snapshot to make the full lightcone halo catalog.  We use the halo lightcone to provide the locations of the selected galaxies for stacking the kSZ effect and galaxy-galaxy signals in the maps.  We have verified that the profiles extracted from the lightcone-based \healpix\ maps using positions from the constructed halo lightcone precisely match profiles constructed directly from the snapshots using the snapshot halo catalogs.

\subsection{Observational data}

\subsubsection{KiDS 1000 + DES Y3 lensing measurements}

The amplitude of the kSZ effect scales approximately with the halo mass of a system, so it is important to match the halo masses of observed and simulated systems to enable a like-with-like comparison of the kSZ profiles.  As halo masses are not directly observable, a common approach is to select galaxies from hydrodynamical simulations based on e.g., stellar mass.  However, the mapping between such observables and halo mass may not be fully realistic in the simulations, potentially resulting in a halo mis-match.  Since halo mass is the key physical quantity that dictates the amplitude of the SZ signal, our approach is to use weak lensing data to ensure that the mean halo mass of the observed and simulated galaxy populations are aligned.  Note that the approximate linear scaling of the kSZ signal with halo mass means that a stack of the kSZ effect of many systems will primarily be sensitive to the mean halo mass of the sample.  Stacked galaxy-galaxy lensing is directly sensitive to the mean halo mass of the sample.  Thus, by adjusting our simulated galaxy selection to match the stacked lensing profiles of the LOWZ and CMASS samples, we can make meaningful predictions for the kSZ effect for the BOSS samples.

We use stacked galaxy-galaxy lensing measurements from \citet{Amon2023} of the BOSS LOWZ and CMASS samples with KiDS 1000 and DES Y3 data.  Here we briefly describe these measurements, referring the reader to \citet{Amon2023} for a more complete description. The LOWZ and CMASS data of the SDSS BOSS Data Release 12 was divided into two distinct lens samples by redshift, with bounds:
\begin{eqnarray*}
    {\rm L1: LOWZ} \,\, z=0.15-0.31 \,\, & \&  \,\, & {\rm L2: LOWZ} \,\, z=0.31-0.43 
    \\
    {\rm C1: CMASS}  \,\, z=0.43-0.54  & \& &
    {\rm C2: CMASS}  \,\, z=0.54-0.70
\end{eqnarray*}  

Lensing measurements were made for each sample, using both KiDS and DES. These were shown to be statistically consistent and a combined DES Y3 + KiDS 1000 measurement was produced by taking the inverse-variance weighted average of the derived lensing profiles. Note that \citet{Schaan2021} present a stacked kSZ profile for each of LOWZ and CMASS (i.e., two redshift bins, as opposed to four in \citealt{Amon2023}). As we will show in Section \ref{sec:results}, the lensing analysis yields consistent minimum stellar masses and halo masses for the two bins within LOWZ and CMASS samples, allowing us to jointly fit the two bins to yield a single galaxy selection (each) for the LOWZ and CMASS samples. 

The galaxy-galaxy lensing signal is most often expressed in terms of the excess surface density, $\Delta \Sigma$, which is defined as the difference between the mean surface mass density interior to a radius and the surface mass density at that radius.  It can be related to the average tangential shear $\langle \gamma_{\rm t} (\theta) \rangle$ as
\begin{equation}
\Delta \Sigma = \frac{\langle \gamma_{\rm t} (\theta) \rangle}{\overline{\Sigma_{\rm c}^{-1}}} \ \ \ , 
\label{eq:ds}
\end{equation}
at a projected separation $\theta=R/\chi(z_{\rm l})$, where $\chi(z_{\rm l})$ is the comoving distance to the lens.  For a source redshift distribution $n(z_s)$ the average inverse critical density is given by
\begin{equation}
\overline{\Sigma_{\rm c}^{-1}}(z_{\rm l}) = \frac{4\pi G (1+z_{\rm l}) \chi(z_{\rm l})}{c^2}\int_{z_{\rm l}}^\infty \mathrm{d}z_{\rm s} n(z_{\rm s}) \frac{\chi(z_{\rm l},z_{\rm s})}{\chi(z_{\rm s})} \ \ \ ,
\end{equation}
where the source redshift distribution is computed for a given lens redshift $z_l$ and normalised such that $\int_0^{\infty} n(z_s) dz_s = 1$. The division by the inverse critical density in eqn.~\ref{eq:ds} removes the dependence of $\Delta \Sigma$ on the background source redshift distribution, $n(z_s)$, as the tangential shear also (implicitly) contains the same geometric factor.

We refer the reader to \citet{Amon2023} for the further details of the estimators of eqn.~\ref{eq:ds} used for the DES and KiDS data, including the treatment of additive and multiplicative biases and lens-source pair weightings and the estimation of uncertainties.  

\subsubsection{\textit{Planck} + ACT kSZ effect measurements}

This work uses the kSZ effect measurements presented in \citet{Schaan2021}, of the ACT DR5 and \textit{Planck} CMB temperature maps stacked at the locations of galaxies in the BOSS LOWZ and CMASS samples and using their reconstructed velocities as weights.  \citet{Schaan2021} presented results at the two ACT frequencies, 90 GHz and 150 GHz.  We focus on the 150 GHz measurements here, given their higher sensitivity and angular resolution (1.3 arcmin, compared to 2.1 arcmin for the 90 GHz channel), although we have checked that none of our conclusions are sensitive to this choice.  

The kSZ effect induces a fluctuation in the temperature of the CMB, $\Delta T _{\rm kSZ}$, that arises from the Doppler shift of the CMB photons with respect to the bulk motion of the ionised gas (e.g., in groups and clusters) off which CMB photons have scattered.  The CMB maintains its black body spectral form but with a fractional temperature change with respect to the mean CMB temperature:
\begin{equation}
\frac{\Delta T _{\rm kSZ} (\pmb{\theta})}{T_{\rm CMB}} 
= 
- \sigma_\textrm{T} \int  \ n_\textrm{e}(\pmb{\theta}, z) \frac{v_\textrm{e, r}(\pmb{\theta}, z)}{c} \frac{d \chi}{1+z}  \ \ \ ,
\label{eq:kSZdef}
\end{equation}

\noindent where the integration is along the observer's line of sight at given angular coordinates $\pmb{\theta}$, $\chi$ is the comoving radial distance, $n_\textrm{e}$ is the physical free electron density, $v_\textrm{e, r}$ is the free electron peculiar radial velocity, and  $\sigma_\textrm{T}$ is the Thomson scattering cross-section.

The stacking analysis selectively extracts the kSZ effect associated with galaxies/groups using their reconstructed velocities, so that
\beq
\frac{\Delta T _{\rm kSZ} (\pmb{\theta})}{T_{\rm CMB}}
=
- \tau_{\rm gal}(\pmb{\theta}) \; \left(\frac{v_\textrm{e, r, gal}}{c} \right) \ \ \ ,
\eeq
where $v_\textrm{e, r, gal}$ is the galaxy's bulk velocity and $\tau(\pmb{\theta})$ is the so-called optical depth to Thomson scattering, which is defined as:
\beq
\tau_{\rm gal}(\pmb{\theta}) \equiv \sigma_\textrm{T} \int n_\textrm{e}(\pmb{\theta}, z) \frac{d \chi}{1+z}  \ \ \ .
\label{eq:tau_def}
\eeq
For each galaxy, \citet{Schaan2021} apply compensated aperture photometry (CAP) filtering to effectively measure a cumulative kSZ effect profile as a function of an angular disk radius, $\theta_\textrm{d}$:
\begin{equation}
\label{eq:ap}
\mathcal{T}(\theta_\textrm{d})=
\int d^2\theta \, \Delta T_{\rm kSZ}(\theta) \, W_{\theta_\textrm{d}}(\theta) \ \ \ ,
\end{equation}
where the CAP filter $W_{\theta_\textrm{d}}$ is defined as:
\begin{equation}
W_{\theta_\textrm{d}}(\theta) =
\left\{
\begin{aligned}
1& &  &\text{for} \, \theta < \theta_\textrm{d} \,, \\
-1& &  &\text{for} \, \theta_\textrm{d} \leq \theta \leq \sqrt{2}\theta_\textrm{d} \,, \\
0& & &\text{otherwise}. \\
\end{aligned}
\right.
\end{equation}
This corresponds to measuring the integrated temperature fluctuation in a disk with radius $\theta_\textrm{d}$ and subtracting the signal measured in a concentric ring of the same area around the disk.  As the disk radius $\theta_\textrm{d}$ is increased, the CAP filter output resembles a cumulative profile: for small disk radii, the output vanishes; for large radii, where all the gas profile is included inside the disk, the output is equal to the integrated gas profile.  Note that since the filter is compensated (i.e. $W$ integrates over area to zero), it has the desirable property that fluctuations with wavelength longer than the filter size will cancel in the subtraction.  This helps to significantly reduce noise from larger scale CMB fluctuations and the correlation of the measurements between different $\theta_\textrm{d}$ bins.

\citet{Schaan2021} note that the minimum-variance unbiased linear estimator is the velocity weighted, inverse-variance weighted mean and thus stack the profiles according to:
\begin{equation}
    \hat{T}_{\rm kSZ}(\theta_\textrm{d}) = -
    \frac{1}{r_v}
    \frac{ v_{\rm rms}^{\rm rec}}{  c}
    \frac{\sum_i \mathcal{T}_i(\theta_\textrm{d}) (v_{{\rm rec}, i}/c) / \sigma_i^2}{\sum_i (v_{{\rm rec}, i}/c)^2 / \sigma_i^2}
    \label{eq:kSZ_est}
\end{equation}
where $v_{\rm rms}^{\rm rec}$ refers to the rms of the radial component of the reconstructed velocity, $\sigma_i^2$ is the noise variance for the CAP filter on galaxy $i$, and the $r_v^{-1}$ factor (discussed below) ensures that the estimator is not biased by the imperfections in the velocity reconstruction.  The velocity weighting is key as without it the kSZ signal would cancel in the numerator, since it is linear in the galaxy radial velocities, which are equally likely to be pointing away or towards us if we subtract our motion relative to the CMB.  With the velocity weighting, both numerator and denominator scale as the mean squared velocity, avoiding the cancellation and selectively extracting the kSZ signal.

As discussed in \citet{Schaan2021}, peculiar radial velocities for the BOSS galaxies are reconstructed using their 3D clustering densities and solving the linearised continuity equation.  The reconstruction is not perfect, however, due to non-linear effects, shot noise, and finite volume effects.  Applying their techniques to BOSS mock galaxy catalogs produced using dissipationless cosmological simulations, \citet{Schaan2021} compute the kSZ bias factor, $r_v$, defined as:
\begin{equation}
    r_v = \frac{\langle v_{\rm true} v_{\rm rec} \rangle}{v_{\rm rms}^{\rm true} \ v_{\rm rms}^{\rm rec}} \,,
\end{equation}
where $v_{\rm rms}^{\rm true}$ and $v_{\rm rms}^{\rm rec}$ are the standard deviations of the true and reconstructed galaxy radial velocities, respectively.  They estimate a value $r_v$ = 0.7 and use this to correct their stacked kSZ profiles to account for the imperfect velocity reconstruction (see also \citealt{Ried2024,Hadzhiyska2024a}).  Note that $r_v$ is a constant correction factor that scales the amplitude of the derived kSZ temperature profiles.  As we will show later, feedback can strongly affect the amplitude of the profile and will thus be degenerate at some level with uncertainties in the velocity reconstruction.  It would be interesting to apply the velocity reconstruction technique on BOSS mocks derived from the \flamingo\ hydrodynamical simulations as an independent estimate of $r_v$, but we leave this exercise for future work, retaining the fiducial estimate from \citet{Schaan2021} for the bias factor.  Note that for kSZ temperature profiles derived from the simulations, the value of $r_v$ is 1 (i.e., unbiased), since we use the true radial velocities rather than reconstructed velocities.

\section{Synthetic observables from \flamingo}
\label{sec:analysis}

\subsection{Galaxy selection}
\label{sec:boss_sel}

The BOSS LOWZ sample primarily selects red galaxies while the BOSS CMASS sample targets galaxies at higher redshifts with a surface density of roughly 120 deg$^{-2}$ and a roughly constant minimum stellar mass of a few $10^{11}$ M$_\odot$.  In this study, we do not attempt to implement the precise BOSS selection criteria for the selection of galaxies in the simulations.  Such an exercise would be non-trivial, as the mapping from intrinsic stellar properties in the simulations (mass, age, abundances) to observed luminosities requires stellar population synthesis modelling, a treatment of dust and radiative transfer, and would be sensitive to the adopted theoretical nucleosynthetic yields, which have considerable uncertainties. Instead, our approach is to ensure that the simulated sample has a stacked galaxy-galaxy lensing signal that is compatible with the BOSS samples. This ensures that the mean halo masses of the simulated and observed samples are aligned, enabling a fair comparison of the observed and predicted stacked kSZ effect profiles.  We discuss below some tests that we have performed to ensure that our selection is realistic.

Our fiducial approach to select the simulated galaxies is to apply a simple minimum stellar mass cut, where we define the stellar mass as the total bound stellar mass within 50 kpc of the most bound particle.  For comparison to the four redshift bins in the \citet{Amon2023} lensing analysis, we select simulated `lens' galaxies from the lightcone shell that is nearest to the mean redshift of the actual (observed) bins.  Specifically, we select the shells with $0.225 < z < 0.275$, $0.325 < z < 0.375$, $0.475 < z < 0.525$, and $0.575 < z < 0.625$ for the L1, L2, C1, and C2 bins, respectively, noting that the mean redshifts of the observed bins are: 0.240, 0.365, 0.496, and 0.592.  For the kSZ effect comparison, we select galaxies in the shells with $0.275 < z < 0.325$ and $0.525 < z < 0.575$ for LOWZ and CMASS, respectively, noting that the mean redshifts for the two observed samples are $z=0.31$ and $z=0.54$, respectively.

As the genuine BOSS LOWZ and CMASS samples do not explicitly exclude satellite galaxies, we should include them in our selection so long as their stellar mass exceeds the minimum stellar mass.  We compare the results with a central-only sample to deduce the role that satellites play in derived lensing and kSZ effect profile in Section \ref{sec:satellite_dep}.  We construct multiple simulated samples by varying the minimum stellar mass cut and compare the derived galaxy-galaxy lensing profiles to the measurements of \citet{Amon2023} to determine the minimum stellar mass cut which best reproduces the lensing measurements.  We propagate the uncertainties in the best-fitting minimum stellar mass through to our kSZ effect analysis.

A simple stellar mass cut is unlikely to yield a galaxy sample that matches all aspects of the CMASS and LOWZ samples.  However, our selection is constrained to match the mean halo masses of the CMASS and LOWZ samples and, as already noted, the amplitude of the stacked kSZ effect should be mostly sensitive to the mean halo mass of the stack.  Nevertheless, as a check, we have explored simultaneous cuts in stellar mass and specific star formation rate, so that we select preferentially `red' galaxies.  In Appendix \ref{sec:ssfr_selection}, we show that our main results and conclusions are unaltered by adopting this more complex selection function, suggesting that the matching the mean halo mass is sufficient for our purposes.  Another test we have performed is to use much narrower bins in the stellar mass selection of 0.1dex width whose bin centre is adjusted to match the same mean halo mass as our fiducial selection.  We find virtually identical kSZ effect predictions for these two cases.  In addition, we have compared the predicted and observed projected clustering of the BOSS galaxies for our fiducial selection methodology.  For this comparison we used the large-scale clustering measurements presented in \citet{Amon2023} and we adopted the same methodology to derive the clustering of the simulated galaxies selected on stellar mass.  In short, we find excellent agreement between the minimum stellar masses and mean halo masses derived from our galaxy-galaxy lensing analysis and our clustering analysis, again suggesting that the simulated galaxy sample is realistic.  We leave a detailed presentation of the clustering results for a future paper.

When selecting galaxies we use the true stellar mass predicted by the simulations.  Observationally measured stellar masses have uncertainties, however, and an interesting question is whether our results would be impacted by folding in such uncertainties in the selection of our simulated galaxies.  For the same reasons argued above, we do not expect such uncertainties to impact our results since the selection of galaxies (with or without observational uncertainties factored in) is forced to match the galaxy-galaxy lensing measurements.  Nevertheless, we explicitly test this hypothesis below.

Note that an alternative possibility for selection would be to select systems based on their halo masses rather than their stellar masses, and to constrain the selected halo mass range based on fits to the galaxy-galaxy lensing measurements as described above.  But such a selection would essentially correspond to a central-only population, since virtually all satellites would have halo masses well below the scale of interest here ($\sim10^{13} \textrm{M}_\odot$) due to tidal stripping by their more massive host and would therefore either not be included in the selection or make no meaningful contribution to the stack.  Selection by stellar mass, on the other hand, is closer to the real observational selection function in BOSS and allows us to naturally include both centrals and satellites in the selected simulation population.

\subsection{Galaxy-galaxy lensing profiles}
\label{sec:delta_sigma_profs}

A nice property of galaxy-galaxy lensing $\Delta \Sigma$ profiles is that they do not depend on the source redshift distribution of the background galaxies used to measure the profiles.  This is in contrast to measurements of cosmic shear.  It is thus relatively straightforward to compute predictions from the simulations, using the mass distribution around haloes.  In particular, we start from the \healpix\ total mass maps, which provide the sum of all of the mass components (gas, DM, stars, BHs, and neutrinos) in pixels in a given redshift shell.  We convert the total mass maps to maps of comoving surface mass density by converting the pixel area (in steradians) into a comoving surface area using the comoving radial distance to the shell centre from the observer and then dividing the total mass maps by this comoving surface area.  For a given galaxy included in our selection, we select all the pixels within a certain angular distance which, following \citet{Amon2023}, we convert to comoving transverse distances assuming a flat cosmology with $\Omega_\textrm{m}=0.3$ and $h=0.7$.  The $\Delta \Sigma$ profile of a given galaxy is calculated by ordering the pixels by projected comoving distance and then subtracting the surface mass density of a given pixel from the mean surface mass density from all of the pixels interior to it\footnote{A convenient feature of the \healpix\ maps is that the pixels are of equal area, which implies that the mean surface mass density interior to a given pixel can be simply estimated as the mean value of all interior pixels.  For non-equal area pixels, a more cumbersome route of computing a $\Sigma$ profile, integrating it to the angular radius in question, and then dividing it by the enclosed surface area, is required.}.

We compute stacked $\Delta \Sigma$ profiles by defining a set of projected comoving radial bins (logarithmically spaced) and computing the mean $\Delta \Sigma$ in those bins by simply summing the profiles of each galaxy in those bins and dividing by the number of galaxies in the stack.  The centre of the bin is not the midpoint between the upper and lower bin bounds but is computed as the $\Delta \Sigma$-weighted radius of all pixels that fall within the radial bin.  We have found that such a weighting scheme is more robust to changes in the radial binning strategy.

The lensing profiles of \citet{Amon2023} span a wide radial range ($R\sim 0.1-100$ Mpc/$h$), and extracting the pixels for large numbers of haloes from our fiducial high-resolution \healpix\ ($N_\text{side} = 16384$) maps is computationally expensive.  We therefore adopt a hybrid approach where we use the high-resolution map to extract the signal on small comoving scales ($R < 2$ Mpc/$h$) and a downsampled version with with $N_\text{side} = 2048$ to retrieve the signal on large scales.  We have verified that there is excellent convergence between the different resolution maps on intermediate scales.  Furthermore, we have parallelised the analysis, allowing us to produce lensing profiles for large numbers of haloes at the same time. 

\subsection{kSZ effect profiles}
\label{sec:ksz_profs}

We construct $\Delta T_{\rm kSZ}$ effect maps and profiles as follows.  As described in \citet{Schaye2023}, when a gas particle crosses the lightcone, we compute its dimensionless Doppler B, $b$, parameter:
\begin{equation}
b = \frac{n_\textrm{e} m_\textrm{g} \sigma_\textrm{T} v_\textrm{r}}{\Omega_\text{pixel} d_\text{A}^2 \rho c} ,
\label{eq:doppler_b}
\end{equation}
where $m_\textrm{g}$, $\rho$, and $v_\textrm{r}$ are the mass, mass density, and radial velocity of the particle, respectively, $\Omega_\text{pixel}$ is the solid angle of a \healpix pixel and $d_\text{A}$ is the angular diameter distance to the observer.  Particles crossing the lightcone in a given redshift shell are accumulated to the corresponding \healpix map.  

Visual inspection of eqn.~\ref{eq:doppler_b} reveals a close relation to eqn.~\ref{eq:kSZdef}.  Indeed, the quantity $m_g / [\Omega_\text{pixel} \ d_\text{A}^2 \ \rho]$ in eqn.~\ref{eq:doppler_b} is a discretised (per particle) estimate of the physical path length $\chi/(1+z)$, noting that $m_g / \rho$ is the volume associated with the particle and $\Omega_\text{pixel} d_\text{A}^2$ is the physical area of the pixel in which the particle is deposited at the distance of the particle.  Aside from this discretisation difference, the mapping between the Doppler B is simple: $\Delta T_{\rm kSZ}/T_{\rm CMB} = -b$, which is independent of observing frequency.

Thus, the production of a $\Delta T_{\rm kSZ}$ map from the \flamingo\ Doppler B maps is trivial, requiring only the multiplication of a factor of $-T_{\rm CMB}$ and the summation of the individual maps (shells) along the line of sight.  In practice, to achieve convergent results for the stacked kSZ effect profiles, we find that stacking along the full line of sight ($z=0$ to $3$) of the lightcones is unnecessary, since most of the line of sight will be uncorrelated with the selected galaxy in the stack.  Indeed, we find that using only three shells (each having width $\Delta z = 0.05$, i.e., a padding shell on each side of the shell that contains the galaxy) is sufficient to achieve convergent results.  The same holds true for the $\Delta \Sigma$ profiles.

As noted previously, we downsample our full resolution $N_\text{side} = 16384$ kSZ effect maps to $N_\text{side} = 8192$ so that we can then smooth the maps with ACT beam.  \citet{Schaan2021} used coadded ACT DR5 (day+night) maps in their study.  We therefore retrieve the corresponding measured ACT beam for this dataset from the NASA Lambda website\footnote{\url{https://lambda.gsfc.nasa.gov/product/act/actpol_dr5_coadd_maps_info.html}} and convolve it with the simulated maps in multipole space using the \textsc{Healpy} function sphtfunc.almxfl.

Apart from convolution with a realistic beam, our maps are idealised, in that they do not contain realistic noise.  We further assume that the kSZ effect signal can be perfectly recovered in the observational measurements, which would appear to be a strong assumption in the face of significant foreground and background contaminant signals, including the primary CMB, the tSZ effect, radio sources, the CIB, and so forth.  However, in general these sources of contamination are not expected to correlate with the velocity field of the selected galaxies and are therefore suppressed in the velocity stacking process (see discussion in \citealt{Schaan2021}).  Nevertheless, some contamination may still arise from the sources that are truly correlated with the selected haloes (i.e., part of the 2-halo term) and it will be important to evaluate their potential effects by constructing realistic mocks from the hydro simulations that contain all relevant signals.  We leave this for future work (T.~Yang, in prep).

We extract and stack the kSZ effect profiles using the same methodology as applied to the genuine \textit{Planck} + ACT data (i.e., eqns.~\ref{eq:ap}-\ref{eq:kSZ_est}) using the true radial velocities of the galaxies in the lightcone (i.e., we set $r_v = 1$).  Also, as our simulated kSZ effect maps are noiseless, the inverse-variance weighting applied for the observations is not applicable for the simulations.  We therefore set $\sigma_i=1$ in eqn.~\ref{eq:kSZ_est} when stacking the simulated kSZ effect profiles.

\begin{figure*}
    \includegraphics[width=\columnwidth]{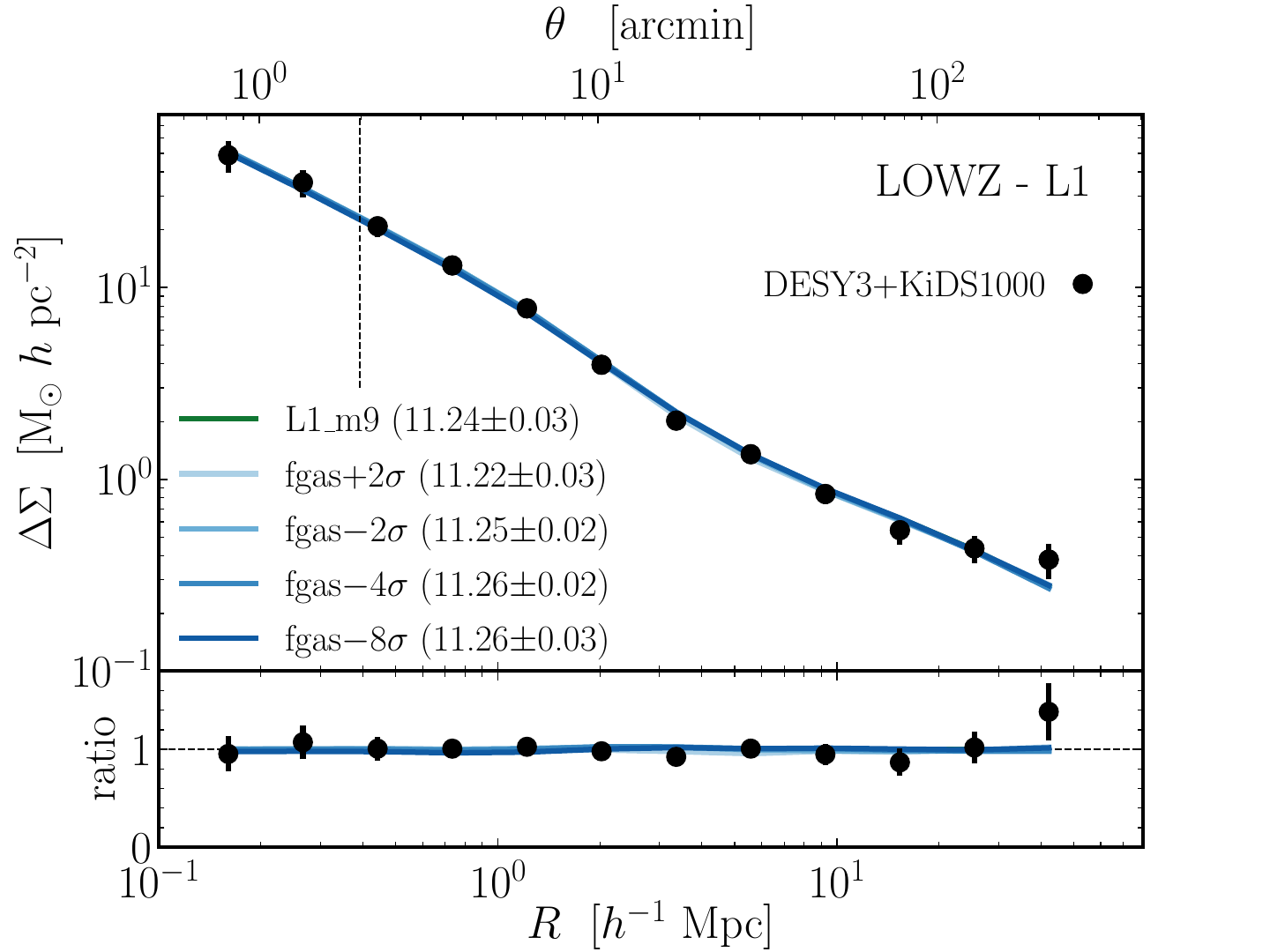}
    \includegraphics[width=\columnwidth]{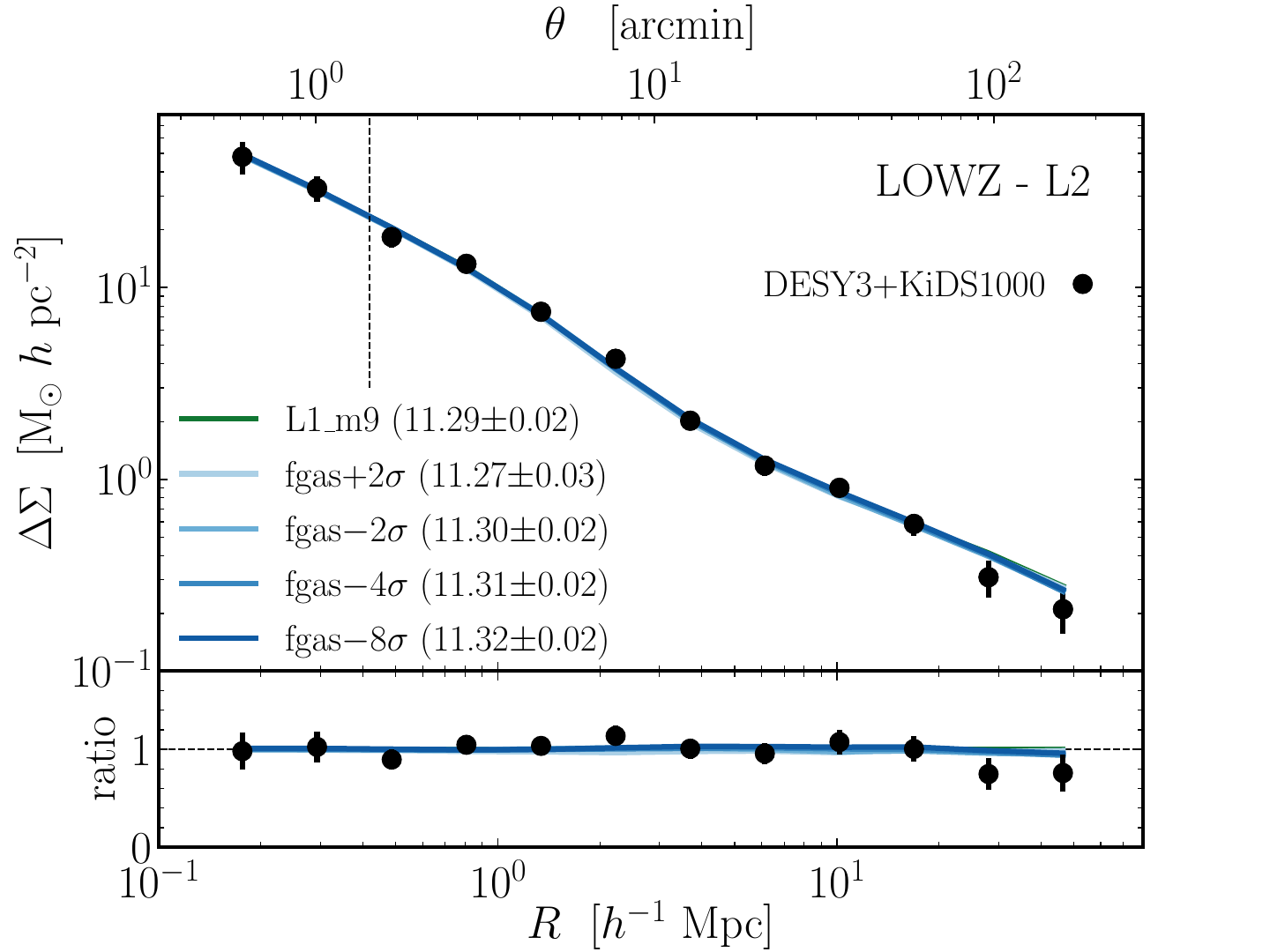}\\
    \includegraphics[width=\columnwidth]{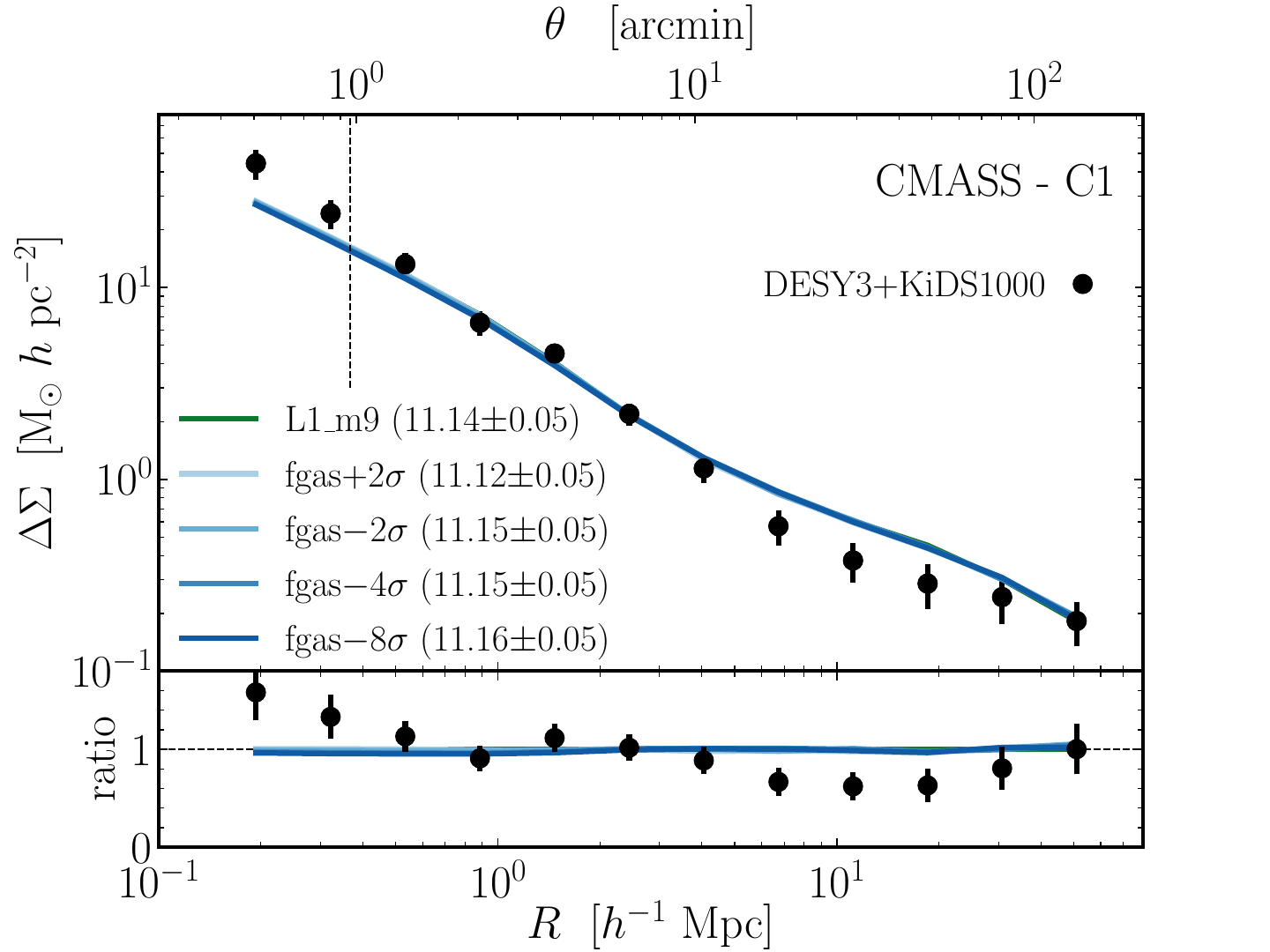}
    \includegraphics[width=\columnwidth]{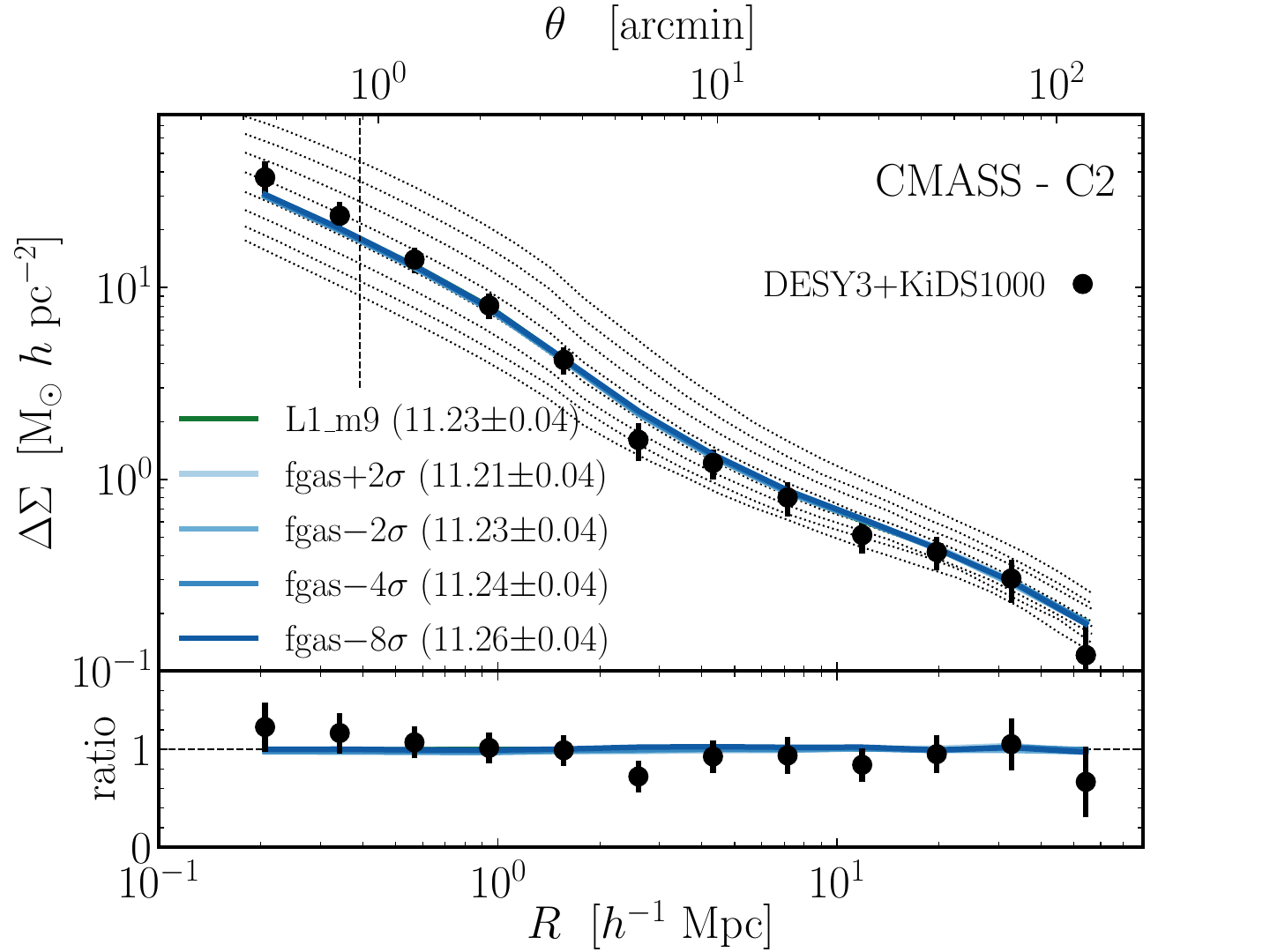}\\    
    \caption{Comparison of the \citet{Amon2023} DES Y3+KiDS 1000 galaxy-galaxy lensing $\Delta \Sigma$ profiles of BOSS LOWZ and CMASS galaxies (data points with $1\sigma$ error bars) with predictions from the fiducial \flamingo\ run (L1\_m9) and the fgas feedback variants.  There are four main panels corresponding to the four BOSS samples (two redshift bins each for LOWZ and CMASS).  The bottom x-axis shows the the comoving projected radius while the top x-axis shows the angular scale (in arcmins) at the mean redshift of each of the redshift bins.  The vertical dashed lines indicate the location of $R_\textrm{500c}$ derived from the mean halo mass of each redshift bin (see Section \ref{sec:mhalo_fgas}).  The solid curves show the best-fitting profiles for each of the \flamingo\ runs, which fall nearly on top of each other.  The legend provides the best-fitting ($\log_{10}$) minimum stellar mass (in solar units) of the simulated galaxy selection for each of the runs, with the error bars reflecting the $2\sigma$ uncertainties (95\% confidence) on the best-fitting minimum mass given the uncertainties on the lensing measurements.  The bottom sub-panels in each of the main panels show the ratio of the fgas feedback variants and the observational data with respect to the fiducial feedback model.  For reference, the dotted curves in the bottom right panel show the predicted $\Delta \Sigma$ profiles for the fiducial \flamingo\ run with minimum stellar mass $\log_{10}[M_\text{star}/\textrm{M}_\odot]$ values ranging between 10.9 and 11.6 in steps of 0.1.  Overall, the simulations reproduce the lensing measurements very well over a wide range of radii, with a simple stellar mass cut (per feedback variant) being sufficient to match the LOWZ ($\log_{10}[M_\text{star}/\textrm{M}_\odot] \approx 11.3$) and CMASS ($\log_{10}[M_\text{star}/\textrm{M}_\odot] \approx 11.2$) measurements.}
    \label{fig:gg_lensing_fgas}
\end{figure*}

\section{Results}
\label{sec:results}

In this section, we explore the feedback and cosmology dependencies of the kSZ effect predictions (Section \ref{sec:feedback_dep}), as well as the role that satellite galaxies play (Section \ref{sec:satellite_dep}).  As already described, we optimise the selection of galaxies from the simulations by fitting to the galaxy-galaxy lensing profiles of BOSS galaxies, thus ensuring that our selection has the correct underlying mean halo mass.  In Section \ref{sec:mhalo_fgas} we discuss the implied halo masses and radii of the BOSS samples.

\subsection{Dependence on feedback and cosmology}
\label{sec:feedback_dep}

Fig.~\ref{fig:gg_lensing_fgas} shows the best-fitting galaxy-galaxy lensing profiles for the comoving excess surface mass density, $\Delta \Sigma$, for the fiducial \flamingo\ run (L1\_m9) and the fgas feedback variants.  There are four panels corresponding to the four BOSS redshift bins from \citet{Amon2023} (two bins each for LOWZ and CMASS).  We show measurement as a function of the comoving projected radius, $R$, on the bottom x-axis, but we also show the angular scale at the median redshift of the sample on the top x-axis, in order to facilitate comparisons with the kSZ effect measurements.  

The solid curves show the best-fitting profiles for each of the \flamingo\ runs, which fall nearly on top of each other.  The legend provides the best-fitting ($\log_{10}$) minimum stellar mass (in solar masses) of the simulated galaxy selection for each of the runs.  For example, in the LOWZ-L1 bin, the \flamingo\ fiducial feedback run with a minimum stellar mass of $\log_{10}[M_\text{star}/\textrm{M}_\odot] = 11.24 \pm 0.03$ provides the best fit to the DES Y3 + KiDS 1000 $\Delta \Sigma$ measurements.  The error bars reflect the $2\sigma$ uncertainties (95\% confidence) on the best-fitting minimum mass given the uncertainties on the lensing measurements.  The best-fitting minimum stellar mass and its uncertainties are estimated by first calculating the $\Delta \Sigma$ profiles at different values of the minimum stellar mass and then interpolating to obtain the best fit result.  The bottom right panel of Fig.~\ref{fig:gg_lensing_fgas} shows the process, where the dotted curves correspond to the predicted $\Delta \Sigma$ profiles for the fiducial \flamingo\ run with $\log_{10}[M_\text{star}/\textrm{M}_\odot]$ values ranging between 10.9 and 11.6 in steps of 0.1 dex.  We then interpolate the $\Delta \Sigma$ values to a much finer grid of $\log_{10}{M_\text{star}}$ 
Specifically, we linearly interpolate $\log_{10} \Delta \Sigma$ in stellar mass bins of width 0.01 dex at projected radius, $R$.  The best-fitting minimum stellar mass is determined through $\chi^2$ minimisation with respect to the DES Y3 + KiDS1000 measurements and their uncertainties from \citet{Amon2023}.  Note that \citet{Amon2023} quote diagonal uncertainties for their combined DES Y3 + KiDS 1000 measurements only; i.e., the radial bins are assumed to be uncorrelated.  Calculation of the off-diagonal elements of covariance matrix for the combined dataset is non-trivial and beyond the scope of this work.

It is interesting that a simple stellar mass cut applied to the various \flamingo\ runs is capable of yielding excellent fits to the lensing measurements over approximately 2.5 decades in radius, spanning both the 1-halo and 2-halo regimes (the transition between the two regimes is clearly visible at $R\approx3-4$ Mpc/$h$). Varying the stellar mass has the impact of varying the mean halo mass of the simulated galaxy sample which affects the amplitude of the predictions.  Thus, the fact that the simulations reproduce the amplitude of the observed galaxy-galaxy lensing signal is not surprising.  But the \textit{shape} of the profile is a genuine prediction of the simulations and $\Lambda$CDM generally, and the fact that the profiles accurately match these precise measurements over a very large range of radii is remarkable.  Note that the detailed shape is expected to be cosmology dependent in $\Lambda$CDM, since the 1-halo and 2-halo terms themselves have different cosmology dependencies.  The main cosmological dependence of the 1-halo regime is through the halo concentration (e.g., \citealt{Bullock2001,Eke2001,Correa2015,Diemer2015,Brown2022}), whereas at large radii (2-halo) it is via the halo bias (e.g., \citealt{Sheth1999,Tinker2010}).

In a given redshift bin (e.g., LOWZ-L1), all of the runs yield similarly good fits to the data and the different runs prefer only slightly different stellar masses.  This is not unexpected, since the fiducial \flamingo\ run and the fgas feedback variants in Fig.~\ref{fig:gg_lensing_fgas} have each been independently calibrated to reproduce the local galaxy stellar mass function.  Thus, the mapping between stellar mass and halo mass is expected to be nearly the same in each of the runs.  The fact that the stronger feedback variants prefer a slightly higher stellar mass cut is likely because the halo masses themselves have been reduced slightly more through baryon ejection in the stronger feedback variants.  Thus, to get back to the same mean halo mass required to match the lensing data, a slightly higher stellar mass cut is required in the models with stronger feedback.  If the measurements could be extended to smaller projected radii ($R < 0.1$ Mpc/$h$), it is possible that the lensing measurements themselves could be used to place constraints on the feedback models, through deviations in the profile shapes on small scales (e.g., fig.~6 of \citealt{Velliscig2014}). In the present study, the lensing measurements are used to constrain the galaxy selection, so that more sensitive kSZ effect measurements can be used to discriminate between the different feedback models.

\begin{figure*}
    \includegraphics[width=\columnwidth]{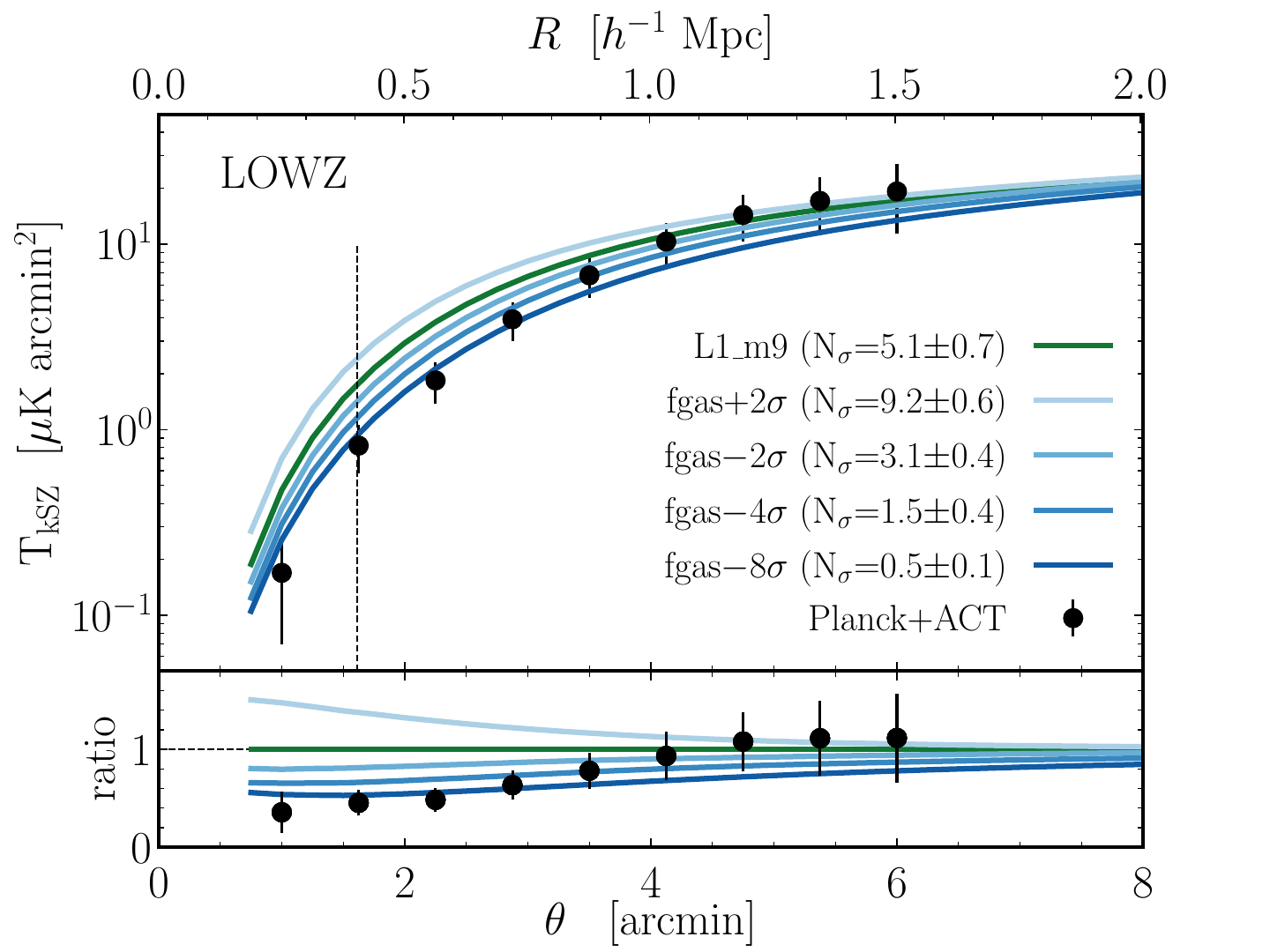}
    \includegraphics[width=\columnwidth]{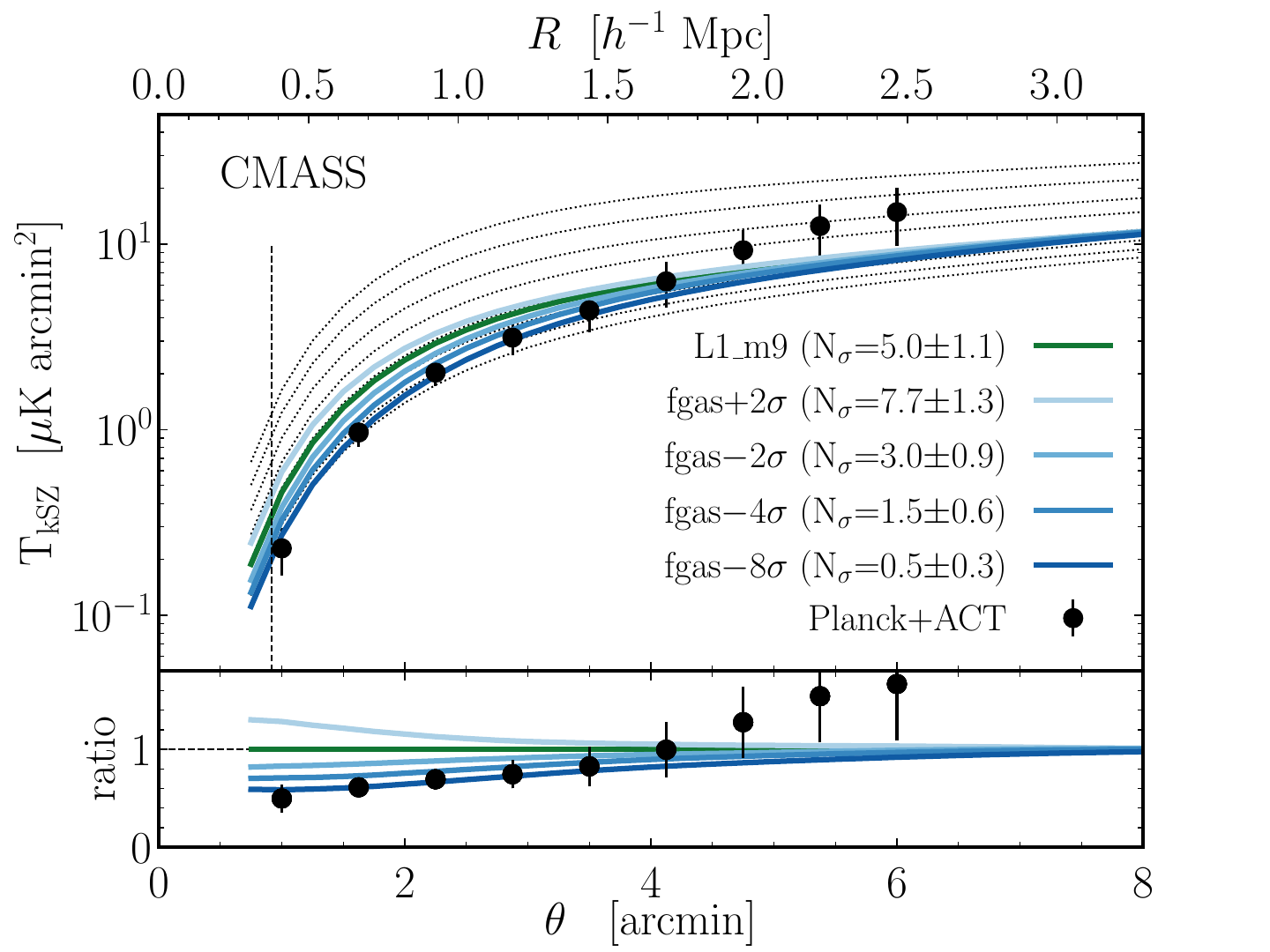}
    \caption{Comparison of the 150 GHz \textit{Planck}+ACT kSZ effect temperature profiles of BOSS LOWZ and CMASS galaxies from \citet{Schaan2021} (data points with $1\sigma$ error bars) with predictions from the fiducial \flamingo\ run (L1\_m9) and the fgas feedback variants.  The bottom x-axis shows the angular scale (in arcmins) while the top x-axis shows the comoving projected radius at $z=0.31$ and $z=0.54$ for the LOWZ and CMASS panels, respectively.  The vertical dashed lines indicate the location of $R_\textrm{500c}$ derived from the mean halo mass of each redshift bin (see Section \ref{sec:mhalo_fgas}).
    The solid curves represent the predictions for the best-fitting minimum stellar mass (fitted to the lensing, see Fig.~\ref{fig:gg_lensing_fgas}). The legend provides the number of standard deviations that the model deviates from the observational measurements with the error bars corresponding to the propagated uncertainties on the stellar mass cut given the uncertainties on the galaxy-galaxy lensing measurements.  Note that the number of standard deviations is computed using the full covariance matrices of the data, which is important since the outermost bins are highly correlated.  The error bars on the data points reflect only the diagonal elements of the covariance matrices.  The bottom sub-panels in each of the main panels show the ratio of the fgas feedback variants and the observational data with respect to the fiducial  model.  For reference, the dotted curves in the right panel show the predicted $\Delta T_{\rm kSZ}$ profiles for the fiducial \flamingo\ run with minimum stellar mass $\log_{10}[M_\text{star}/\textrm{M}_\odot]$ values ranging between 10.9 and 11.6 in steps of 0.1. The fiducial calibrated \flamingo\ run is statistically ruled out by the kSZ effect measurements at about the $5\sigma$ level, for both the LOWZ and CMASS samples which are independent.  Only the two strongest feedback models in this comparison (fgas-$4\sigma$, fgas-$8\sigma$) are formally consistent with the measurements.}
    \label{fig:kSZ_fgas}
\end{figure*}

Comparing the fits across the two LOWZ bins (top two panels), for the majority of the cases the preferred minimum stellar masses are consistent within a few sigma for a given run.  For example, the fiducial \flamingo\ model prefers a minimum stellar mass of $\log_{10}[M_\text{star}/\textrm{M}_\odot] = 11.24 \pm 0.03$ for the LOWZ-L1 bin and $\log_{10}[M_\text{star}/\textrm{M}_\odot] = 11.29 \pm 0.02$ for the LOWZ-L2 bin.  As the two LOWZ bins are consistent with a single LOWZ selection for a given model, we jointly fit the LOWZ L1 and L2 bins to determine the minimum stellar mass cut for the kSZ effect predictions.  The same is true for the higher redshift CMASS bins (C1 and C2, with $\log_{10}[M_\text{star}/\textrm{M}_\odot] = 11.14 \pm 0.05$ and $\log_{10}[M_\text{star}/\textrm{M}_\odot] = 11.23 \pm 0.04$, respectively), although we note that for the CMASS-C1 bin the shape of the best-fitting simulated profiles do not match the measurements perfectly.  In particular, we note that the 1-halo regime dominates the fit (in terms of signal to noise), and the simulations provide a good match to the data there, but they predict a signal that is too large compared to the measurements at larger radii, in the 2-halo regime.  The CMASS-C1 bin was also identified as an outlier in \citet{Amon2023}, who found that even their flexible HOD framework (with 5 free parameters) was unable to reproduce the lensing measurements in detail.  Nevertheless, the best-fitting stellar masses are consistent within a few sigma between the C1 and C2 bins for all the runs and we therefore jointly fit them to derive a single stellar mass cut (for each feedback variant) for the kSZ effect CMASS analysis.  Furthermore, we highlight that the kSZ effect measurements mostly probe relatively small radii, in the 1-halo regime, where the simulation predictions match the CMASS lensing measurements well even for the C1 bin.

An alternative to approach to handling the two redshift bins in the LOWZ and CMASS samples of \citet{Amon2023} would be to predict the kSZ effect profiles for each bin given their respective best-fit minimum stellar masses and then to average the profiles (e.g., by inverse-variance weighting).  In practice, we find that this gives nearly identical results to our default method of combining the bins.  Indeed, even if we uniformly applied the lower or higher of the two stellar mass estimates to the whole sample, our general conclusions with regards to the strength of feedback required to match the kSZ effect measurements would be unchanged.

While we have elected to constrain our simulated galaxy selection based on the galaxy-galaxy lensing signal rather than a more observable quantity such as stellar mass, it is interesting to note that our best-fitting minimum stellar masses, which range from $\log_{10}[M_\text{star}/\textrm{M}_\odot] \approx 11.15$ to $11.3$, align remarkably well with the actual observed peak and mean stellar masses of the BOSS LOWZ and CMASS samples.  In particular, \citet{Maraston2013} find mean $\log_{10}[M_\text{star}/\textrm{M}_\odot]$ values of 11.33 at $0.2 \la z \la 0.4$ (LOWZ), 11.27 at $0.4 \la z \la 0.5$ (CMASS), and 11.26 at $0.5 \la z \la 0.6$ (CMASS).  Thus, the calibrated \flamingo\ runs have realistic stellar mass to halo mass ratios at the typical stellar mass scale probed by the BOSS survey.

Armed with strong constraints on the galaxy selection (minimum stellar mass) from the galaxy-galaxy lensing comparisons, we compare the kSZ effect profiles predicted by the fiducial \flamingo\ run and the fgas feedback variants with the 150 GHz stacking measurements of \citet{Schaan2021} in Fig.~\ref{fig:kSZ_fgas}.  The solid curves represent the predictions for the best-fitting minimum stellar masses.  The legend provides the number of standard deviations that the model deviates from the observational measurements with the error bars corresponding to the propagated uncertainties on the stellar mass cut given the uncertainties on the galaxy-galaxy lensing measurements.  Note that the number of standard deviations is computed using the full covariance matrices of the data, which is important since the outermost bins are highly correlated.  Taking into account this correlation, the strongest constraint on the goodness of fit comes from the innermost three or four radial bins.

We see visually from Fig.~\ref{fig:kSZ_fgas} and from the computed number of standard deviations that the fiducial calibrated \flamingo\ run is statistically ruled out by the kSZ effect measurements, at $\approx$5 sigma, for both the LOWZ and CMASS samples, which are independent.  Only the two strongest feedback models in this comparison, fgas-$4\sigma$ and fgas-$8\sigma$, are formally consistent with the measurements, with $\approx1.5\sigma$ and $\approx0.5\sigma$ deviations from the measurements, respectively. Note that in computing the number of standard deviations we have not marginalised over uncertainties in cosmology (which we expect to be small, as discussed below) or uncertainties in the velocity reconstruction used in the observational measurements.  Thus, the quoted level of tension may be slightly overestimated.

Note that because the kSZ effect is proportional to the gas mass, it will be affected by any physical process that alters the gas mass fractions of groups and clusters.  Ejection of gas due to AGN feedback is believed to be the main mechanism for altering the gas fractions, but gas is also removed via radiative cooling leading to neutral gas and star formation.  An important aspect of the \flamingo\ simulations shown in Figs.~\ref{fig:gg_lensing_fgas} and \ref{fig:kSZ_fgas} is that they have all been calibrated to reproduce the observed galaxy stellar mass function.  Thus, the differences between the models in Fig.~\ref{fig:kSZ_fgas} are due entirely to differences in the level of gas ejection, and the relatively low observed kSZ signal indicates that a relatively high level of ejection is required.  A caveat, of course, is if there are significant biases in the observed stellar masses (e.g., due to uncertainties in stellar population synthesis modelling or the extrapolation of surface brightness profiles) this would in turn bias the estimates of the required level of feedback.

Our results are qualitatively consistent with \citet{Bigwood2024}, who used the baryonification formalism to model the kSZ effect jointly with cosmic shear, in the sense that the kSZ effect data appears to prefer stronger feedback relative to that inferred by modelling X-ray-based baryon fractions of galaxy groups. However, given the use of self-consistent full cosmological hydrodynamical simulations, galaxy-galaxy lensing to strongly pin down the mass scale of the BOSS samples, inclusion of satellites and mis-centring effects, and the use of both the LOWZ and CMASS samples, our quantitative results are more robust.

\begin{figure*}
    \includegraphics[width=\columnwidth]{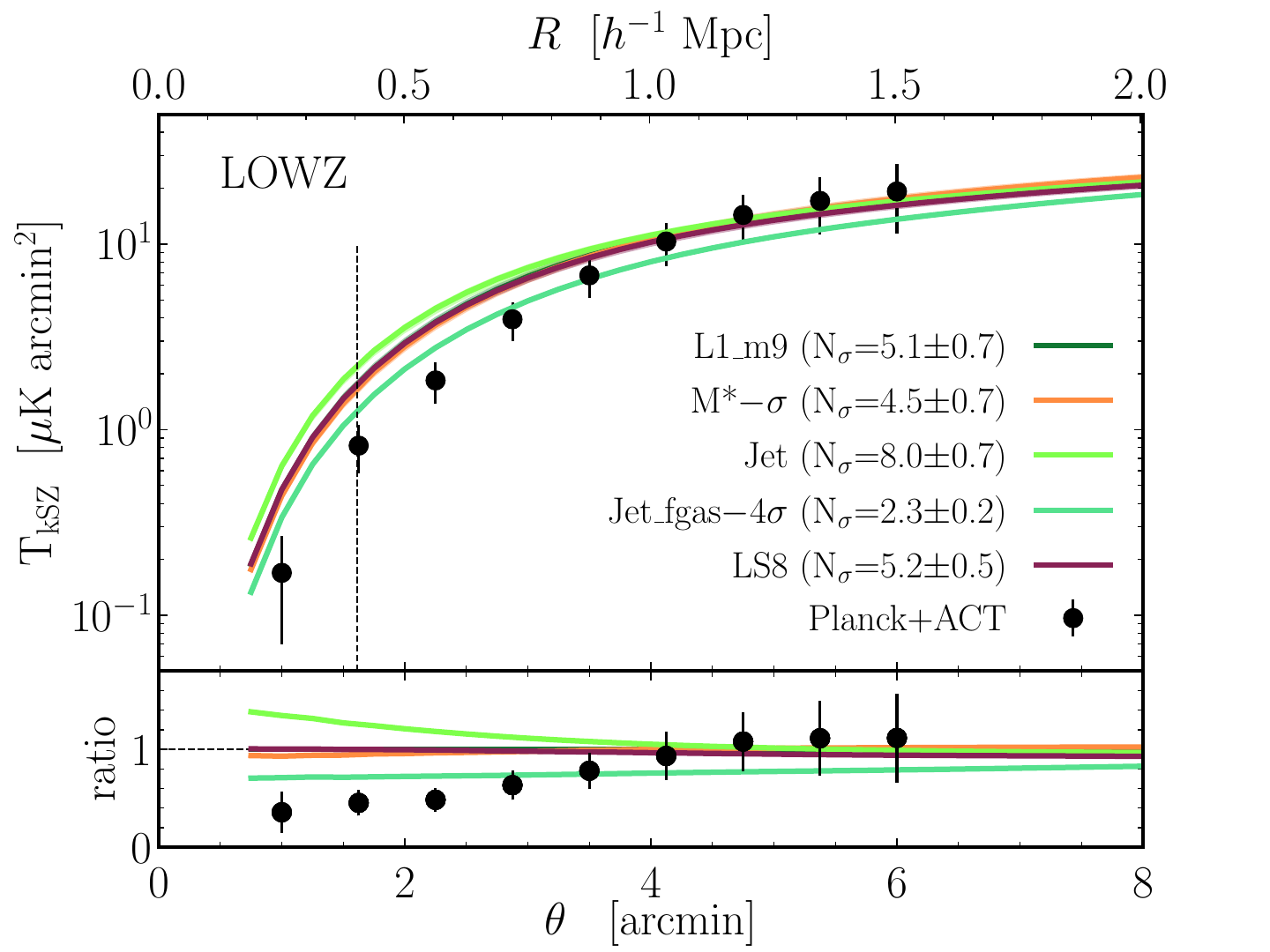}
    \includegraphics[width=\columnwidth]{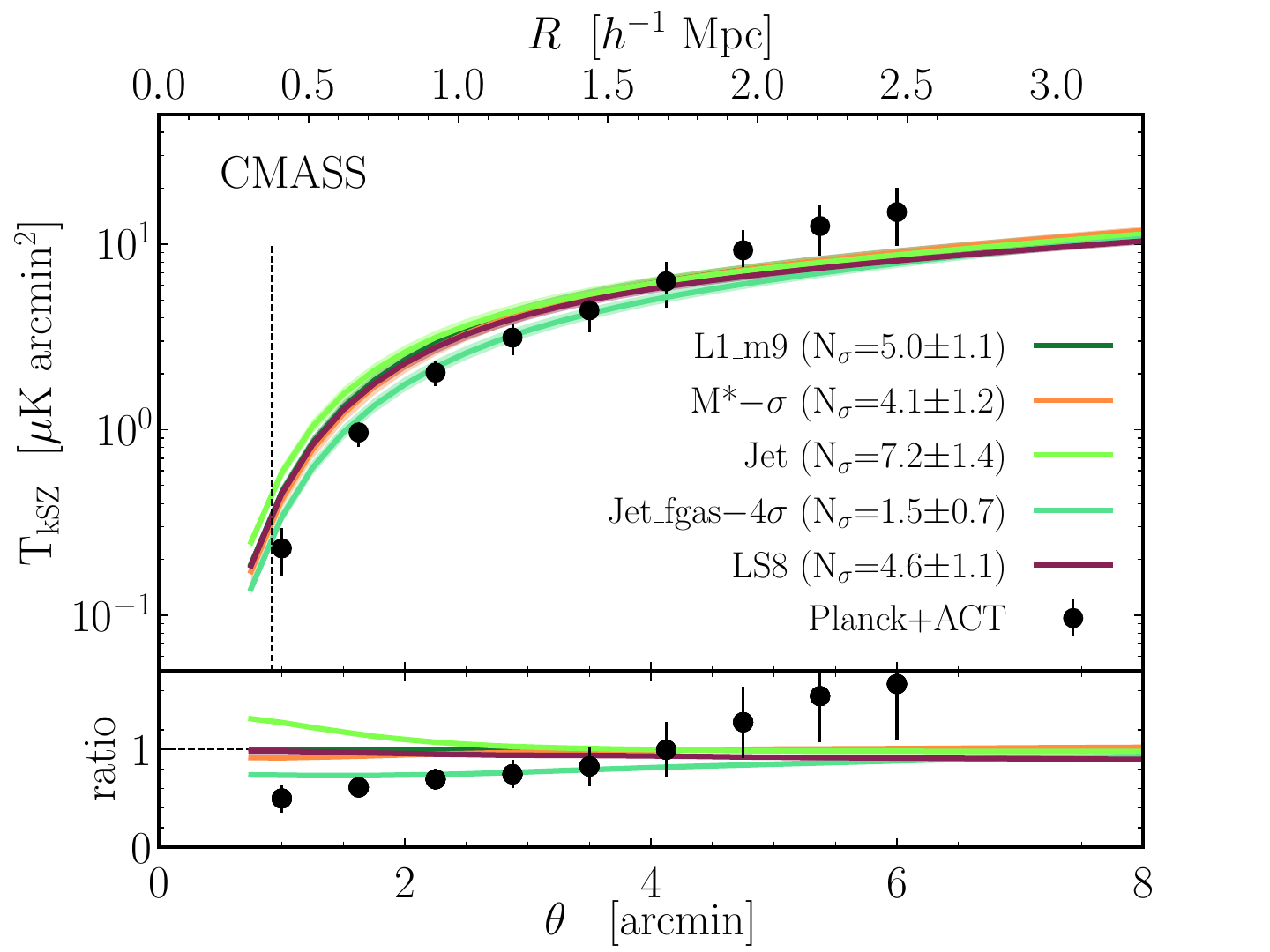}
    \caption{As Fig.~\ref{fig:kSZ_fgas}, but showing the dependence of the predicted kSZ effect profiles on other model variations, namely variations in the galaxy stellar mass function and the fiducial and strong jet models of AGN feedback, as well as the lensing (LS8) cosmology run with fiducial feedback.  The solid coloured curves correspond to the different \flamingo\ simulations, with the number of standard deviations from the data provided in the legend.  The best performing model shown is the Jet model with reduced gas fractions (Jet\_fgas$-4\sigma$), which is consistent with the findings from Fig.~\ref{fig:kSZ_fgas} that enhanced gas ejection is required by the measurements.}
    \label{fig:kSZ_other_var}
\end{figure*}

The dotted curves in the right panel of Fig.~\ref{fig:kSZ_fgas} illustrate the importance of the galaxy-galaxy lensing measurements in pinning down the mass scale of the BOSS galaxies and therefore the required level of feedback.  These correspond to the same variations in the minimum stellar mass as shown in the bottom right panel of Fig.~\ref{fig:gg_lensing_fgas} for the fiducial feedback model.  With the kSZ effect data alone to go by, we would not be able to easily distinguish the fiducial feedback scenario with a lower minimum stellar mass, of $\log_{10}[M_\text{star}/\textrm{M}_\odot] \approx 10.9$, from a stronger feedback scenario with higher stellar mass (noting that it is the innermost 3 or 4 bins which dominate the fit).  The lensing data are therefore crucial to break the degeneracy between halo mass and feedback-driven gas ejection.

We note that the measurements of \citet{Schaan2021} extend out to $\approx 2$ Mpc/$h$ which is the scale where the 2-halo term becomes visible in the galaxy-galaxy lensing measurements in Fig.~\ref{fig:gg_lensing_fgas}.  The 2-halo term is easily visible in the lensing measurements given their precision, the large dynamic range of the measurements, and that the signal is a differential one.  In the case of the kSZ profiles, the measurement is effectively a cumulative signal which will partially mask the transition to the 2-halo regime.  Nevertheless, the correlated clustering of nearby haloes is expected to modestly contribute to the outermost bins in these measurements (see also \citealt{Amodeo2021}) and the profiles should therefore not be regarded as being due solely to the selected galaxies in the stack.

As discussed in Section \ref{sec:boss_sel}, observational stellar masses have measurement uncertainties and it is interesting to ask what the effect of such uncertainties might be if we applied them to the simulated galaxies and repeated the above analysis.  \citet{Behroozi2019} find that a log-normal scatter with a standard deviation $\sigma(\log_{10} M_\text{star}) = $min$(0.070 + 0.071z, 0.3)$ dex describes typical random measurement errors in the observed stellar masses.  As a test, we have applied this scatter to the true simulated stellar masses to mimic an observed stellar mass (note that we also applied this scatter during the calibration of \flamingo; see \citealt{Schaye2023}).  We then analysed the simulations as described above, by determining the best-fit minimum stellar mass cut required to match the galaxy-galaxy lensing measurements and then predicting the kSZ profiles for this selection.

We find that the best-fit minimum stellar mass required to match the galaxy-galaxy lensing measurements typically increases by 0.04-0.05 dex with respect to our fiducial analysis with no measurement scatter.  This reflects the fact there are more lower mass objects than high mass objects, thus leading to a slight net up-scattering.  To reproduce the galaxy-galaxy lensing signal in the presence of lower stellar mass galaxies entering the selection, a slightly higher minimum stellar mass cut is therefore required to match the stacked lensing profiles, compared to the case with no scatter.  Nevertheless, the best-fit mean halo mass is virtually identical to the no-scatter case, as is the predicted stacked kSZ profile.  We therefore conclude that measurement uncertainties in the observed stellar masses do not significantly impact our results or conclusions, by virtue of the fact that the selection is constrained to match the lensing signal.

We have also tested the sensitivity of our results and conclusions to numerical resolution, by analysing the high-resolution calibrated \flamingo\ model (L1\_m8) and comparing the results with the fiducial resolution, calibrated model (L1\_m9).  This comparison is presented in Appendix \ref{sec:res_dependence}.  In short we find that the high-resolution run displays a similar level of tension with the observed kSZ effect measurements of \citet{Schaan2021} compared to the fiducial resolution, suggesting that our conclusions are robust to changes in resolution.

In Fig.~\ref{fig:kSZ_other_var} we explore kSZ effect predictions using several other feedback variations in the \flamingo\ suite at the fiducial resolution, namely variations in the galaxy stellar mass function at the fiducial gas fraction (M*$-\sigma$) and the fiducial and strong jet models of AGN feedback (Jet and Jet\_fgas$-4\sigma$). The corresponding galaxy-galaxy lensing profiles are shown in Appendix \ref{sec:galgal_other} (see Fig.~\ref{fig:gg_lensing_other_var}). We find that none of the models provide an acceptable fit to the data. The best performing model is the jet model with reduced gas fractions (Jet\_fgas$-4\sigma$), which is consistent with the findings in Fig.~\ref{fig:kSZ_fgas} that enhanced gas ejection is required by the measurements. The lensing LS8 cosmology run, which gives a similarly poor fit to the measurements as the fiducial L1\_m9 run.  This is likely because both models adopt the same feedback model and that, intrinsically, the kSZ effect profile is not expected to depend significantly on cosmology.  The underlying matter profile depends weakly on cosmology through the dependence of the concentration on cosmology, but this dependence will likely be even weaker when dealing with the hot gas due to the effects of non-gravitational processes such as cooling and feedback.  The kSZ effect would be expected to scale with the universal baryon fraction, $f_\textrm{b} \equiv \Omega_\textrm{b}/\Omega_\textrm{m}$, as this quantity dictates the fraction of baryons that haloes can accrete (at least in the absence of feedback).  However, $f_\textrm{b}$ is precisely determined from several cosmological probes and all of our runs have very similar values of $f_\textrm{b}$.  Lastly, we have also analysed the \flamingo\ runs that vary the summed neutrino mass (which also adopt the fiducial feedback model) and find that they, too, yield similarly poor fits to the kSZ effect measurements, but we do not show them here for brevity.

\begin{figure*}
    \includegraphics[width=0.9\columnwidth]{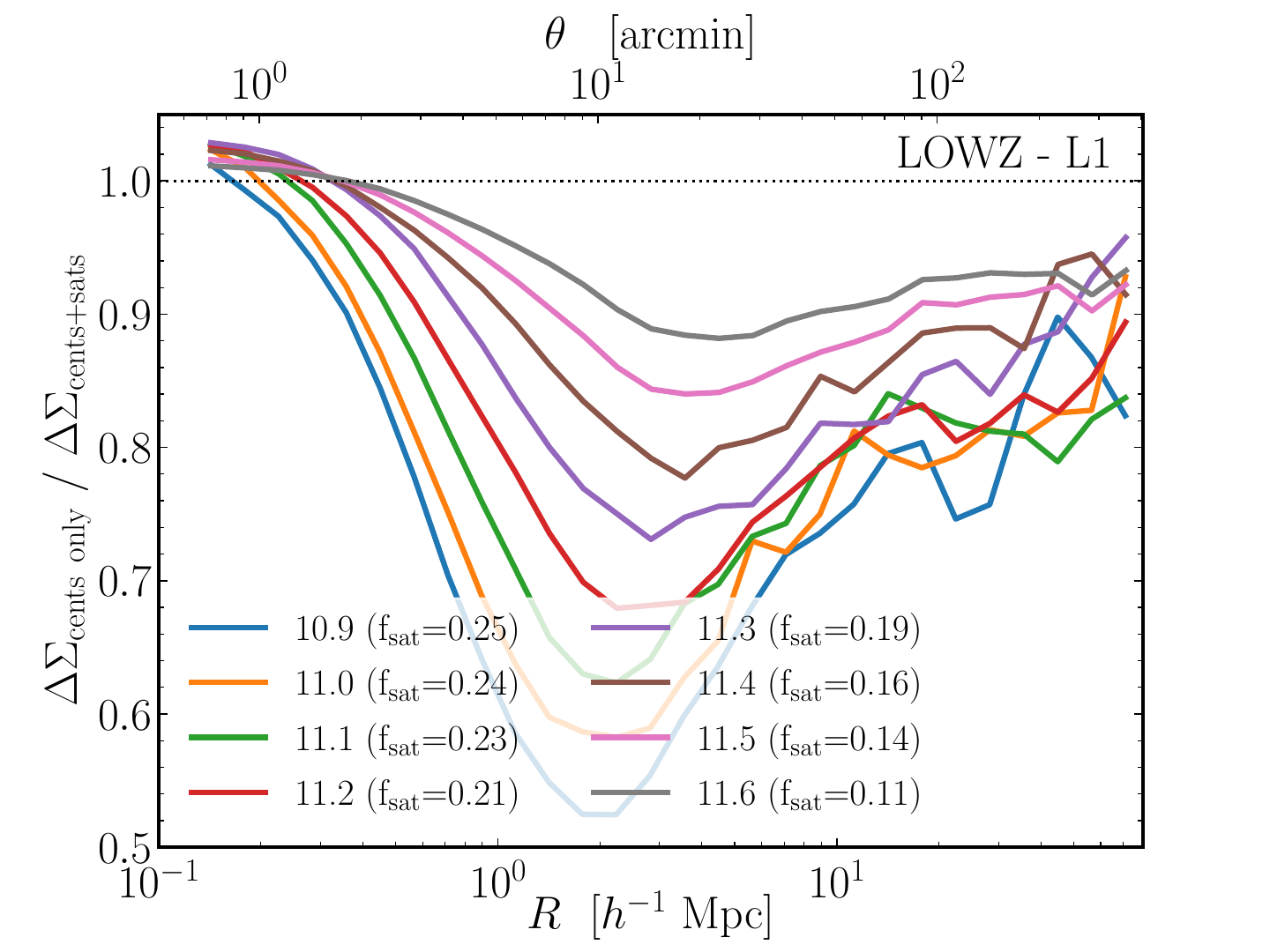}
    \includegraphics[width=0.9\columnwidth]{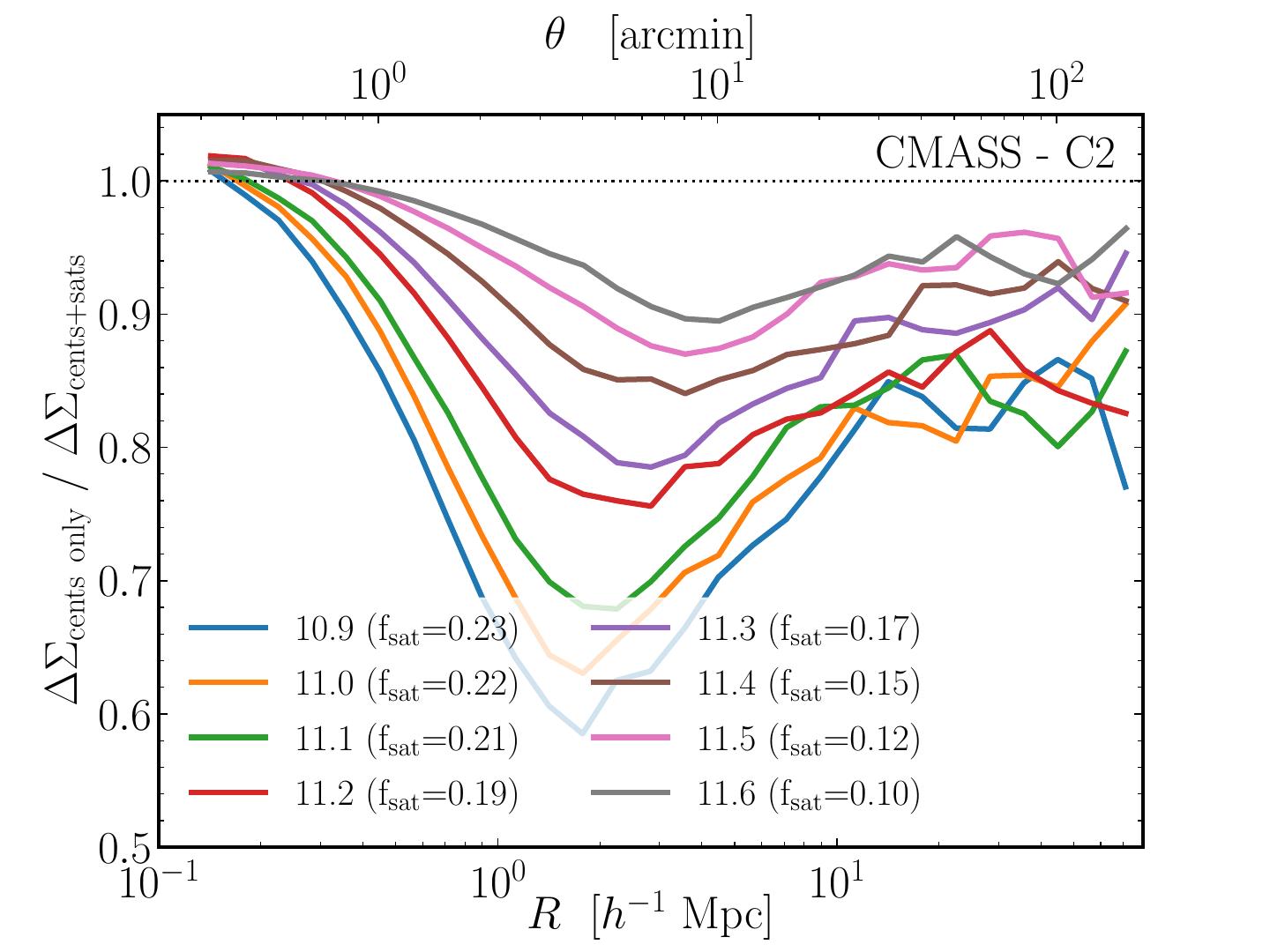}\\
    \includegraphics[width=0.9\columnwidth]{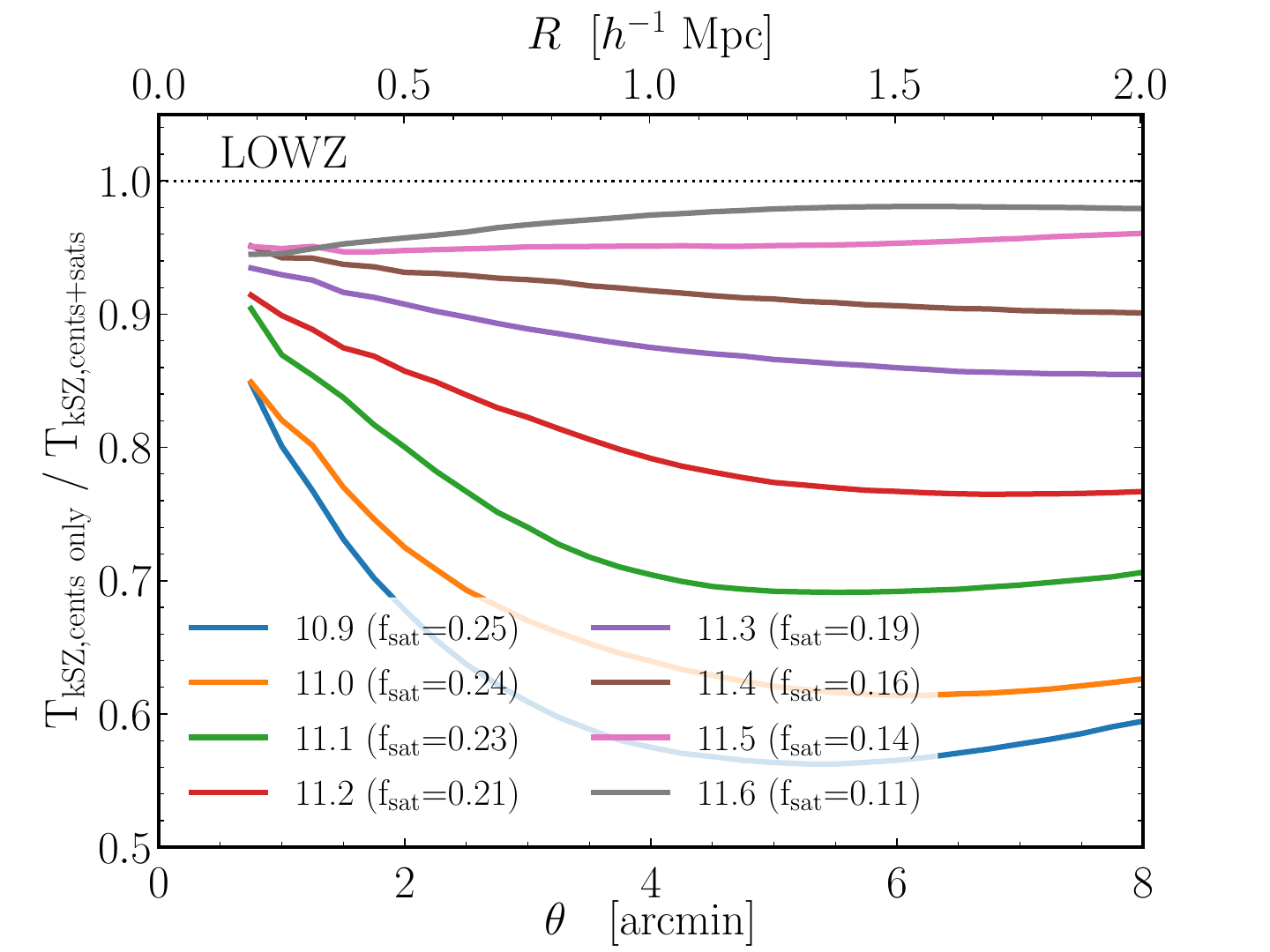}
    \includegraphics[width=0.9\columnwidth]{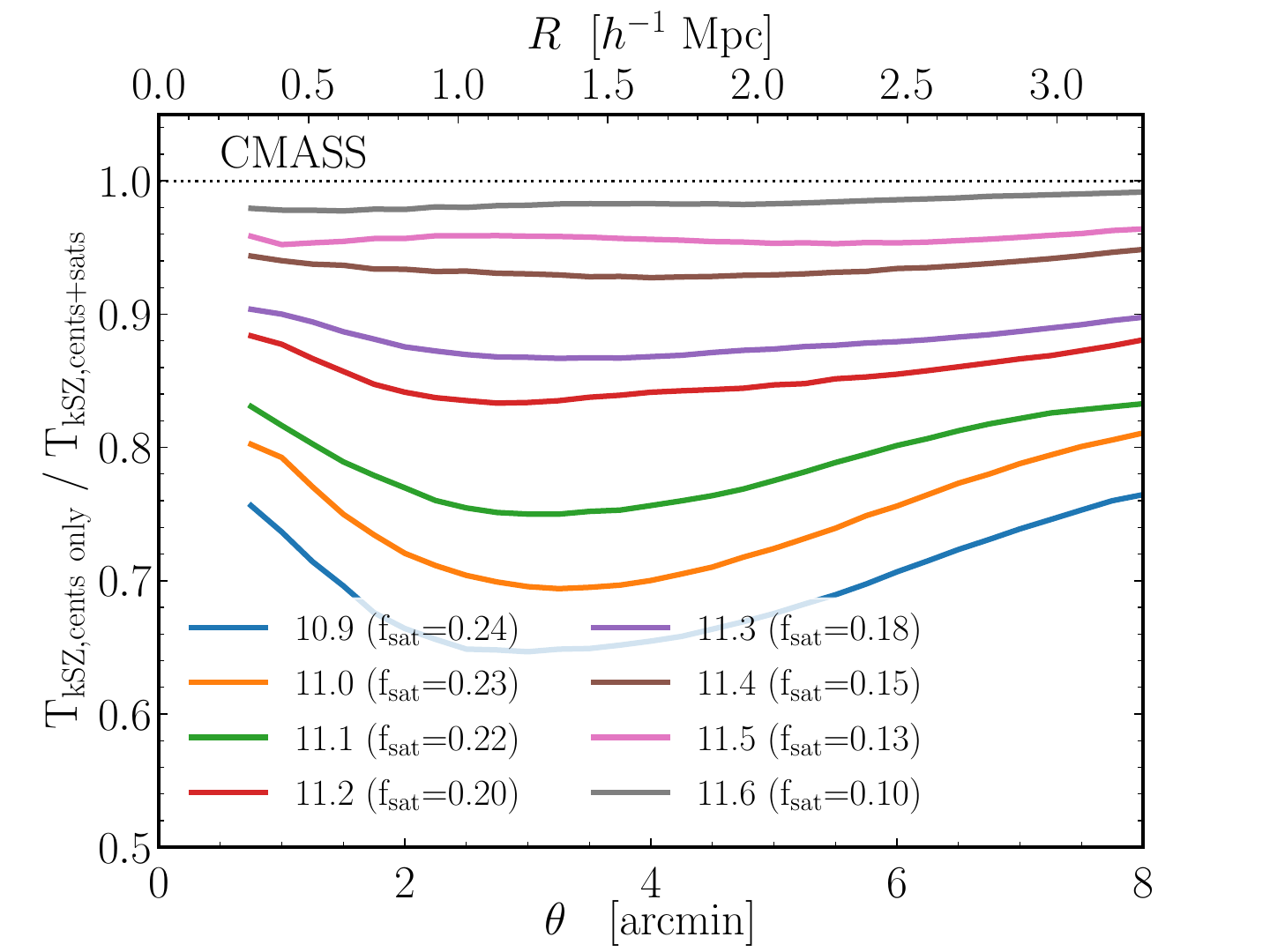}    
    \caption{Ratios of the stacked lensing and kSZ signals for a central-only sample with respect to the fiducial sample for different choices of the minimum stellar mass cut.  For the lensing comparison (top panels) we show only the L1 and C2 bins, for brevity, noting that the L2 and C1 bins give similar results.  The solid curves show the ratios for the fiducial \flamingo\ feedback model and correspond to different minimum stellar masses $\log_{10}[M_\text{star}/\textrm{M}_\odot]$ values ranging between 10.9 and 11.6 in steps of 0.1.  The effects of including satellites tend to be stronger for selections including lower mass galaxies.  For minimum stellar masses of $\approx11.2$, appropriate for the LOWZ and CMASS selections, the inclusion of satellites can boost the lensing and kSZ effect signals by up to $\approx30\%$ and $\approx20\%$, respectively.}
    \label{fig:sats}
\end{figure*}

\subsection{The role of satellite galaxies}
\label{sec:satellite_dep}

Some fraction of the galaxies comprising the BOSS LOWZ and CMASS samples will be satellites.  Previous attempts to model the kSZ effect measurements have not accounted for the impact of satellites, but whether this is a significant omission is unclear.  On the one hand, satellites will be mis-centred with respect to their host haloes and this might be expected to lead to a reduced kSZ signal compared to a galaxy that is centred on the hot gas distribution.  On the other hand, a stellar mass-based selection implies that the satellites will typically be in hosts that are more massive than a host which has a central of similar stellar mass.  This will tend to boost the kSZ signal.

Here we use the \flamingo\ simulations and associated subhalo catalogs to explore the impact of the inclusion of satellite galaxies on the derived stacked lensing and kSZ effect profiles.  In Fig.~\ref{fig:sats} we compare the ratios of the stacked lensing and kSZ signals for a central-only sample with that for the fiducial sample for different choices of the minimum stellar mass cut.  For the lensing comparison (top panels) we show only the L1 and C2 bins for brevity, noting that the L2 and C1 bins give similar results.  The solid curves show the ratios for the fiducial \flamingo\ feedback model and correspond to different minimum stellar masses: $\log_{10}[M_\text{star}/\textrm{M}_\odot]$ values ranging between 10.9 and 11.6 in steps of 0.1 and in the legend we quote the satellite fractions corresponding to these minimum stellar mass selections.

It is first worth noting that all of the curves in Fig.~\ref{fig:sats} are below 1, meaning that the fiducial selection including satellites leads to boosted mean lensing and kSZ signals relative to a central-only selection. This implies that of the two effects discussed above; i.e., mis-centring vs.~satellites living in a higher-mass host, it is the latter that dominates.  It also implies that analyses that do not account for satellites will likely tend to overestimate the halo mass required to match the lensing and kSZ measurements and therefore potentially underestimate the impact of feedback.

From our previous analysis, we found that minimum stellar masses in the range of $\approx 11.15$-$11.25$ best describe the selection for the BOSS LOWZ and CMASS samples.  For these selections, Fig.~\ref{fig:sats} implies that the inclusion of satellites can boost the lensing and kSZ effect signals by up to $\approx30\%$ and $\approx20\%$, respectively.  The effect is scale dependent, which is likely due to the mis-centring of satellites with respect to their hosts.  One can therefore potentially constrain the satellite fraction through detailed measurements of the shape of the lensing and kSZ effect profiles.

While the effects of satellites are not large enough to alter our conclusion that stronger feedback is required to match the kSZ effect measurements relative to an X-ray-based calibration strategy, they nevertheless should be factored in for quantitative analyses. Furthermore, according to Fig.~\ref{fig:sats}, as observations push to lower stellar masses, the role of satellites will become more significant (given the large satellite fractions) and will need a careful accounting.

\subsection{Implied halo masses}
\label{sec:mhalo_fgas}

Using the simulated galaxy selection that best fits the lensing measurements it is straightforward to compute a mean halo mass.  We quote the mean halo masses in terms of $M_\textrm{500c}$ in order to place the them in the context of X-ray samples.  Note that for satellite galaxies we use the $M_\textrm{500c}$ value associated with the FOF group in which the satellite resides.  Weighting each selected galaxy equally yields mean halo masses of $\log_{10}[M_\textrm{500c}/\textrm{M}_\odot] = 13.53 \pm 0.02$ and $13.34 \pm 0.04$ for the LOWZ and CMASS samples, respectively.  Thus, the lensing data yields a $\la 10\%$ constraint on the halo mass at $2\sigma$ uncertainty.  In the above we have combined the L1 and L2 (C1 and C2) constraints into a single mean halo mass using inverse-variance weighting for LOWZ (CMASS), given that the individual bins are consistent (within $2\sigma$).  We find that these constraints on the mean halo mass are virtually independent of which \flamingo\ model we use to compute the mean halo masses, which we attribute to the fact that we adjust the minimum stellar mass for each model to refit the lensing data.  This includes runs which vary the galaxy stellar mass function, which require a significantly different minimum stellar mass to match the lensing data. 

We note that if we apply the same minimum stellar mass cuts as in the fiducial selection but limit our analysis to central galaxies only, we find lower mean halo masses\footnote{We find consistent constraints with the central galaxy only mean halo masses if, instead of computing the mean of the selected central galaxies, we simply fit a composite NFW profile + 2-halo term (using the halo bias model of \citealt{Tinker2010}) to the lensing data, using the \textsc{Colossus} package \citep{Diemer2018}.} of $\log_{10}[M_\textrm{500c}/\textrm{M}_\odot] = 13.32 \pm 0.03$ and $13.13 \pm 0.06$ for the LOWZ and CMASS samples, respectively.  This confirms that the inclusion of satellite galaxies in the selection boosts the mean halo mass of the sample and by consequence also the stacked lensing and kSZ effect signals.  If we were instead to re-fit the minimum stellar mass for the central-only selection, the derived mean halo mass would increase, as expected.  But note that a central-only selection does not provide a statistically good fit to the lensing data, so the derived mean halo mass would have questionable value.

\section{Discussion}
\label{sec:discuss}

\subsection{Implications for feedback models}
\label{sec:feedback}

Here we discuss the interpretation of our findings in the context of feedback modelling. This work makes crucial steps in confirming the preference for stronger feedback from WL+kSZ without the reliance on a baryonification model. The origin of the difference in the implied required strength of feedback from the kSZ effect measurements and the X-ray-based baryon fraction measurements is unclear. Taking the observations at face value and as discussed previously by \citet{Bigwood2024}, the discrepancy between X-ray and kSZ effect constraints on the strength of feedback can be explained by: (1) differences between the real and simulated mass- and/or redshfit-dependencies of feedback; (2) differences between the real and simulated scale-dependence of feedback effects; and/or (3) the possibility of unaccounted for systematics or selection effects in the measurements. \\

\subsubsection{Mass and redshift dependence of feedback models}

One way to potentially reconcile the kSZ effect profiles with X-ray-based gas fraction constraints is to appeal to the different halo masses and redshifts that they probe.  For example, the BOSS CMASS sample has a mean redshift of $z\approx0.54$ and a halo mass of $M_\textrm{500c} \approx 2.2\times10^{13}$ M$_\odot$ (see Section \ref{sec:mhalo_fgas}), whereas the X-ray measurements of gas fractions are generally confined to groups at $z\la0.3$ with halo masses of $10^{14}$ M$_\odot$, although with considerable variation about these typical values.  Therefore, if the effective halo mass and/or redshift dependencies of feedback in the simulations differ from those in nature (such that the simulated feedback has a weaker dependence on mass and/or redshift than in reality), one could potentially understand the difference between the kSZ effect and group gas fractions calibrations.  

However, such an explanation is made more difficult by the inclusion of the LOWZ sample here, since its mean redshift is lower at $z\approx0.31$ and its mean halo mass is slightly higher at $3.4\times10^{13}$ M$_\odot$. These have some overlap with the X-ray samples used to calibrate the simulations, including \flamingo.

\subsubsection{Scale dependence of feedback}

The mean halo masses of the LOWZ and CMASS samples imply that the angular scales corresponding to $R_\textrm{500c}$ at $z=0.31$ (LOWZ) and $z=0.54$ (CMASS) are $\theta_\textrm{500c} \approx 1.61$ and $0.92$ arcmins, respectively.  These scales are indicated by vertical dashed lines in Fig.~\ref{fig:kSZ_fgas}, for example.  For LOWZ the first data corresponds roughly to $0.65 R_\textrm{500c}$, whereas for CMASS the first data point is $\approx 1.1 R_\textrm{500c}$.  Most of the constraining power on the feedback models is therefore coming from $\approx 1.5$-$3 \ R_\textrm{500c}$ for LOWZ and from $\approx 2.5$-$5 \ R_\textrm{500c}$ for CMASS.  The X-ray-constrained baryon fractions, by contrast, are measured within $R_\textrm{500c}$.  

Therefore, a way to potentially reconcile the X-ray and kSZ measurements is to invoke more steeply declining gas density profiles at large radii (beyond $R_\textrm{500c}$) than are seen in the simulations.  What physical mechanism could result in the required steepening is unclear.  In addition, it would be interesting to calculate if the required steepening is consistent with the constraint that, according to \citet{Planck2013}, the tSZ effect within $5 R_\textrm{500c}$ scales self-similarly with halo mass over a very wide range of masses, implying that haloes are fully baryon loaded within that aperture, at least at low redshift.  But we leave this question for future work.

\subsubsection{Unaccounted for systematics in the measurements}

It is possible that, instead of there being a fundamental issue with the feedback in the simulations, there could be unaccounted for (or mischaracterised) systematic errors in the gas fraction, kSZ effect, or galaxy-galaxy lensing measurements.  For example, there are considerable uncertainties in the X-ray selection function of galaxy groups (e.g., \citealt{Pearson2017,Giles2022,Marini2025}) and some studies make no attempt to account for selection effects.  For the kSZ effect measurements, a significant underestimate of the bias in the velocity reconstruction, or from tSZ effect leakage in the kSZ effect stacking analysis, could potentially reconcile the measurements with the simulation predictions using X-ray-calibrated feedback.  Ideally, we would use the same theory (hydrodynamical simulations) to select our systems, analyse them in precisely the same way as done for the real systems, and compare the processed observables in a like-with-like fashion.  While we have taken important steps in this direction in the present study, the issues of the X-ray selection function and kSZ effect velocity reconstruction (and/or tSZ effect leakage) remain open questions.  Finally, we have used galaxy-galaxy lensing measurements to constrain the selection of simulated galaxies and therefore any biases present in the lensing measurements will impact our feedback conclusions.  However, as previously noted, we find excellent consistency in the derived mean halo masses from the lensing with that inferred from fitting to the large-scale projected clustering, suggesting any biases in the lensing masses are likely to be small.

\subsection{Implications for the $S_8$ tension}
\label{sec:s8_tension}

Our analyses of the independent LOWZ and CMASS samples suggests that feedback stronger than adopted in the fiducial \flamingo\ model is required to match the stacked kSZ effect measurements of \citet{Schaan2021}.  The strong \flamingo\ fgas-$8\sigma$ variant yields a reasonably good fit to the kSZ effect profiles. We examine the impact of the stronger feedback in this model on the 3D matter power spectrum as well as on other observable measures of clustering.

\begin{figure*}
    \centering
    \includegraphics[width=\columnwidth]{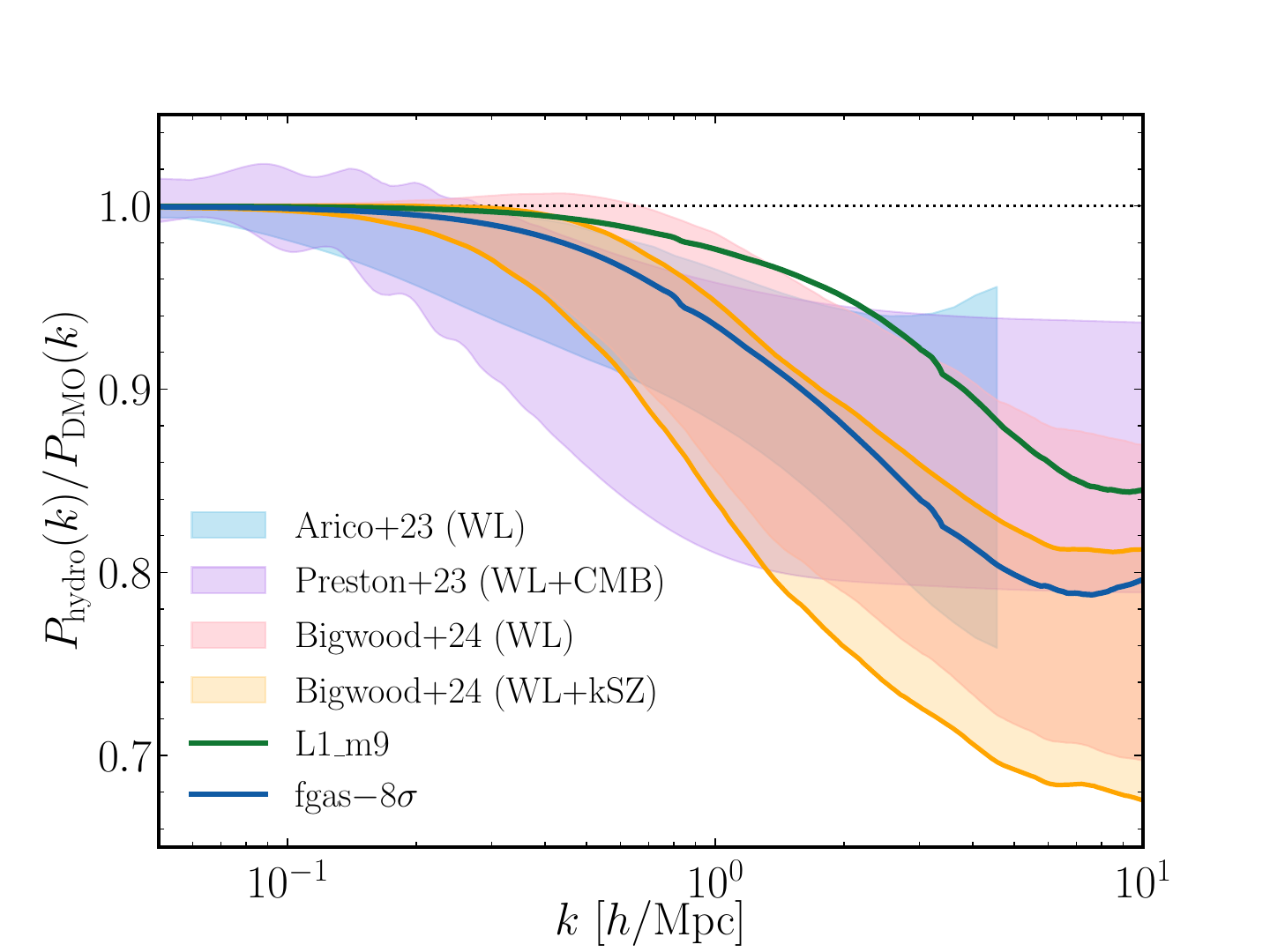}
    \includegraphics[width=\columnwidth]{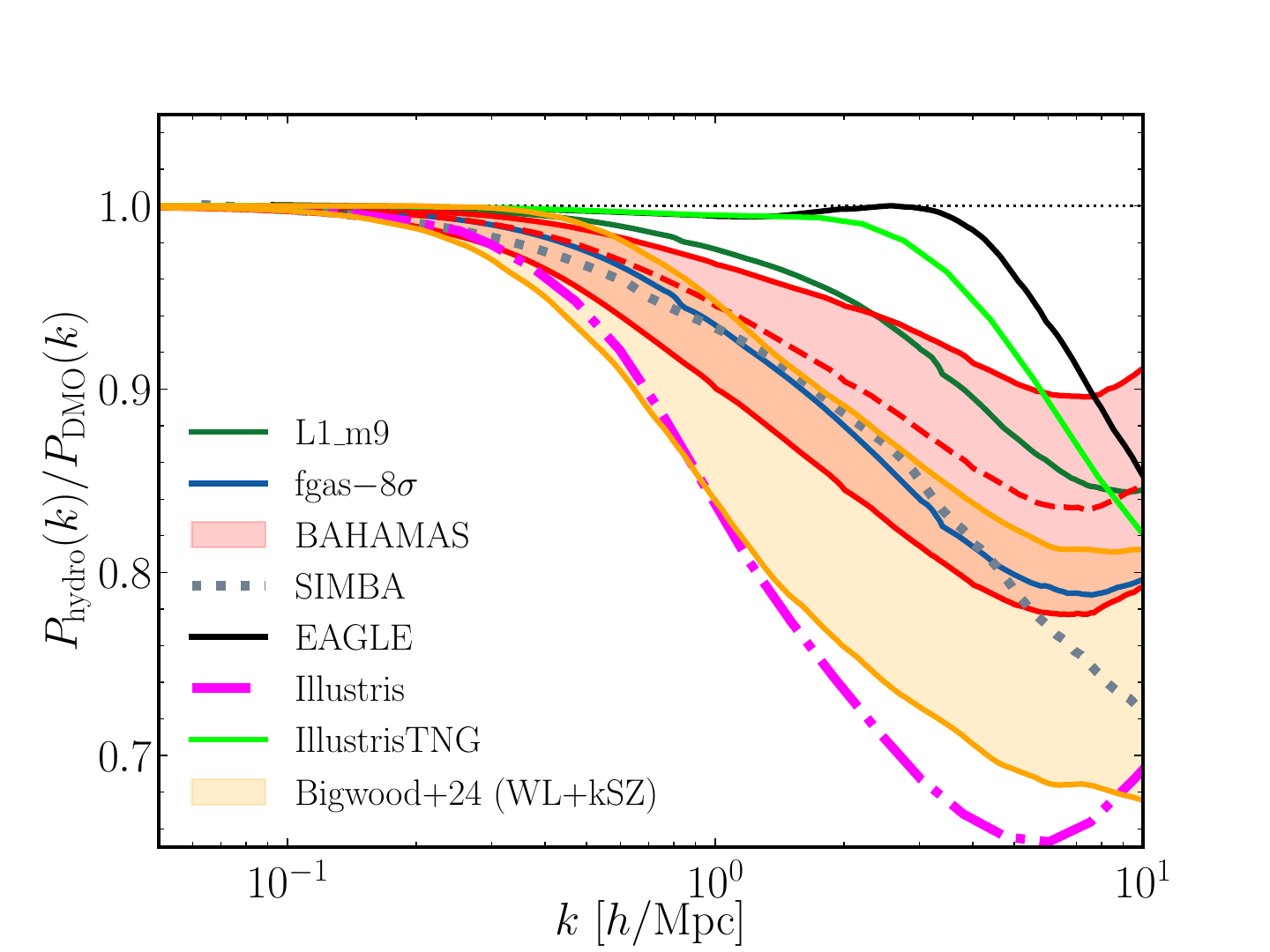}
    \caption{Suppression of the 3D matter power spectrum at $z=0$ due to baryon feedback. {\it Left:}  Comparison of the fiducial \flamingo\ feedback model and the strong fgas-$8\sigma$ variant with recent observational constraints on the suppression.  The solid curves shows the suppression predicted by the \flamingo\ models.  The shaded regions correspond to constraints from: \citet{Preston2023}, who jointly model the DES Y3 cosmic shear and \textit{Planck} 2018 primary CMB data; \citet{Arico2023} and \citet{Bigwood2024} (WL), who re-analyse DES Y3 data including small-scale shear measurements (see text); and \citet{Bigwood2024} (WL+kSZ), who perform a joint analysis of cosmic shear and the kSZ effect CMASS measurements of \citet{Schaan2021}. {\it Right:} Comparison with other recent hydrodynamical simulations, including: BAHAMAS \citep{McCarthy2017,McCarthy2018} (the red shaded region encloses the low-AGN and hi-AGN variants, with the dashed red curve corresponding to the fiducial BAHAMAS model); SIMBA \citep{Dave2019}, EAGLE \citep{Schaye2015}, the original Illustris simulations \citep{Vogelsberger2014}, IllustrisTNG \citep {Pillepich2018TNGmethod}.  Recent WL-based constraints favour stronger suppressions and are compatible with the predictions of the \flamingo\ fgas-$8\sigma$ variant, as well as SIMBA and BAHAMAS hi-AGN.}
    \label{fig:pk_sup}
\end{figure*}

\subsubsection{Cosmic shear / 3D matter power spectrum}

In Fig.~\ref{fig:pk_sup} (left panel), we compare the predictions of the suppression of the matter power spectrum from fiducial \flamingo\ model and the strong fgas-$8\sigma$ variant with the recent constraints from \citet{Bigwood2024}, \citet{Arico2023}, and \citet{Preston2023} (see also \citealt{Schaller2024b}).  The \citet{Preston2023} constraint comes from a joint analysis of the DES Y3 data and the \textit{Planck} 2018 primary CMB.  The constraints shown in Fig.~\ref{fig:pk_sup} correspond to their best-fitting empirical `Amod' model, which is intended to parameterise possible modifications of the non-linear part of the matter power spectrum due to baryonic physics and/or non-standard dark matter.  The two constraints from \citet{Bigwood2024} come from an analysis of DES Y3 cosmic shear data (labelled WL) or from a joint analysis of the DES Y3 cosmic shear and the kSZ effect CMASS profile of \citet{Schaan2021} (labelled WL+kSZ), modelled using the BCEmu baryonification formalism \citep{Schneider2019,Giri2021}.  Note that \citet{Bigwood2024} do not adopt the small scale cuts employed in the fiducial DES Y3 cosmic shear analysis (e.g., \citealt{Krause2022,Amon2023b,Secco2022}) that were designed to mitigate the impact of baryonic physics, but instead include the small scale measurements and use the baryonification formalism to model the baryonic effects, aided by the inclusion of kSZ effect data.  \citet{Arico2023} use the BACCO baryonification emulator\footnote{We note that the BCEmu and BACCO model differences have been studied and shown to impact the shape of the predicted power spectrum \citep{Grandis2024}. A primary difference is that the former displaces formally all the particles in the simulation, whereas the latter displaces only particles in haloes.} \citep{Arico2021emulator} to model the DES Y3 cosmic shear data (including small scale measurements).

Examining Fig.~\ref{fig:pk_sup}, we see that the fgas-$8\sigma$ variant is in good agreement with the suppression constraints from \citet{Bigwood2024}, for both their WL only and WL+kSZ analysis, as well as those from \citet{Arico2023} .  There is also reasonable agreement with the constraints from \citet{Preston2023}, with a slight mismatch on intermediate scales ($0.1 \la k [h/\textrm{Mpc}] \la 1$), such that the fgas-$8\sigma$ variant predicts a bit too much power.  As noted above, a difference between the constraints of the other studies and those of \citet{Preston2023} is that the latter also fit to the primary CMB measurements.  Nevertheless, the agreement with the other constraints is generally good.  It is worth highlighting that the observational constraints shown in Fig.~\ref{fig:pk_sup} are likely to be model dependent, particularly on small scales ($k$ greater than a few $h$/Mpc) which current cosmic shear measurements are less sensitive to.

\citet{Bigwood2024} derive a value of $S_8=0.818^{+0.017}_{-0.024}$ from their cosmic shear-only analysis and $S_8=0.823^{+0.019}_{-0.020}$ from their cosmic shear+kSZ effect analysis, which are consistent with the \textit{Planck} primary CMB estimate of $S_8 = 0.832 \pm 0.013$ \citep{Planck2020cosmopars} at the $0.7\sigma$ and $0.4\sigma$ levels, respectively (i.e., no significant tension).  Using the same DES Y3 cosmic shear data, \citet{Arico2023} find a slightly lower value  of $S_8 = 0.795^{+0.015}_{-0.017}$. Note that even prior to the inclusion of small scales and associated baryonic modelling, the DES Y3 cosmic shear results were not in strong tension with the \textit{Planck} primary CMB (typically $2\sigma$ level).  The inclusion of small scales, updated modelling of intrinsic alignments, and inclusion of baryonic effects has weakened this tension.  The KiDS 1000 constraints, by contrast, typically show stronger levels of tension ($\ga3\sigma$), which would require feedback in excess of that implied by the kSZ effect measurements studied here (e.g., \citealt{Heymans2021,Amon2022,Schneider2022,McCarthy2023}). 

In the right panel of Fig.~\ref{fig:pk_sup} we compare the fiducial \flamingo\ model and the fgas-$8\sigma$ variant with the predictions of other recent hydrodynamical simulations, including: BAHAMAS \citep[$L_\textrm{box} = 571$ Mpc]{McCarthy2017,McCarthy2018} (the red shaded region encloses the low-AGN and hi-AGN variants, with the dashed red curve corresponding to the fiducial BAHAMAS model); SIMBA \citep[$L_\textrm{box} = 147$ Mpc]{Dave2019}, EAGLE \citep[$L_\textrm{box} = 100$ Mpc]{Schaye2015}, the original Illustris simulations \citep[$L_\textrm{box} = 107$ Mpc]{Vogelsberger2014}, andIllustrisTNG \citep[$L_\textrm{box} = 300$ Mpc]{Pillepich2018TNGmethod}.    Recent WL-based constraints favour stronger suppressions and are compatible with the predictions of the \flamingo\ fgas-$8\sigma$ variant, as well as SIMBA and BAHAMAS hi-AGN.

\subsubsection{Thermal SZ power spectrum}
Moving on from cosmic shear, previous studies have shown that various measures of the tSZ effect are in tension with the standard model fit to the primary CMB (e.g., \citealt{Planck2014,McCarthy2014,Planck2016,Planck2016_tSZ_PS,Bolliet2018,McCarthy2018}).  Note that the tSZ effect (e.g., its PDF, power spectrum, and number counts) is typically sensitive to the most massive clusters and is therefore expected to be more sensitive to parameters such as $\sigma_8$ and $\Omega_\textrm{m}$ than cosmic shear.  The `catch' is that the tSZ effect needs to be cleanly separated from various other signals present in the CMB temperature maps, including the primary CMB, the kSZ effect, radio sources, the CIB, etc., which is non-trivial.

We revisit the tSZ effect power spectrum and the tSZ effect-shear cross-spectrum examined for \flamingo\ in \citet{McCarthy2023}.  We refer the reader to that study for a full description of how the cosmological observables are calculated.  

\begin{figure}
    \includegraphics[width=\columnwidth]{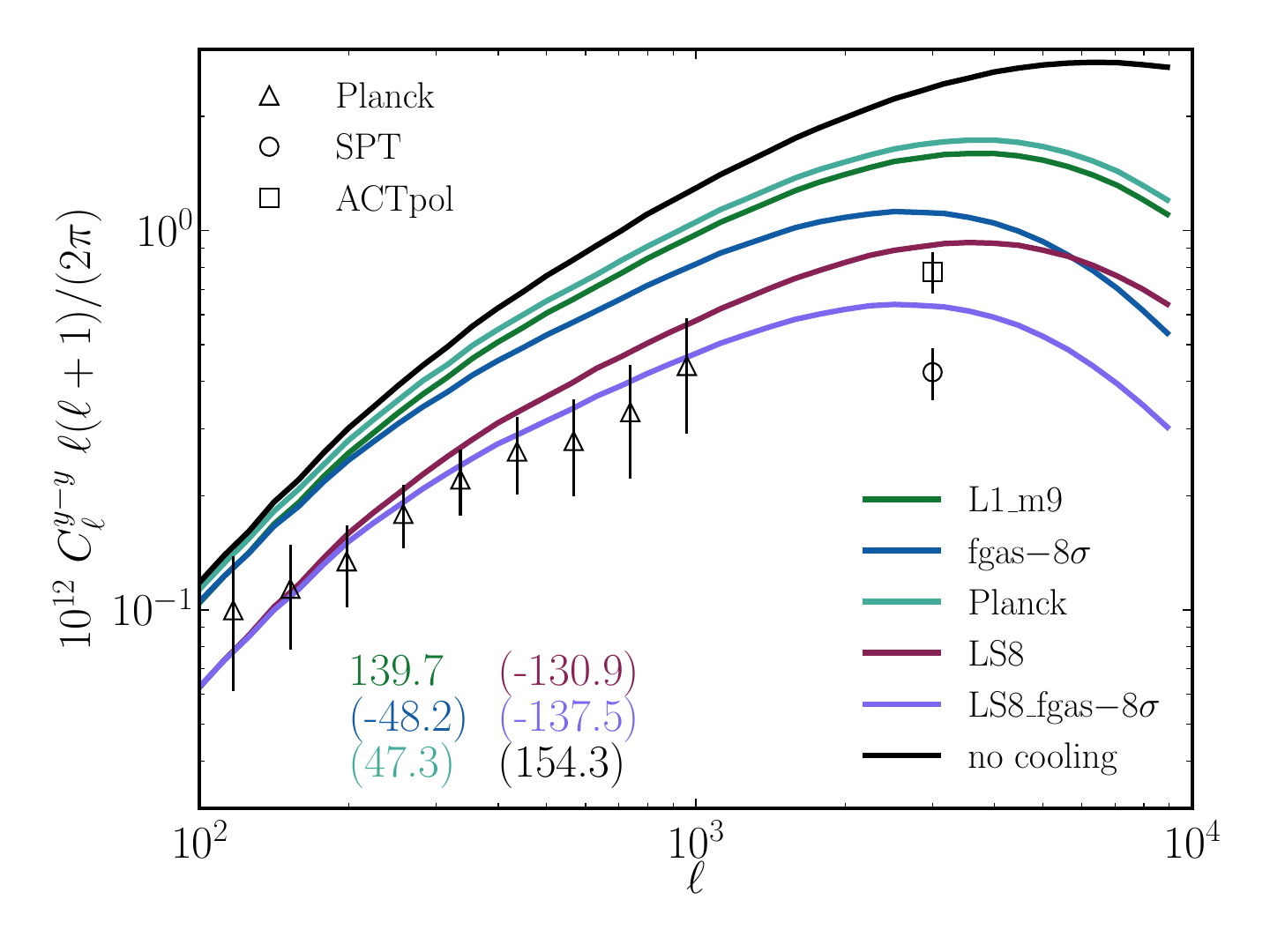}
    \caption{The tSZ effect angular power spectrum.  The open triangles correspond to the \textit{Planck} tSZ measurements of \citet{Bolliet2018} while the open circle and open square correspond to the SPT and ACTpol measurements of \citet{Reichardt2021} and \citet{Choi2020}, respectively.  Following the same colour coding, the numbers in the bottom left indicate $\chi^2$ for the L1\_m9 run and the $\Delta \chi^2$ (in parentheses) of the other runs with respect to L1\_m9.  Note that the $\chi^2$ is computed here with respect to the Planck tSZ measurements only, using the diagonal elements of the covariance matrix.  A negative value for the $\Delta \chi^2$ indicates a better match to the data.  A lensing LS8 cosmology yields a significantly better fit to the data relative to the fiducial D3A cosmology, whereas a \textit{Planck} CMB cosmology yields a worse fit.  Increasing the efficiency of feedback (fgas-$8\sigma$), as preferred by kSZ effect results in the present study, improves the fit on small angular scales.  But the offset on large scales cannot be reconciled through feedback alone.}
    \label{fig:tSZ_powerspec}
\end{figure}

In Fig.~\ref{fig:tSZ_powerspec} we compare selected \flamingo\ runs to the latest tSZ effect power spectrum measurements, namely the \textit{Planck}-based measurements reported in \citet{Bolliet2018}, the SPT measurement from \citet{Reichardt2021}, and the ACTpol measurement from \citet{Choi2020}.  Note that \citet{Bolliet2018} present an improved re-analysis of the \textit{Planck} 2015 tSZ data set from \citet{Planck2016_tSZ_PS}, by taking into account the tri-spectrum in the covariance matrix and placing physical constraints on the amplitudes of foreground contaminants (particularly radio and infrared point sources and the clustered infrared background, or CIB).  The ACTpol measurement was not considered in \citet{McCarthy2023} as we were unaware of it at the time.  The measurement from ACTpol is significantly higher in amplitude compared to previous ACT measurements (e.g., \citealt{Sievers2013}) and compared to the SPT measurement of \citet{Reichardt2021}. The origin of the differences between the ACTpol measurements and the previous ACT measurements was not discussed in \citet{Choi2020}.

Comparing the simulations to the observational measurements, we see that both the fiducial feedback model (L1\_m9) and fgas-$8\sigma$ variant in the fiducial D3A cosmology are strongly in tension with the \textit{Planck} tSZ power spectrum measurements on large scales, which is sensitive mainly to very massive clusters that are generally unaffected by feedback (e.g, \citealt{McCarthy2014}).  At the smaller scales probed by SPT and ACT, increased feedback yields a qualitatively better match to the observational measurements though the discrepancy between the SPT and ACTpol measurements, prevents a quantitative assessment of the goodness of fit.  A \textit{Planck} primary CMB cosmology with fiducial feedback yields a slightly worse match to the measurements than the D3A cosmology.  The LS8 cosmology with fiducial feedback, by contrast, yields a significantly improved fit to the \textit{Planck} tSZ power spectrum measurements on large scales and with the ACTpol measurements on small scales.  Increasing the feedback in an LS8 cosmology (LS8\_fgas-$8\sigma$) further improves the fit, though the improvement is small compared to the improvement that resulted from lowering $S_8$.

\subsubsection{Thermal SZ--cosmic shear cross-spectrum}

\begin{figure*}
    \includegraphics[width=0.75\textwidth]{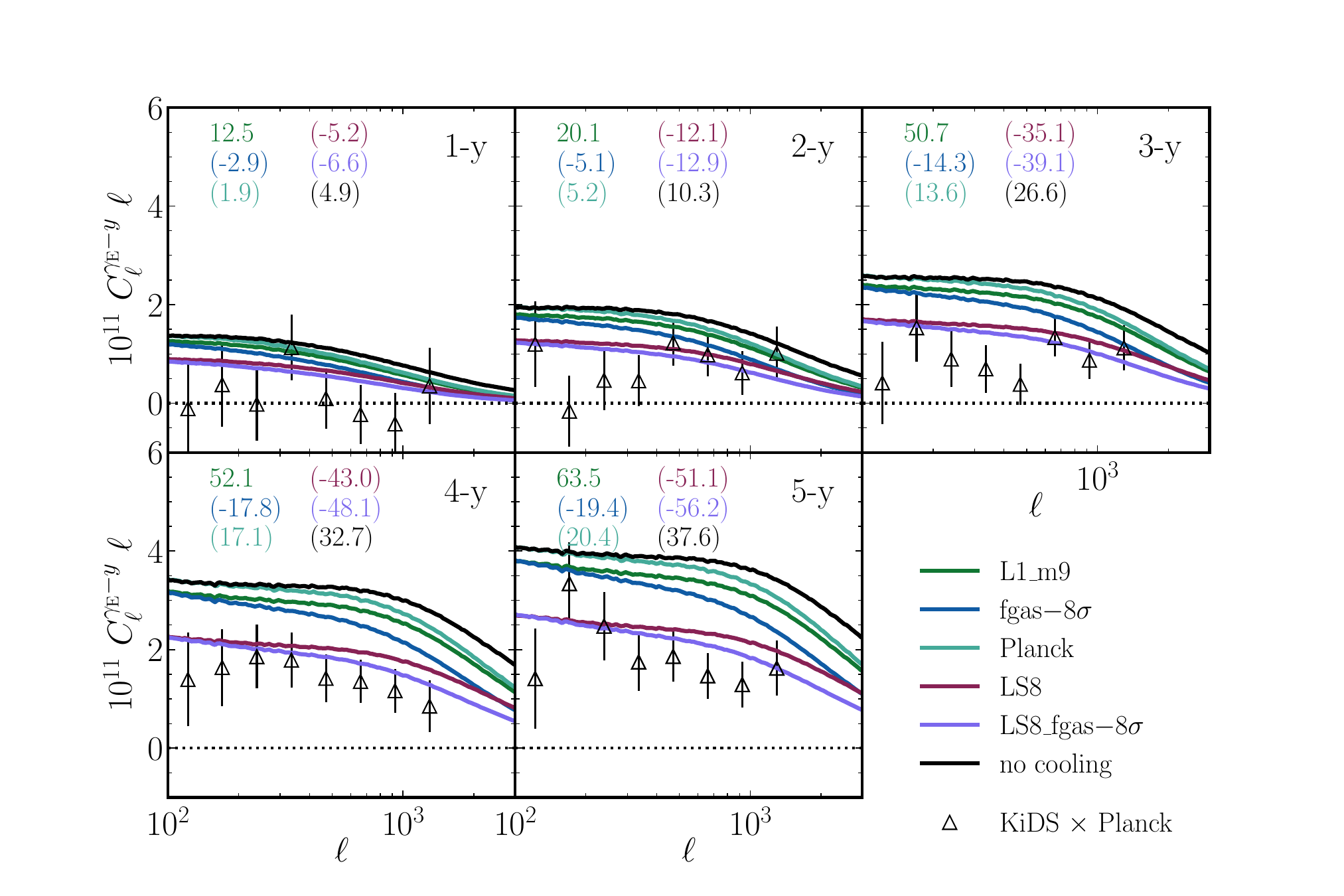}
    \caption{The cosmic shear--tSZ effect angular cross-power spectrum.  The open triangles correspond to the KiDS1000 $\times$ \textit{Planck} measurements of \citet{Troster2022}.  The different panels correspond to the cross-spectrum between different KiDS tomographic bins (1-5) and the tSZ data.  Similar to case of the tSZ power spectrum (see Fig.~\ref{fig:tSZ_powerspec}), a lensing LS8 cosmology yields a significantly better fit to the data relative to the fiducial D3A cosmology, whereas a Planck CMB cosmology yields a worse fit.  Increasing the efficiency of feedback (fgas-$8\sigma$ and LS8-fgas-$8\sigma$), improves the fit on small angular scales.  But the offset on large scales cannot be reconciled through feedback alone.
    }
    \label{fig:tSZ_shear_powerspec}
\end{figure*}

In Fig.~\ref{fig:tSZ_shear_powerspec} we compare selected \flamingo\ runs to the tSZ effect-cosmic shear cross-spectrum measurements of \citet{Troster2022}, who cross-correlated the 5 tomographic KiDS1000 bins with tSZ effect maps constructed from the \textit{Planck} 2015 data set \citep{Planck2016_tSZ_PS}.  In agreement with the tSZ effect power spectrum analysis, we see that both the fiducial feedback model and fgas-$8\sigma$ variant in the fiducial D3A cosmology are strongly in tension with the measurements on large scales.  At smaller scales ($\ell \ga 700$), increased feedback yields a better match to the observational measurements.  A \textit{Planck} primary CMB cosmology with fiducial feedback yields a slightly worse match to the measurements than the D3A cosmology.  The LS8 cosmology with fiducial feedback, by contrast, provides a significantly better match on all scales compared to the fiducial model in either the D3A or \textit{Planck} cosmologies.  Increasing the feedback in the LS8 cosmology slightly improves the fit relative to the fiducial feedback model in the LS8 cosmology.

\subsubsection{Summary}

The above comparisons indicate that when adopting a \textit{Planck}-like cosmology (such as D3A or the \textit{Planck} maximum likelihood cosmology), increasing the strength of baryonic feedback generally serves to improve the match to various observed clustering statistics, primarily through reducing the power on small scales.  Importantly, though, on large scales the tSZ effect power spectrum and tSZ effect-shear cross-spectrum are insensitive to the effects of feedback even when the feedback is increased significantly to match the kSZ effect measurements examined here.  Increasing the feedback improves the match on small scales in a \textit{Planck}-like cosmology but only marginally improves it in an LS8 cosmology.

As the tSZ effect measurements appear strongly in tension with the standard model fit to the primary CMB and on the largest scales are generally insensitive to baryonic effects, it will be important to further scrutinise the robustness of these measurements, including the robustness of the component separation techniques employed to separate the tSZ effect from the other signals present in the data such as the CIB and radio sources.  In this regard, cross correlations with other signals, such as cosmic shear, are invaluable since they will be differently affected by any residual contamination than is the case for the tSZ effect power spectrum.  Relativistic effects should also be included in future analyses (e.g., \citealt{Remazeilles2019,Remazeilles2025}).  Furthermore, we note that the baryonification formalism has recently been extended to tSZ effect modelling by \citet{Arico2024} and it would be interesting to compare constraints from this methodology with those derived from the hydrodynamical simulations.

\section{Conclusions}
\label{sec:conclusions}

In this work we have used the state-of-the-art \flamingo\ suite of cosmological hydrodynamical simulations to compare to kSZ effect stacking measurements of \citet{Schaan2021}, based on a sample selection derived from fitting galaxy-galaxy lensing measurements \citep{Amon2023}.  This is the first time the powerful combination of kSZ effect and galaxy-galaxy lensing measurements has been used as a benchmark for feedback modelling in cosmological hydrodynamical simulations.  

We used associated full-sky lightcone-based \healpix maps and catalogs to perform a like-with-like comparison with these observations for the first time in the context of self-consistent hydro simulations.  We explored the dependence of the predictions on feedback efficiency and implementation as well as on cosmology. The use of cosmological simulations also allowed us to quantitatively assess the impact that satellite galaxies included in the stack have on the resulting lensing and kSZ effect profiles.

We have highlighted that a quantitative interpretation of the kSZ effect measurements requires precise knowledge of the halo mass scale of the BOSS samples, since the amplitude of the kSZ effect scales approximately with the halo mass.  A meaningful comparison to the predictions of simulations (whatever the feedback implementation) requires that we select simulated systems that have an implicit mean halo mass matching that of the observed systems in the kSZ effect stacking.  To this end, a unique aspect of this work is that we have employed high-precision measurements of the stacked galaxy-galaxy lensing profiles of the BOSS LOWZ and CMASS from \citet{Amon2023} using DES Y3 + KiDS1000 data to better than $10\%$ mass constraints ($2\sigma$).  We have used the lensing measurements to select the simulated galaxies required for a fair comparison of the kSZ effect predictions to the measurements of \citet{Schaan2021}.

Our main finding is that the kSZ effect measurements imply that more aggressive feedback is required in the simulations compared to that inferred from X-ray cluster observations and that this goes some way to alleviating the difference in the observed and predicted clustering on small scales.  However, we have shown that the offsets on large scales between measurements and predictions of the tSZ power spectrum and its cross spectrum with cosmic shear cannot be resolved through increased feedback. In more detail, our findings may be summarised as follows:
\begin{itemize}
    \item A simple minimum stellar mass-based selection employed on the simulations yields galaxy-galaxy lensing profiles which match the measurements of \citet{Amon2023} over the wide range of observed radii, with the best-fitting minimum stellar masses, but not the best-fitting haloes masses, depending slightly on the feedback model (Figs.~\ref{fig:gg_lensing_fgas} and \ref{fig:gg_lensing_other_var}) and the adopted cosmology.  Higher minimum stellar masses (and halo masses) are required to match the LOWZ sample compared to the CMASS sample.  Invoking more complicated selection functions that depend on both stellar mass and specific star formation rate yields results consistent with our fiducial analysis (see Appendix \ref{sec:ssfr_selection}).  The derived mean halo masses for the CMASS and LOWZ samples are $\log_{10}[M_\textrm{500c}/\textrm{M}_\odot] = 13.53 \pm 0.02$ and $13.34 \pm 0.04$ (at $2\sigma$ uncertainty), respectively.  
    \item Using the lensing measurements as a strong constraint on the selection for the kSZ effect predictions, we find that the fiducial \flamingo\ feedback model, which was calibrated on the X-ray gas fractions of low-redshift galaxy groups and clusters, is ruled out at the $5\sigma$ level by both the BOSS LOWZ and CMASS samples (Fig.~\ref{fig:kSZ_fgas}).  The strongest feedback variant considered in the \flamingo\ suite, the so-called fgas-$8\sigma$ model, provides a good match to the measurements, though is inconsistent with the local X-ray measurements.
    \item Our conclusions are robust to the details of the feedback implementation, in that we find similar results for the jet-based AGN feedback implementation as we do for the fiducial thermal (isotropic) implementation when both models are calibrated to the same gas fraction data (Fig.~\ref{fig:kSZ_other_var}).  The main determining factor of the predicted kSZ signal is the fraction of baryons that are retained by haloes.  The kSZ effect results are also generally insensitive to reasonable variations in the cosmological parameters.
    \item Both the galaxy-galaxy lensing and kSZ effect profiles are boosted by the inclusion of satellite galaxies included in the stacking analyses.  For the BOSS selection, as implied by the lensing measurements, the boost is up to $\approx30\%$ and $\approx20\%$, respectively (Fig.~\ref{fig:sats}).  This effect increases with decreasing galaxy stellar mass and will therefore need to be carefully modelled as observations push to lower masses.
    \item We have discussed various ways to reconcile the differing constraints on feedback modelling from X-ray and kSZ effect measurements (see Section~\ref{sec:feedback}), including (1) that there could be differences between the real and simulated halo mass- or redshfit-dependencies of feedback; (2) differences between the real and simulated scale-dependence of feedback effects; and/or (3) unaccounted for systematics or selection effects in either the X-ray or kSZ effect measurements.  We have used lensing-based halo mass measurements to highlight that the kSZ effect measurements of the BOSS galaxies typically probe gas at several virial radii, whereas the X-ray measurements are confined to smaller scales.
    \item In light of the finding that stronger feedback is required to reproduce the kSZ effect measurements, we have reassessed the impact of feedback on several clustering statistics, including the matter power spectrum, relevant for cosmic shear (Fig.~\ref{fig:pk_sup}), the tSZ effect power spectrum (Fig.~\ref{fig:tSZ_powerspec}) and the tSZ effect cross-spectrum (Fig.~\ref{fig:tSZ_shear_powerspec}).  When adopting a \textit{Planck}-like cosmology (such as D3A or the \textit{Planck} maximum likelihood cosmology), increasing the strength of baryonic feedback generally serves to improve the match to various observed clustering statistics, primarily through reducing the power on small scales, in general agreement with the findings of \citet{McCarthy2023} and previous studies employing dark matter only simulations with corrections for baryonic effects (e.g., \citealt{Amon2022,Preston2023}). The level of tension between a \textit{Planck}-like cosmology and the DES Y3 cosmic shear measurements was already relatively low in the case of standard feedback modelling and the inclusion of stronger feedback reduces the tension further.  However, the tSZ effect power spectrum and tSZ effect-shear cross-spectrum are dominated by more massive haloes and are insensitive to the effects of feedback on large scales even when the feedback is increased significantly to match the kSZ effect measurements examined here.  Increasing the feedback improves the match on small scales in a \textit{Planck}-like cosmology but the improvement is only marginal in the `lensing' cosmology (LS8).
\end{itemize}

Whilst in the final stages of preparation of the present study, \citet{Hadzhiyska2024b} posted to the arxiv a new \textit{Planck}+ACT kSZ effect stacking analysis of galaxies in the DESI Legacy Imaging Survey, using photometric redshifts in the velocity reconstruction.  The use of photometric redshifts leads to a larger bias in the reconstructed velocities, which has been estimated using DESI mocks \citep{Ried2024,Hadzhiyska2024a}.  \citet{Hadzhiyska2024b} define four tomographic bins spanning $0.4 \la z \la 1.0$ (i.e., similar to the range probed by BOSS LOWZ and CMASS) and compare their stacking results with the Illustris-TNG \citep[$L_\textrm{box} = 300$ Mpc]{Pillepich2018TNGmethod} and the original Illustris \citep[$L_\textrm{box} = 107$ Mpc]{Vogelsberger2014} simulations, although without using light cones and adopting a simplified abundance matching approach for the selection of simulated galaxies.  Note that the original Illustris simulation has considerably stronger feedback than Illustris-TNG and that the gas fractions of the former are considerably below what is observed for X-ray-selected galaxy groups (whereas the latter lie considerably above the X-ray measurements).  Consistent with the present study, \citet{Hadzhiyska2024b} find that strong feedback is required to match their stacked profiles and that the original Illustris simulation provides a good fit to the data, whereas Illustris-TNG is ruled out at many sigma.  

Upcoming LSS surveys demand an accurate and precise feedback model that is consistent with a wide range of observations. Toward this goal, X-ray measurements remain a valuable source of information on the hot gas properties of groups/clusters with new insights coming from eROSITA data. Understanding the impact of X-ray selection effects will be crucial (e.g., \citealt{Popesso2024,Kugel2024}).  kSZ effect measurements binned by mass and redshift (e.g., via stacking on DESI galaxies) will be a rich source of information that is complementary to that of the X-ray measurements in that it can probe larger scales and higher redshifts.  For more nearby and massive galaxies, there will also be the opportunity to directly compare the X-ray and kSZ effect measurements to test for consistency.
Comparisons with complementary probes of the hot gas, such as through radio-based dispersion measure data (e.g., \citealt{Macquart2020}) and so-called `patchy screening' of the CMB (e.g., \citealt{Coulton2024}) should also provide important insights.

\section*{Acknowledgements}
We thank the referee for a helpful report that improved the paper.
This work was supported by the Science and Technology Facilities Council (grant number ST/Y002733/1).
This project has received funding from the European Research Council (ERC) under the European Union’s Horizon 2020 research and innovation programme (grant agreement No 769130). REA acknowledges support from the Spanish Ministry of Science under grant number PID2021-128338NB-I00.  CSF acknowledges support from the European Research Council (ERC) through Advanced Investigator grant DMIDAS (GA 786910).  This work used the DiRAC@Durham facility managed by the Institute for Computational Cosmology on behalf of the STFC DiRAC HPC Facility (\url{www.dirac.ac.uk}). The equipment was funded by BEIS capital funding via STFC capital grants ST/K00042X/1, ST/P002293/1, ST/R002371/1 and ST/S002502/1, Durham University and STFC operations grant ST/R000832/1. DiRAC is part of the National e-Infrastructure.  

\section*{Data Availability}

The data supporting the plots within this article are available on reasonable request to the corresponding author. The \flamingo\ simulation data will eventually be made publicly available, though we note that the data volume (several petabytes) may prohibit us from simply placing the raw data on a server. In the meantime, people interested in using the simulations are encouraged to contact the corresponding author.



\bibliographystyle{mnras}
\bibliography{references} 

\begin{thebibliography}{}
\makeatletter
\relax
\def\mn@urlcharsother{\let\do\@makeother \do\$\do\&\do\#\do\^\do\_\do\%\do\~}
\def\mn@doi{\begingroup\mn@urlcharsother \@ifnextchar [ {\mn@doi@} {\mn@doi@[]}}
\def\mn@doi@[#1]#2{\def\@tempa{#1}\ifx\@tempa\@empty \href {http://dx.doi.org/#2} {doi:#2}\else \href {http://dx.doi.org/#2} {#1}\fi \endgroup}
\def\mn@eprint#1#2{\mn@eprint@#1:#2::\@nil}
\def\mn@eprint@arXiv#1{\href {http://arxiv.org/abs/#1} {{\tt arXiv:#1}}}
\def\mn@eprint@dblp#1{\href {http://dblp.uni-trier.de/rec/bibtex/#1.xml} {dblp:#1}}
\def\mn@eprint@#1:#2:#3:#4\@nil{\def\@tempa {#1}\def\@tempb {#2}\def\@tempc {#3}\ifx \@tempc \@empty \let \@tempc \@tempb \let \@tempb \@tempa \fi \ifx \@tempb \@empty \def\@tempb {arXiv}\fi \@ifundefined {mn@eprint@\@tempb}{\@tempb:\@tempc}{\expandafter \expandafter \csname mn@eprint@\@tempb\endcsname \expandafter{\@tempc}}}

\bibitem[\protect\citeauthoryear{{Abbott} et~al.,}{{Abbott} et~al.}{2022}]{Abbott2022}
{Abbott} T.~M.~C.,  et~al., 2022, \mn@doi [\prd] {10.1103/PhysRevD.105.023520}, \href {https://ui.adsabs.harvard.edu/abs/2022PhRvD.105b3520A} {105, 023520}

\bibitem[\protect\citeauthoryear{{Aghanim} et~al.,}{{Aghanim} et~al.}{2019}]{Aghanim2019}
{Aghanim} N.,  et~al., 2019, \mn@doi [\aap] {10.1051/0004-6361/201935271}, \href {https://ui.adsabs.harvard.edu/abs/2019A&A...632A..47A} {632, A47}

\bibitem[\protect\citeauthoryear{{Ahn} et~al.,}{{Ahn} et~al.}{2014}]{Ahn2014}
{Ahn} C.~P.,  et~al., 2014, \mn@doi [\apjs] {10.1088/0067-0049/211/2/17}, \href {https://ui.adsabs.harvard.edu/abs/2014ApJS..211...17A} {211, 17}

\bibitem[\protect\citeauthoryear{{Akino} et~al.,}{{Akino} et~al.}{2022}]{Akino2022}
{Akino} D.,  et~al., 2022, \mn@doi [\pasj] {10.1093/pasj/psab115}, \href {https://ui.adsabs.harvard.edu/abs/2022PASJ...74..175A} {74, 175}

\bibitem[\protect\citeauthoryear{{Amodeo} et~al.,}{{Amodeo} et~al.}{2021}]{Amodeo2021}
{Amodeo} S.,  et~al., 2021, \mn@doi [\prd] {10.1103/PhysRevD.103.063514}, \href {https://ui.adsabs.harvard.edu/abs/2021PhRvD.103f3514A} {103, 063514}

\bibitem[\protect\citeauthoryear{{Amon} \& {Efstathiou}}{{Amon} \& {Efstathiou}}{2022}]{Amon2022}
{Amon} A.,  {Efstathiou} G.,  2022, \mn@doi [\mnras] {10.1093/mnras/stac2429}, \href {https://ui.adsabs.harvard.edu/abs/2022MNRAS.516.5355A} {516, 5355}

\bibitem[\protect\citeauthoryear{{Amon} et~al.,}{{Amon} et~al.}{2022}]{Amon2023b}
{Amon} A.,  et~al., 2022, \mn@doi [\prd] {10.1103/PhysRevD.105.023514}, \href {https://ui.adsabs.harvard.edu/abs/2022PhRvD.105b3514A} {105, 023514}

\bibitem[\protect\citeauthoryear{{Amon} et~al.,}{{Amon} et~al.}{2023}]{Amon2023}
{Amon} A.,  et~al., 2023, \mn@doi [\mnras] {10.1093/mnras/stac2938}, \href {https://ui.adsabs.harvard.edu/abs/2023MNRAS.518..477A} {518, 477}

\bibitem[\protect\citeauthoryear{{Aric{\`o}} \& {Angulo}}{{Aric{\`o}} \& {Angulo}}{2024}]{Arico2024}
{Aric{\`o}} G.,  {Angulo} R.~E.,  2024, \mn@doi [\aap] {10.1051/0004-6361/202451055}, \href {https://ui.adsabs.harvard.edu/abs/2024A&A...690A.188A} {690, A188}

\bibitem[\protect\citeauthoryear{{Aric{\`o}}, {Angulo}, {Contreras}, {Ondaro-Mallea}, {Pellejero-Iba{\~n}ez}  \& {Zennaro}}{{Aric{\`o}} et~al.}{2021}]{Arico2021emulator}
{Aric{\`o}} G.,  {Angulo} R.~E.,  {Contreras} S.,  {Ondaro-Mallea} L.,  {Pellejero-Iba{\~n}ez} M.,   {Zennaro} M.,  2021, \mn@doi [\mnras] {10.1093/mnras/stab1911}, \href {https://ui.adsabs.harvard.edu/abs/2021MNRAS.506.4070A} {506, 4070}

\bibitem[\protect\citeauthoryear{{Aric{\`o}}, {Angulo}, {Zennaro}, {Contreras}, {Chen}  \& {Hern{\'a}ndez-Monteagudo}}{{Aric{\`o}} et~al.}{2023}]{Arico2023}
{Aric{\`o}} G.,  {Angulo} R.~E.,  {Zennaro} M.,  {Contreras} S.,  {Chen} A.,   {Hern{\'a}ndez-Monteagudo} C.,  2023, \mn@doi [\aap] {10.1051/0004-6361/202346539}, \href {https://ui.adsabs.harvard.edu/abs/2023A&A...678A.109A} {678, A109}

\bibitem[\protect\citeauthoryear{{Behroozi}, {Wechsler}, {Hearin}  \& {Conroy}}{{Behroozi} et~al.}{2019}]{Behroozi2019}
{Behroozi} P.,  {Wechsler} R.~H.,  {Hearin} A.~P.,   {Conroy} C.,  2019, \mn@doi [\mnras] {10.1093/mnras/stz1182}, \href {https://ui.adsabs.harvard.edu/abs/2019MNRAS.488.3143B} {488, 3143}

\bibitem[\protect\citeauthoryear{{Bigwood} et~al.,}{{Bigwood} et~al.}{2024}]{Bigwood2024}
{Bigwood} L.,  et~al., 2024, \mn@doi [\mnras] {10.1093/mnras/stae2100}, \href {https://ui.adsabs.harvard.edu/abs/2024MNRAS.534..655B} {534, 655}

\bibitem[\protect\citeauthoryear{{Bleem} et~al.,}{{Bleem} et~al.}{2022}]{Bleem2022}
{Bleem} L.~E.,  et~al., 2022, \mn@doi [\apjs] {10.3847/1538-4365/ac35e9}, \href {https://ui.adsabs.harvard.edu/abs/2022ApJS..258...36B} {258, 36}

\bibitem[\protect\citeauthoryear{{Bolliet}, {Comis}, {Komatsu}  \& {Mac{\'\i}as-P{\'e}rez}}{{Bolliet} et~al.}{2018}]{Bolliet2018}
{Bolliet} B.,  {Comis} B.,  {Komatsu} E.,   {Mac{\'\i}as-P{\'e}rez} J.~F.,  2018, \mn@doi [\mnras] {10.1093/mnras/sty823}, \href {https://ui.adsabs.harvard.edu/abs/2018MNRAS.477.4957B} {477, 4957}

\bibitem[\protect\citeauthoryear{{Borrow}, {Schaller}, {Bower}  \& {Schaye}}{{Borrow} et~al.}{2022}]{Borrow2022sphenix}
{Borrow} J.,  {Schaller} M.,  {Bower} R.~G.,   {Schaye} J.,  2022, \mn@doi [\mnras] {10.1093/mnras/stab3166}, \href {https://ui.adsabs.harvard.edu/abs/2022MNRAS.511.2367B} {511, 2367}

\bibitem[\protect\citeauthoryear{{Brown}, {McCarthy}, {Stafford}  \& {Font}}{{Brown} et~al.}{2022}]{Brown2022}
{Brown} S.~T.,  {McCarthy} I.~G.,  {Stafford} S.~G.,   {Font} A.~S.,  2022, \mn@doi [\mnras] {10.1093/mnras/stab3394}, \href {https://ui.adsabs.harvard.edu/abs/2022MNRAS.509.5685B} {509, 5685}

\bibitem[\protect\citeauthoryear{{Bulbul} et~al.,}{{Bulbul} et~al.}{2024}]{Bulbul2024}
{Bulbul} E.,  et~al., 2024, \mn@doi [\aap] {10.1051/0004-6361/202348264}, \href {https://ui.adsabs.harvard.edu/abs/2024A&A...685A.106B} {685, A106}

\bibitem[\protect\citeauthoryear{{Bullock}, {Kolatt}, {Sigad}, {Somerville}, {Kravtsov}, {Klypin}, {Primack}  \& {Dekel}}{{Bullock} et~al.}{2001}]{Bullock2001}
{Bullock} J.~S.,  {Kolatt} T.~S.,  {Sigad} Y.,  {Somerville} R.~S.,  {Kravtsov} A.~V.,  {Klypin} A.~A.,  {Primack} J.~R.,   {Dekel} A.,  2001, \mn@doi [\mnras] {10.1046/j.1365-8711.2001.04068.x}, \href {https://ui.adsabs.harvard.edu/abs/2001MNRAS.321..559B} {321, 559}

\bibitem[\protect\citeauthoryear{{Castro}, {Borgani}, {Dolag}, {Marra}, {Quartin}, {Saro}  \& {Sefusatti}}{{Castro} et~al.}{2021}]{Castro2021}
{Castro} T.,  {Borgani} S.,  {Dolag} K.,  {Marra} V.,  {Quartin} M.,  {Saro} A.,   {Sefusatti} E.,  2021, \mn@doi [\mnras] {10.1093/mnras/staa3473}, \href {https://ui.adsabs.harvard.edu/abs/2021MNRAS.500.2316C} {500, 2316}

\bibitem[\protect\citeauthoryear{{Chisari} et~al.,}{{Chisari} et~al.}{2019}]{Chisari2019}
{Chisari} N.~E.,  et~al., 2019, \mn@doi [The Open Journal of Astrophysics] {10.21105/astro.1905.06082}, \href {https://ui.adsabs.harvard.edu/abs/2019OJAp....2E...4C} {2, 4}

\bibitem[\protect\citeauthoryear{{Choi} et~al.,}{{Choi} et~al.}{2020}]{Choi2020}
{Choi} S.~K.,  et~al., 2020, \mn@doi [\jcap] {10.1088/1475-7516/2020/12/045}, \href {https://ui.adsabs.harvard.edu/abs/2020JCAP...12..045C} {2020, 045}

\bibitem[\protect\citeauthoryear{{Correa}, {Wyithe}, {Schaye}  \& {Duffy}}{{Correa} et~al.}{2015}]{Correa2015}
{Correa} C.~A.,  {Wyithe} J. S.~B.,  {Schaye} J.,   {Duffy} A.~R.,  2015, \mn@doi [\mnras] {10.1093/mnras/stv1363}, \href {https://ui.adsabs.harvard.edu/abs/2015MNRAS.452.1217C} {452, 1217}

\bibitem[\protect\citeauthoryear{{Coulton} et~al.,}{{Coulton} et~al.}{2024}]{Coulton2024}
{Coulton} W.~R.,  et~al., 2024, \mn@doi [arXiv e-prints] {10.48550/arXiv.2401.13033}, \href {https://ui.adsabs.harvard.edu/abs/2024arXiv240113033C} {p. arXiv:2401.13033}

\bibitem[\protect\citeauthoryear{{Dav{\'e}}, {Angl{\'e}s-Alc{\'a}zar}, {Narayanan}, {Li}, {Rafieferantsoa}  \& {Appleby}}{{Dav{\'e}} et~al.}{2019}]{Dave2019}
{Dav{\'e}} R.,  {Angl{\'e}s-Alc{\'a}zar} D.,  {Narayanan} D.,  {Li} Q.,  {Rafieferantsoa} M.~H.,   {Appleby} S.,  2019, \mn@doi [\mnras] {10.1093/mnras/stz937}, \href {https://ui.adsabs.harvard.edu/abs/2019MNRAS.486.2827D} {486, 2827}

\bibitem[\protect\citeauthoryear{{Debackere}, {Schaye}  \& {Hoekstra}}{{Debackere} et~al.}{2020}]{Debackere2020}
{Debackere} S. N.~B.,  {Schaye} J.,   {Hoekstra} H.,  2020, \mn@doi [\mnras] {10.1093/mnras/stz3446}, \href {https://ui.adsabs.harvard.edu/abs/2020MNRAS.492.2285D} {492, 2285}

\bibitem[\protect\citeauthoryear{{Diemer}}{{Diemer}}{2018}]{Diemer2018}
{Diemer} B.,  2018, \mn@doi [\apjs] {10.3847/1538-4365/aaee8c}, \href {https://ui.adsabs.harvard.edu/abs/2018ApJS..239...35D} {239, 35}

\bibitem[\protect\citeauthoryear{{Diemer} \& {Kravtsov}}{{Diemer} \& {Kravtsov}}{2015}]{Diemer2015}
{Diemer} B.,  {Kravtsov} A.~V.,  2015, \mn@doi [\apj] {10.1088/0004-637X/799/1/108}, \href {https://ui.adsabs.harvard.edu/abs/2015ApJ...799..108D} {799, 108}

\bibitem[\protect\citeauthoryear{{Eckert} et~al.,}{{Eckert} et~al.}{2016}]{Eckert2016}
{Eckert} D.,  et~al., 2016, \mn@doi [\aap] {10.1051/0004-6361/201527293}, \href {https://ui.adsabs.harvard.edu/abs/2016A&A...592A..12E} {592, A12}

\bibitem[\protect\citeauthoryear{{Eke}, {Navarro}  \& {Steinmetz}}{{Eke} et~al.}{2001}]{Eke2001}
{Eke} V.~R.,  {Navarro} J.~F.,   {Steinmetz} M.,  2001, \mn@doi [\apj] {10.1086/321345}, \href {https://ui.adsabs.harvard.edu/abs/2001ApJ...554..114E} {554, 114}

\bibitem[\protect\citeauthoryear{{Elahi}, {Ca{\~n}as}, {Poulton}, {Tobar}, {Willis}, {Lagos}, {Power}  \& {Robotham}}{{Elahi} et~al.}{2019}]{Elahi2019vr}
{Elahi} P.~J.,  {Ca{\~n}as} R.,  {Poulton} R. J.~J.,  {Tobar} R.~J.,  {Willis} J.~S.,  {Lagos} C. d.~P.,  {Power} C.,   {Robotham} A. S.~G.,  2019, \mn@doi [\pasa] {10.1017/pasa.2019.12}, \href {https://ui.adsabs.harvard.edu/abs/2019PASA...36...21E} {36, e021}

\bibitem[\protect\citeauthoryear{{Elbers}, {Frenk}, {Jenkins}, {Li}  \& {Pascoli}}{{Elbers} et~al.}{2021}]{Elbers2021}
{Elbers} W.,  {Frenk} C.~S.,  {Jenkins} A.,  {Li} B.,   {Pascoli} S.,  2021, \mn@doi [\mnras] {10.1093/mnras/stab2260}, \href {https://ui.adsabs.harvard.edu/abs/2021MNRAS.507.2614E} {507, 2614}

\bibitem[\protect\citeauthoryear{{Elbers}, {Frenk}, {Jenkins}, {Li}  \& {Pascoli}}{{Elbers} et~al.}{2022}]{Elbers2022a}
{Elbers} W.,  {Frenk} C.~S.,  {Jenkins} A.,  {Li} B.,   {Pascoli} S.,  2022, \mn@doi [\mnras] {10.1093/mnras/stac2365}, \href {https://ui.adsabs.harvard.edu/abs/2022MNRAS.516.3821E} {516, 3821}

\bibitem[\protect\citeauthoryear{{Elbers} et~al.,}{{Elbers} et~al.}{2025}]{Elbers2025}
{Elbers} W.,  et~al., 2025, \mn@doi [\mnras] {10.1093/mnras/staf093}, \href {https://ui.adsabs.harvard.edu/abs/2025MNRAS.537.2160E} {537, 2160}

\bibitem[\protect\citeauthoryear{{Forouhar Moreno}, {Helly}, {McGibbon}, {Schaye}, {Schaller}, {Han}  \& {Kugel}}{{Forouhar Moreno} et~al.}{2025}]{ForouharMoreno2025}
{Forouhar Moreno} V.~J.,  {Helly} J.,  {McGibbon} R.,  {Schaye} J.,  {Schaller} M.,  {Han} J.,   {Kugel} R.,  2025, \mn@doi [arXiv e-prints] {10.48550/arXiv.2502.06932}, \href {https://ui.adsabs.harvard.edu/abs/2025arXiv250206932F} {p. arXiv:2502.06932}

\bibitem[\protect\citeauthoryear{{Giles} et~al.,}{{Giles} et~al.}{2022}]{Giles2022}
{Giles} P.~A.,  et~al., 2022, \mn@doi [\mnras] {10.1093/mnras/stab3626}, \href {https://ui.adsabs.harvard.edu/abs/2022MNRAS.511.1227G} {511, 1227}

\bibitem[\protect\citeauthoryear{{Giri} \& {Schneider}}{{Giri} \& {Schneider}}{2021}]{Giri2021}
{Giri} S.~K.,  {Schneider} A.,  2021, \mn@doi [\jcap] {10.1088/1475-7516/2021/12/046}, \href {https://ui.adsabs.harvard.edu/abs/2021JCAP...12..046G} {2021, 046}

\bibitem[\protect\citeauthoryear{Gorski, Hivon, Banday, Wandelt, Hansen, Reinecke  \& Bartelmann}{Gorski et~al.}{2005}]{Gorski2005}
Gorski K.~M.,  Hivon E.,  Banday A.~J.,  Wandelt B.~D.,  Hansen F.~K.,  Reinecke M.,   Bartelmann M.,  2005, \mn@doi [\apj] {10.1086/427976}, 622, 759

\bibitem[\protect\citeauthoryear{{Grandis}, {Aric{\`o}}, {Schneider}  \& {Linke}}{{Grandis} et~al.}{2024}]{Grandis2024}
{Grandis} S.,  {Aric{\`o}} G.,  {Schneider} A.,   {Linke} L.,  2024, \mn@doi [\mnras] {10.1093/mnras/stae259}, \href {https://ui.adsabs.harvard.edu/abs/2024MNRAS.528.4379G} {528, 4379}

\bibitem[\protect\citeauthoryear{{Greco}, {Hill}, {Spergel}  \& {Battaglia}}{{Greco} et~al.}{2015}]{Greco2015}
{Greco} J.~P.,  {Hill} J.~C.,  {Spergel} D.~N.,   {Battaglia} N.,  2015, \mn@doi [\apj] {10.1088/0004-637X/808/2/151}, \href {https://ui.adsabs.harvard.edu/abs/2015ApJ...808..151G} {808, 151}

\bibitem[\protect\citeauthoryear{{Hadzhiyska} et~al.,}{{Hadzhiyska} et~al.}{2024a}]{Hadzhiyska2024b}
{Hadzhiyska} B.,  et~al., 2024a, \mn@doi [arXiv e-prints] {10.48550/arXiv.2407.07152}, \href {https://ui.adsabs.harvard.edu/abs/2024arXiv240707152H} {p. arXiv:2407.07152}

\bibitem[\protect\citeauthoryear{{Hadzhiyska}, {Ferraro}, {Ried Guachalla}  \& {Schaan}}{{Hadzhiyska} et~al.}{2024b}]{Hadzhiyska2024a}
{Hadzhiyska} B.,  {Ferraro} S.,  {Ried Guachalla} B.,   {Schaan} E.,  2024b, \mn@doi [\prd] {10.1103/PhysRevD.109.103534}, \href {https://ui.adsabs.harvard.edu/abs/2024PhRvD.109j3534H} {109, 103534}

\bibitem[\protect\citeauthoryear{{Hahn}, {Rampf}  \& {Uhlemann}}{{Hahn} et~al.}{2021}]{Hahn2021}
{Hahn} O.,  {Rampf} C.,   {Uhlemann} C.,  2021, \mn@doi [\mnras] {10.1093/mnras/staa3773}, \href {https://ui.adsabs.harvard.edu/abs/2021MNRAS.503..426H} {503, 426}

\bibitem[\protect\citeauthoryear{{Han}, {Jing}, {Wang}  \& {Wang}}{{Han} et~al.}{2012}]{Han2012}
{Han} J.,  {Jing} Y.~P.,  {Wang} H.,   {Wang} W.,  2012, \mn@doi [\mnras] {10.1111/j.1365-2966.2012.22111.x}, \href {https://ui.adsabs.harvard.edu/abs/2012MNRAS.427.2437H} {427, 2437}

\bibitem[\protect\citeauthoryear{{Han}, {Cole}, {Frenk}, {Benitez-Llambay}  \& {Helly}}{{Han} et~al.}{2018}]{Han2018}
{Han} J.,  {Cole} S.,  {Frenk} C.~S.,  {Benitez-Llambay} A.,   {Helly} J.,  2018, \mn@doi [\mnras] {10.1093/mnras/stx2792}, \href {https://ui.adsabs.harvard.edu/abs/2018MNRAS.474..604H} {474, 604}

\bibitem[\protect\citeauthoryear{{Henden}, {Puchwein}, {Shen}  \& {Sijacki}}{{Henden} et~al.}{2018}]{Henden2018}
{Henden} N.~A.,  {Puchwein} E.,  {Shen} S.,   {Sijacki} D.,  2018, \mn@doi [\mnras] {10.1093/mnras/sty1780}, \href {https://ui.adsabs.harvard.edu/abs/2018MNRAS.479.5385H} {479, 5385}

\bibitem[\protect\citeauthoryear{{Henden}, {Puchwein}  \& {Sijacki}}{{Henden} et~al.}{2020}]{Henden2020}
{Henden} N.~A.,  {Puchwein} E.,   {Sijacki} D.,  2020, \mn@doi [\mnras] {10.1093/mnras/staa2235}, \href {https://ui.adsabs.harvard.edu/abs/2020MNRAS.498.2114H} {498, 2114}

\bibitem[\protect\citeauthoryear{{Heymans} et~al.,}{{Heymans} et~al.}{2021}]{Heymans2021}
{Heymans} C.,  et~al., 2021, \mn@doi [\aap] {10.1051/0004-6361/202039063}, \href {https://ui.adsabs.harvard.edu/abs/2021A&A...646A.140H} {646, A140}

\bibitem[\protect\citeauthoryear{{Krause} et~al.,}{{Krause} et~al.}{2021}]{Krause2022}
{Krause} E.,  et~al., 2021, arXiv e-prints, \href {https://ui.adsabs.harvard.edu/abs/2021arXiv210513548K} {p. arXiv:2105.13548}

\bibitem[\protect\citeauthoryear{{Kugel} et~al.,}{{Kugel} et~al.}{2023}]{Kugel2023}
{Kugel} R.,  et~al., 2023, \mn@doi [\mnras] {10.1093/mnras/stad2540}, \href {https://ui.adsabs.harvard.edu/abs/2023MNRAS.526.6103K} {526, 6103}

\bibitem[\protect\citeauthoryear{{Kugel}, {Schaye}, {Schaller}, {McCarthy}, {Braspenning}, {Helly}, {Moreno}  \& {McGibbon}}{{Kugel} et~al.}{2024}]{Kugel2024}
{Kugel} R.,  {Schaye} J.,  {Schaller} M.,  {McCarthy} I.~G.,  {Braspenning} J.,  {Helly} J.~C.,  {Moreno} V. J.~F.,   {McGibbon} R.~J.,  2024, \mn@doi [\mnras] {10.1093/mnras/stae2218}, \href {https://ui.adsabs.harvard.edu/abs/2024MNRAS.tmp.2195K} {}

\bibitem[\protect\citeauthoryear{{Kukstas} et~al.,}{{Kukstas} et~al.}{2023}]{Kukstas2023}
{Kukstas} E.,  et~al., 2023, \mn@doi [\mnras] {10.1093/mnras/stac3438}, \href {https://ui.adsabs.harvard.edu/abs/2023MNRAS.518.4782K} {518, 4782}

\bibitem[\protect\citeauthoryear{{Le Brun}, {McCarthy}, {Schaye}  \& {Ponman}}{{Le Brun} et~al.}{2014}]{LeBrun2014}
{Le Brun} A. M.~C.,  {McCarthy} I.~G.,  {Schaye} J.,   {Ponman} T.~J.,  2014, \mn@doi [\mnras] {10.1093/mnras/stu608}, \href {https://ui.adsabs.harvard.edu/abs/2014MNRAS.441.1270L} {441, 1270}

\bibitem[\protect\citeauthoryear{{Le Brun}, {McCarthy}  \& {Melin}}{{Le Brun} et~al.}{2015}]{LeBrun2015}
{Le Brun} A. M.~C.,  {McCarthy} I.~G.,   {Melin} J.-B.,  2015, \mn@doi [\mnras] {10.1093/mnras/stv1172}, \href {https://ui.adsabs.harvard.edu/abs/2015MNRAS.451.3868L} {451, 3868}

\bibitem[\protect\citeauthoryear{{Lovisari}, {Reiprich}  \& {Schellenberger}}{{Lovisari} et~al.}{2015}]{Lovisari2015}
{Lovisari} L.,  {Reiprich} T.~H.,   {Schellenberger} G.,  2015, \mn@doi [\aap] {10.1051/0004-6361/201423954}, \href {https://ui.adsabs.harvard.edu/abs/2015A&A...573A.118L} {573, A118}

\bibitem[\protect\citeauthoryear{{Macquart} et~al.,}{{Macquart} et~al.}{2020}]{Macquart2020}
{Macquart} J.~P.,  et~al., 2020, \mn@doi [\nat] {10.1038/s41586-020-2300-2}, \href {https://ui.adsabs.harvard.edu/abs/2020Natur.581..391M} {581, 391}

\bibitem[\protect\citeauthoryear{{Maraston} et~al.,}{{Maraston} et~al.}{2013}]{Maraston2013}
{Maraston} C.,  et~al., 2013, \mn@doi [\mnras] {10.1093/mnras/stt1424}, \href {https://ui.adsabs.harvard.edu/abs/2013MNRAS.435.2764M} {435, 2764}

\bibitem[\protect\citeauthoryear{{Marini} et~al.,}{{Marini} et~al.}{2025}]{Marini2025}
{Marini} I.,  et~al., 2025, \mn@doi [\aap] {10.1051/0004-6361/202452028}, \href {https://ui.adsabs.harvard.edu/abs/2025A&A...694A.207M} {694, A207}

\bibitem[\protect\citeauthoryear{{McCarthy}, {Le Brun}, {Schaye}  \& {Holder}}{{McCarthy} et~al.}{2014}]{McCarthy2014}
{McCarthy} I.~G.,  {Le Brun} A.~M.~C.,  {Schaye} J.,   {Holder} G.~P.,  2014, \mn@doi [\mnras] {10.1093/mnras/stu543}, \href {https://ui.adsabs.harvard.edu/abs/2014MNRAS.440.3645M} {440, 3645}

\bibitem[\protect\citeauthoryear{{McCarthy}, {Schaye}, {Bird}  \& {Le Brun}}{{McCarthy} et~al.}{2017}]{McCarthy2017}
{McCarthy} I.~G.,  {Schaye} J.,  {Bird} S.,   {Le Brun} A. M.~C.,  2017, \mn@doi [\mnras] {10.1093/mnras/stw2792}, \href {https://ui.adsabs.harvard.edu/abs/2017MNRAS.465.2936M} {465, 2936}

\bibitem[\protect\citeauthoryear{{McCarthy}, {Bird}, {Schaye}, {Harnois-Deraps}, {Font}  \& {van Waerbeke}}{{McCarthy} et~al.}{2018}]{McCarthy2018}
{McCarthy} I.~G.,  {Bird} S.,  {Schaye} J.,  {Harnois-Deraps} J.,  {Font} A.~S.,   {van Waerbeke} L.,  2018, \mn@doi [\mnras] {10.1093/mnras/sty377}, \href {https://ui.adsabs.harvard.edu/abs/2018MNRAS.476.2999M} {476, 2999}

\bibitem[\protect\citeauthoryear{{McCarthy} et~al.,}{{McCarthy} et~al.}{2023}]{McCarthy2023}
{McCarthy} I.~G.,  et~al., 2023, \mn@doi [\mnras] {10.1093/mnras/stad3107}, \href {https://ui.adsabs.harvard.edu/abs/2023MNRAS.526.5494M} {526, 5494}

\bibitem[\protect\citeauthoryear{{Mead}, {Tr{\"o}ster}, {Heymans}, {Van Waerbeke}  \& {McCarthy}}{{Mead} et~al.}{2020}]{Mead2020}
{Mead} A.~J.,  {Tr{\"o}ster} T.,  {Heymans} C.,  {Van Waerbeke} L.,   {McCarthy} I.~G.,  2020, \mn@doi [\aap] {10.1051/0004-6361/202038308}, \href {https://ui.adsabs.harvard.edu/abs/2020A&A...641A.130M} {641, A130}

\bibitem[\protect\citeauthoryear{{Melin}, {Bartlett}, {Tarr{\'\i}o}  \& {Pratt}}{{Melin} et~al.}{2021}]{Melin2021}
{Melin} J.~B.,  {Bartlett} J.~G.,  {Tarr{\'\i}o} P.,   {Pratt} G.~W.,  2021, \mn@doi [\aap] {10.1051/0004-6361/202039471}, \href {https://ui.adsabs.harvard.edu/abs/2021A&A...647A.106M} {647, A106}

\bibitem[\protect\citeauthoryear{{Mummery}, {McCarthy}, {Bird}  \& {Schaye}}{{Mummery} et~al.}{2017}]{Mummery2017}
{Mummery} B.~O.,  {McCarthy} I.~G.,  {Bird} S.,   {Schaye} J.,  2017, \mn@doi [\mnras] {10.1093/mnras/stx1469}, \href {https://ui.adsabs.harvard.edu/abs/2017MNRAS.471..227M} {471, 227}

\bibitem[\protect\citeauthoryear{{Pearson} et~al.,}{{Pearson} et~al.}{2017}]{Pearson2017}
{Pearson} R.~J.,  et~al., 2017, \mn@doi [\mnras] {10.1093/mnras/stx1081}, \href {https://ui.adsabs.harvard.edu/abs/2017MNRAS.469.3489P} {469, 3489}

\bibitem[\protect\citeauthoryear{{Peebles}}{{Peebles}}{2024}]{Peebles2024}
{Peebles} P.~J.~E.,  2024, \mn@doi [arXiv e-prints] {10.48550/arXiv.2405.18307}, \href {https://ui.adsabs.harvard.edu/abs/2024arXiv240518307P} {p. arXiv:2405.18307}

\bibitem[\protect\citeauthoryear{{Pillepich} et~al.,}{{Pillepich} et~al.}{2018}]{Pillepich2018TNGmethod}
{Pillepich} A.,  et~al., 2018, \mn@doi [\mnras] {10.1093/mnras/stx2656}, \href {https://ui.adsabs.harvard.edu/abs/2018MNRAS.473.4077P} {473, 4077}

\bibitem[\protect\citeauthoryear{{Planck Collaboration} et~al.,}{{Planck Collaboration} et~al.}{2013}]{Planck2013}
{Planck Collaboration} et~al., 2013, \mn@doi [\aap] {10.1051/0004-6361/201220941}, \href {https://ui.adsabs.harvard.edu/abs/2013A&A...557A..52P} {557, A52}

\bibitem[\protect\citeauthoryear{{Planck Collaboration} et~al.,}{{Planck Collaboration} et~al.}{2014}]{Planck2014}
{Planck Collaboration} et~al., 2014, \mn@doi [\aap] {10.1051/0004-6361/201321522}, \href {https://ui.adsabs.harvard.edu/abs/2014A&A...571A..21P} {571, A21}

\bibitem[\protect\citeauthoryear{{Planck Collaboration} et~al.,}{{Planck Collaboration} et~al.}{2016a}]{Planck2016_tSZ_PS}
{Planck Collaboration} et~al., 2016a, \mn@doi [\aap] {10.1051/0004-6361/201525826}, \href {https://ui.adsabs.harvard.edu/abs/2016A&A...594A..22P} {594, A22}

\bibitem[\protect\citeauthoryear{{Planck Collaboration} et~al.,}{{Planck Collaboration} et~al.}{2016b}]{Planck2016}
{Planck Collaboration} et~al., 2016b, \mn@doi [\aap] {10.1051/0004-6361/201525833}, \href {https://ui.adsabs.harvard.edu/abs/2016A&A...594A..24P} {594, A24}

\bibitem[\protect\citeauthoryear{{Planck Collaboration} et~al.,}{{Planck Collaboration} et~al.}{2020}]{Planck2020cosmopars}
{Planck Collaboration} et~al., 2020, \mn@doi [\aap] {10.1051/0004-6361/201833910}, \href {https://ui.adsabs.harvard.edu/abs/2020A&A...641A...6P} {641, A6}

\bibitem[\protect\citeauthoryear{{Planelles}, {Borgani}, {Fabjan}, {Killedar}, {Murante}, {Granato}, {Ragone-Figueroa}  \& {Dolag}}{{Planelles} et~al.}{2014}]{Planelles2014}
{Planelles} S.,  {Borgani} S.,  {Fabjan} D.,  {Killedar} M.,  {Murante} G.,  {Granato} G.~L.,  {Ragone-Figueroa} C.,   {Dolag} K.,  2014, \mn@doi [\mnras] {10.1093/mnras/stt2141}, \href {https://ui.adsabs.harvard.edu/abs/2014MNRAS.438..195P} {438, 195}

\bibitem[\protect\citeauthoryear{{Popesso} et~al.,}{{Popesso} et~al.}{2024}]{Popesso2024}
{Popesso} P.,  et~al., 2024, \mn@doi [\mnras] {10.1093/mnras/stad3253}, \href {https://ui.adsabs.harvard.edu/abs/2024MNRAS.527..895P} {527, 895}

\bibitem[\protect\citeauthoryear{{Preston}, {Amon}  \& {Efstathiou}}{{Preston} et~al.}{2023}]{Preston2023}
{Preston} C.,  {Amon} A.,   {Efstathiou} G.,  2023, \mn@doi [\mnras] {10.1093/mnras/stad2573}, \href {https://ui.adsabs.harvard.edu/abs/2023MNRAS.525.5554P} {525, 5554}

\bibitem[\protect\citeauthoryear{{Reichardt} et~al.,}{{Reichardt} et~al.}{2021}]{Reichardt2021}
{Reichardt} C.~L.,  et~al., 2021, \mn@doi [\apj] {10.3847/1538-4357/abd407}, \href {https://ui.adsabs.harvard.edu/abs/2021ApJ...908..199R} {908, 199}

\bibitem[\protect\citeauthoryear{{Remazeilles} \& {Chluba}}{{Remazeilles} \& {Chluba}}{2025}]{Remazeilles2025}
{Remazeilles} M.,  {Chluba} J.,  2025, \mn@doi [\mnras] {10.1093/mnras/staf384}, \href {https://ui.adsabs.harvard.edu/abs/2025MNRAS.538.1576R} {538, 1576}

\bibitem[\protect\citeauthoryear{{Remazeilles}, {Bolliet}, {Rotti}  \& {Chluba}}{{Remazeilles} et~al.}{2019}]{Remazeilles2019}
{Remazeilles} M.,  {Bolliet} B.,  {Rotti} A.,   {Chluba} J.,  2019, \mn@doi [\mnras] {10.1093/mnras/sty3352}, \href {https://ui.adsabs.harvard.edu/abs/2019MNRAS.483.3459R} {483, 3459}

\bibitem[\protect\citeauthoryear{{Ried Guachalla}, {Schaan}, {Hadzhiyska}  \& {Ferraro}}{{Ried Guachalla} et~al.}{2024}]{Ried2024}
{Ried Guachalla} B.,  {Schaan} E.,  {Hadzhiyska} B.,   {Ferraro} S.,  2024, \mn@doi [\prd] {10.1103/PhysRevD.109.103533}, \href {https://ui.adsabs.harvard.edu/abs/2024PhRvD.109j3533R} {109, 103533}

\bibitem[\protect\citeauthoryear{{Rogers}, {Hlo{\v{z}}ek}, {Lagu{\"e}}, {Ivanov}, {Philcox}, {Cabass}, {Akitsu}  \& {Marsh}}{{Rogers} et~al.}{2023}]{Rogers2023}
{Rogers} K.~K.,  {Hlo{\v{z}}ek} R.,  {Lagu{\"e}} A.,  {Ivanov} M.~M.,  {Philcox} O. H.~E.,  {Cabass} G.,  {Akitsu} K.,   {Marsh} D. J.~E.,  2023, \mn@doi [\jcap] {10.1088/1475-7516/2023/06/023}, \href {https://ui.adsabs.harvard.edu/abs/2023JCAP...06..023R} {2023, 023}

\bibitem[\protect\citeauthoryear{{Salcido} \& {McCarthy}}{{Salcido} \& {McCarthy}}{2024}]{Salcido2024}
{Salcido} J.,  {McCarthy} I.~G.,  2024, \mn@doi [arXiv e-prints] {10.48550/arXiv.2409.05716}, \href {https://ui.adsabs.harvard.edu/abs/2024arXiv240905716S} {p. arXiv:2409.05716}

\bibitem[\protect\citeauthoryear{{Salcido}, {McCarthy}, {Kwan}, {Upadhye}  \& {Font}}{{Salcido} et~al.}{2023}]{Salcido2023}
{Salcido} J.,  {McCarthy} I.~G.,  {Kwan} J.,  {Upadhye} A.,   {Font} A.~S.,  2023, \mn@doi [\mnras] {10.1093/mnras/stad1474}, \href {https://ui.adsabs.harvard.edu/abs/2023MNRAS.523.2247S} {523, 2247}

\bibitem[\protect\citeauthoryear{{Schaan} et~al.,}{{Schaan} et~al.}{2021}]{Schaan2021}
{Schaan} E.,  et~al., 2021, \mn@doi [\prd] {10.1103/PhysRevD.103.063513}, \href {https://ui.adsabs.harvard.edu/abs/2021PhRvD.103f3513S} {103, 063513}

\bibitem[\protect\citeauthoryear{{Schaller}, {Schaye}, {Kugel}, {Broxterman}  \& {van Daalen}}{{Schaller} et~al.}{2024a}]{Schaller2024b}
{Schaller} M.,  {Schaye} J.,  {Kugel} R.,  {Broxterman} J.~C.,   {van Daalen} M.~P.,  2024a, \mn@doi [arXiv e-prints] {10.48550/arXiv.2410.17109}, \href {https://ui.adsabs.harvard.edu/abs/2024arXiv241017109S} {p. arXiv:2410.17109}

\bibitem[\protect\citeauthoryear{{Schaller} et~al.,}{{Schaller} et~al.}{2024b}]{Schaller2024a}
{Schaller} M.,  et~al., 2024b, \mn@doi [\mnras] {10.1093/mnras/stae922}, \href {https://ui.adsabs.harvard.edu/abs/2024MNRAS.530.2378S} {530, 2378}

\bibitem[\protect\citeauthoryear{{Schaye} et~al.,}{{Schaye} et~al.}{2015}]{Schaye2015}
{Schaye} J.,  et~al., 2015, \mn@doi [\mnras] {10.1093/mnras/stu2058}, \href {https://ui.adsabs.harvard.edu/abs/2015MNRAS.446..521S} {446, 521}

\bibitem[\protect\citeauthoryear{{Schaye} et~al.,}{{Schaye} et~al.}{2023}]{Schaye2023}
{Schaye} J.,  et~al., 2023, \mn@doi [\mnras] {10.1093/mnras/stad2419}, \href {https://ui.adsabs.harvard.edu/abs/2023MNRAS.526.4978S} {526, 4978}

\bibitem[\protect\citeauthoryear{{Schneider} \& {Teyssier}}{{Schneider} \& {Teyssier}}{2015}]{Schneider2015}
{Schneider} A.,  {Teyssier} R.,  2015, \mn@doi [\jcap] {10.1088/1475-7516/2015/12/049}, \href {https://ui.adsabs.harvard.edu/abs/2015JCAP...12..049S} {2015, 049}

\bibitem[\protect\citeauthoryear{{Schneider}, {Teyssier}, {Stadel}, {Chisari}, {Le Brun}, {Amara}  \& {Refregier}}{{Schneider} et~al.}{2019}]{Schneider2019}
{Schneider} A.,  {Teyssier} R.,  {Stadel} J.,  {Chisari} N.~E.,  {Le Brun} A. M.~C.,  {Amara} A.,   {Refregier} A.,  2019, \mn@doi [\jcap] {10.1088/1475-7516/2019/03/020}, \href {https://ui.adsabs.harvard.edu/abs/2019JCAP...03..020S} {2019, 020}

\bibitem[\protect\citeauthoryear{{Schneider}, {Stoira}, {Refregier}, {Weiss}, {Knabenhans}, {Stadel}  \& {Teyssier}}{{Schneider} et~al.}{2020}]{Schneider2020}
{Schneider} A.,  {Stoira} N.,  {Refregier} A.,  {Weiss} A.~J.,  {Knabenhans} M.,  {Stadel} J.,   {Teyssier} R.,  2020, \mn@doi [\jcap] {10.1088/1475-7516/2020/04/019}, \href {https://ui.adsabs.harvard.edu/abs/2020JCAP...04..019S} {2020, 019}

\bibitem[\protect\citeauthoryear{{Schneider}, {Giri}, {Amodeo}  \& {Refregier}}{{Schneider} et~al.}{2022}]{Schneider2022}
{Schneider} A.,  {Giri} S.~K.,  {Amodeo} S.,   {Refregier} A.,  2022, \mn@doi [\mnras] {10.1093/mnras/stac1493}, \href {https://ui.adsabs.harvard.edu/abs/2022MNRAS.514.3802S} {514, 3802}

\bibitem[\protect\citeauthoryear{{Secco}, {Samuroff}  et~al.}{{Secco} et~al.}{2022}]{Secco2022}
{Secco} L.~F.,  {Samuroff} S.,   et~al., 2022, \mn@doi [\prd] {10.1103/PhysRevD.105.023515}, \href {https://ui.adsabs.harvard.edu/abs/2022PhRvD.105b3515S} {105, 023515}

\bibitem[\protect\citeauthoryear{{Semboloni}, {Hoekstra}, {Schaye}, {van Daalen}  \& {McCarthy}}{{Semboloni} et~al.}{2011}]{Semboloni2011}
{Semboloni} E.,  {Hoekstra} H.,  {Schaye} J.,  {van Daalen} M.~P.,   {McCarthy} I.~G.,  2011, \mn@doi [\mnras] {10.1111/j.1365-2966.2011.19385.x}, \href {https://ui.adsabs.harvard.edu/abs/2011MNRAS.417.2020S} {417, 2020}

\bibitem[\protect\citeauthoryear{{Semboloni}, {Hoekstra}  \& {Schaye}}{{Semboloni} et~al.}{2013}]{Semboloni2013}
{Semboloni} E.,  {Hoekstra} H.,   {Schaye} J.,  2013, \mn@doi [\mnras] {10.1093/mnras/stt1013}, \href {https://ui.adsabs.harvard.edu/abs/2013MNRAS.434..148S} {434, 148}

\bibitem[\protect\citeauthoryear{{Sheth} \& {Tormen}}{{Sheth} \& {Tormen}}{1999}]{Sheth1999}
{Sheth} R.~K.,  {Tormen} G.,  1999, \mn@doi [\mnras] {10.1046/j.1365-8711.1999.02692.x}, \href {https://ui.adsabs.harvard.edu/abs/1999MNRAS.308..119S} {308, 119}

\bibitem[\protect\citeauthoryear{{Sievers} et~al.,}{{Sievers} et~al.}{2013}]{Sievers2013}
{Sievers} J.~L.,  et~al., 2013, \mn@doi [\jcap] {10.1088/1475-7516/2013/10/060}, \href {https://ui.adsabs.harvard.edu/abs/2013JCAP...10..060S} {2013, 060}

\bibitem[\protect\citeauthoryear{{Springel} et~al.,}{{Springel} et~al.}{2018}]{Springel2018}
{Springel} V.,  et~al., 2018, \mn@doi [\mnras] {10.1093/mnras/stx3304}, \href {https://ui.adsabs.harvard.edu/abs/2018MNRAS.475..676S} {475, 676}

\bibitem[\protect\citeauthoryear{{Sun}, {Voit}, {Donahue}, {Jones}, {Forman}  \& {Vikhlinin}}{{Sun} et~al.}{2009}]{Sun2009}
{Sun} M.,  {Voit} G.~M.,  {Donahue} M.,  {Jones} C.,  {Forman} W.,   {Vikhlinin} A.,  2009, \mn@doi [\apj] {10.1088/0004-637X/693/2/1142}, \href {https://ui.adsabs.harvard.edu/abs/2009ApJ...693.1142S} {693, 1142}

\bibitem[\protect\citeauthoryear{{Tinker}, {Robertson}, {Kravtsov}, {Klypin}, {Warren}, {Yepes}  \& {Gottl{\"o}ber}}{{Tinker} et~al.}{2010}]{Tinker2010}
{Tinker} J.~L.,  {Robertson} B.~E.,  {Kravtsov} A.~V.,  {Klypin} A.,  {Warren} M.~S.,  {Yepes} G.,   {Gottl{\"o}ber} S.,  2010, \mn@doi [\apj] {10.1088/0004-637X/724/2/878}, \href {https://ui.adsabs.harvard.edu/abs/2010ApJ...724..878T} {724, 878}

\bibitem[\protect\citeauthoryear{{Tr{\"o}ster} et~al.,}{{Tr{\"o}ster} et~al.}{2022}]{Troster2022}
{Tr{\"o}ster} T.,  et~al., 2022, \mn@doi [\aap] {10.1051/0004-6361/202142197}, \href {https://ui.adsabs.harvard.edu/abs/2022A&A...660A..27T} {660, A27}

\bibitem[\protect\citeauthoryear{{\VAN{Van}{Van}{van}}~Daalen, {Schaye}, {Booth}  \& {Dalla Vecchia}}{{\VAN{Van}{Van}{van}}~Daalen et~al.}{2011}]{VanDaalen2011}
{\VAN{Van}{Van}{van}}~Daalen M.~P.,  {Schaye} J.,  {Booth} C.~M.,   {Dalla Vecchia} C.,  2011, \mn@doi [\mnras] {10.1111/j.1365-2966.2011.18981.x}, \href {https://ui.adsabs.harvard.edu/abs/2011MNRAS.415.3649V} {415, 3649}

\bibitem[\protect\citeauthoryear{{\VAN{Van}{Van}{van}}~Daalen, {McCarthy}  \& {Schaye}}{{\VAN{Van}{Van}{van}}~Daalen et~al.}{2020}]{VanDaalen2020}
{\VAN{Van}{Van}{van}}~Daalen M.~P.,  {McCarthy} I.~G.,   {Schaye} J.,  2020, \mn@doi [\mnras] {10.1093/mnras/stz3199}, \href {https://ui.adsabs.harvard.edu/abs/2020MNRAS.491.2424V} {491, 2424}

\bibitem[\protect\citeauthoryear{{Velliscig}, {van Daalen}, {Schaye}, {McCarthy}, {Cacciato}, {Le Brun}  \& {Dalla Vecchia}}{{Velliscig} et~al.}{2014}]{Velliscig2014}
{Velliscig} M.,  {van Daalen} M.~P.,  {Schaye} J.,  {McCarthy} I.~G.,  {Cacciato} M.,  {Le Brun} A. M.~C.,   {Dalla Vecchia} C.,  2014, \mn@doi [\mnras] {10.1093/mnras/stu1044}, \href {https://ui.adsabs.harvard.edu/abs/2014MNRAS.442.2641V} {442, 2641}

\bibitem[\protect\citeauthoryear{{Vogelsberger} et~al.,}{{Vogelsberger} et~al.}{2014}]{Vogelsberger2014}
{Vogelsberger} M.,  et~al., 2014, \mn@doi [\mnras] {10.1093/mnras/stu1536}, \href {https://ui.adsabs.harvard.edu/abs/2014MNRAS.444.1518V} {444, 1518}

\bibitem[\protect\citeauthoryear{{Yang}, {Cai}, {Cui}, {Dav{\'e}}, {Peacock}  \& {Sorini}}{{Yang} et~al.}{2022}]{Yang2022}
{Yang} T.,  {Cai} Y.-C.,  {Cui} W.,  {Dav{\'e}} R.,  {Peacock} J.~A.,   {Sorini} D.,  2022, \mn@doi [\mnras] {10.1093/mnras/stac2505}, \href {https://ui.adsabs.harvard.edu/abs/2022MNRAS.516.4084Y} {516, 4084}

\bibitem[\protect\citeauthoryear{{van Daalen} \& {Schaye}}{{van Daalen} \& {Schaye}}{2015}]{VanDaalen2015}
{van Daalen} M.~P.,  {Schaye} J.,  2015, \mn@doi [\mnras] {10.1093/mnras/stv1456}, \href {https://ui.adsabs.harvard.edu/abs/2015MNRAS.452.2247V} {452, 2247}

\makeatother
\end{thebibliography}



\appendix

\section{Galaxy selection including sSFR cuts}
\label{sec:ssfr_selection}

As described in Section \ref{sec:boss_sel}, our fiducial analysis employs a simple minimum stellar mass cut for the simulated galaxies.  We vary the minimum stellar mass to determine the cut which results in the best match to the galaxy-galaxy lensing measurements.  The real BOSS selection functions employ colour cuts, such that the CMASS and (particularly) the LOWZ samples are composed primarily of red galaxies that lack significant ongoing star formation.  As a check of whether our fiducial stellar mass cut-only selection method is sufficient, we have also explored selections which apply cuts in both stellar mass and sSFR.  In this appendix we show the impact of this more complicated `quenched' selection function for the fiducial \flamingo\ feedback model.

\begin{figure*}
    \includegraphics[width=0.85\columnwidth]{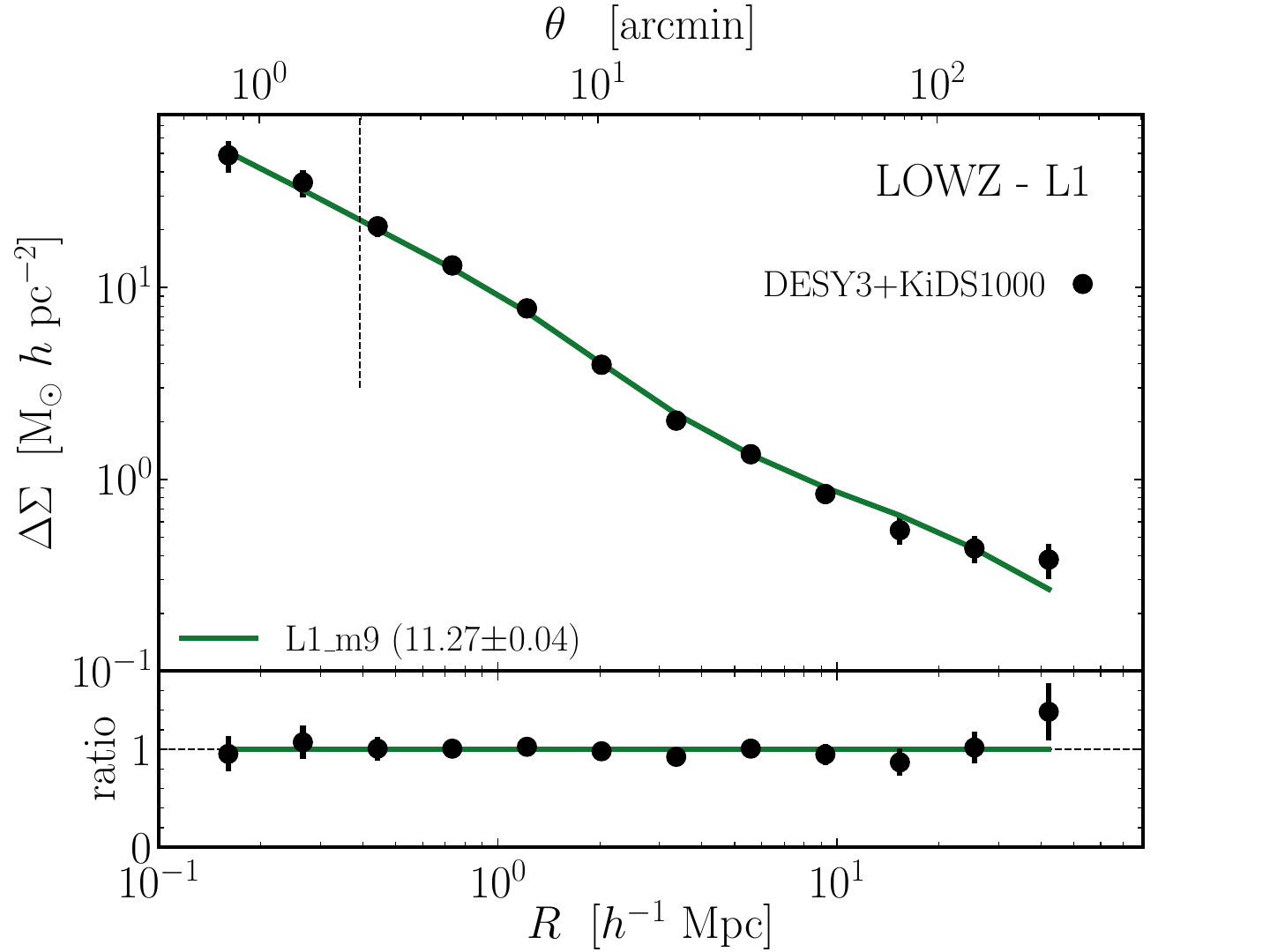}
    \includegraphics[width=0.85\columnwidth]{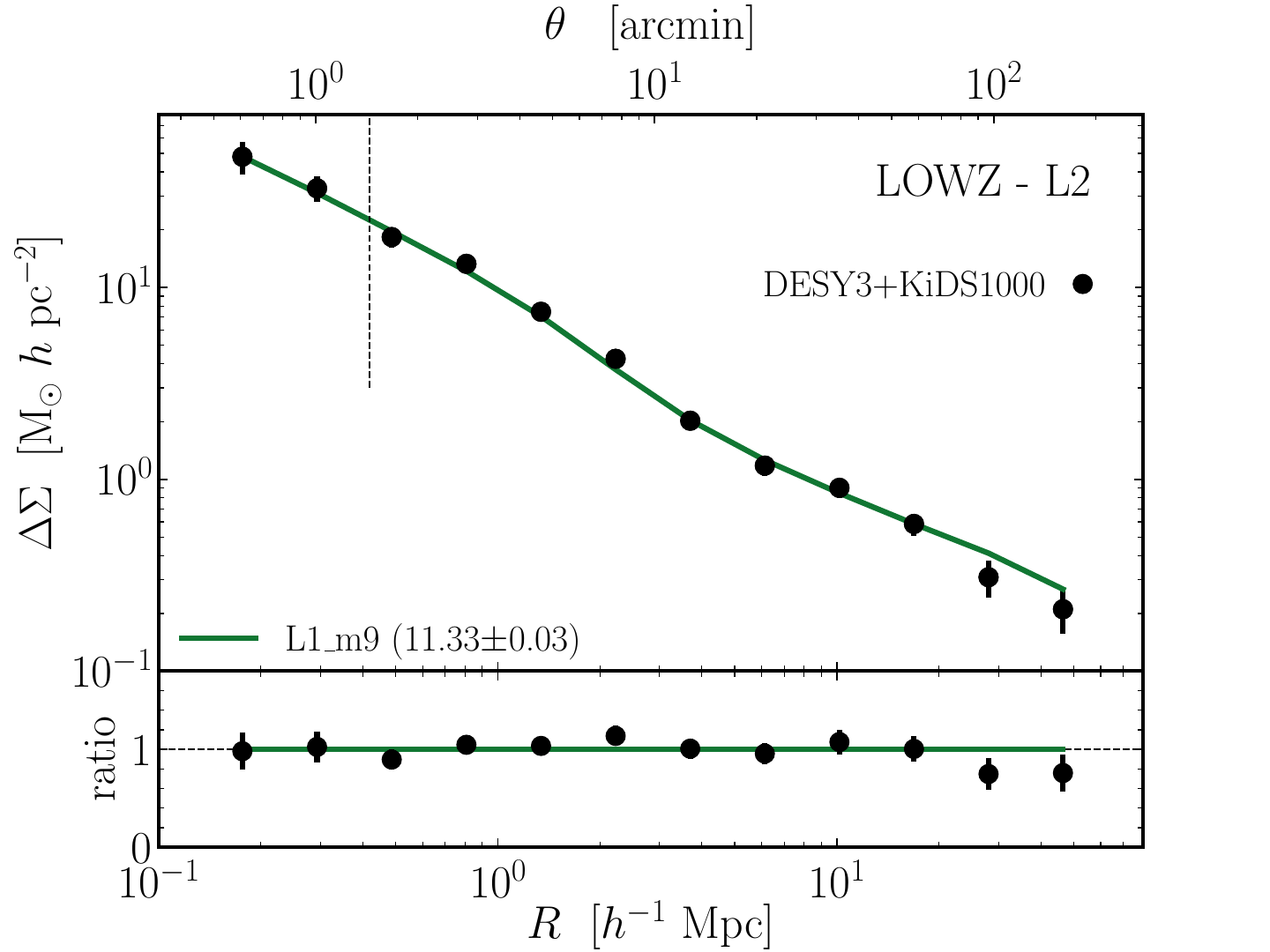}\\
    \includegraphics[width=0.85\columnwidth]{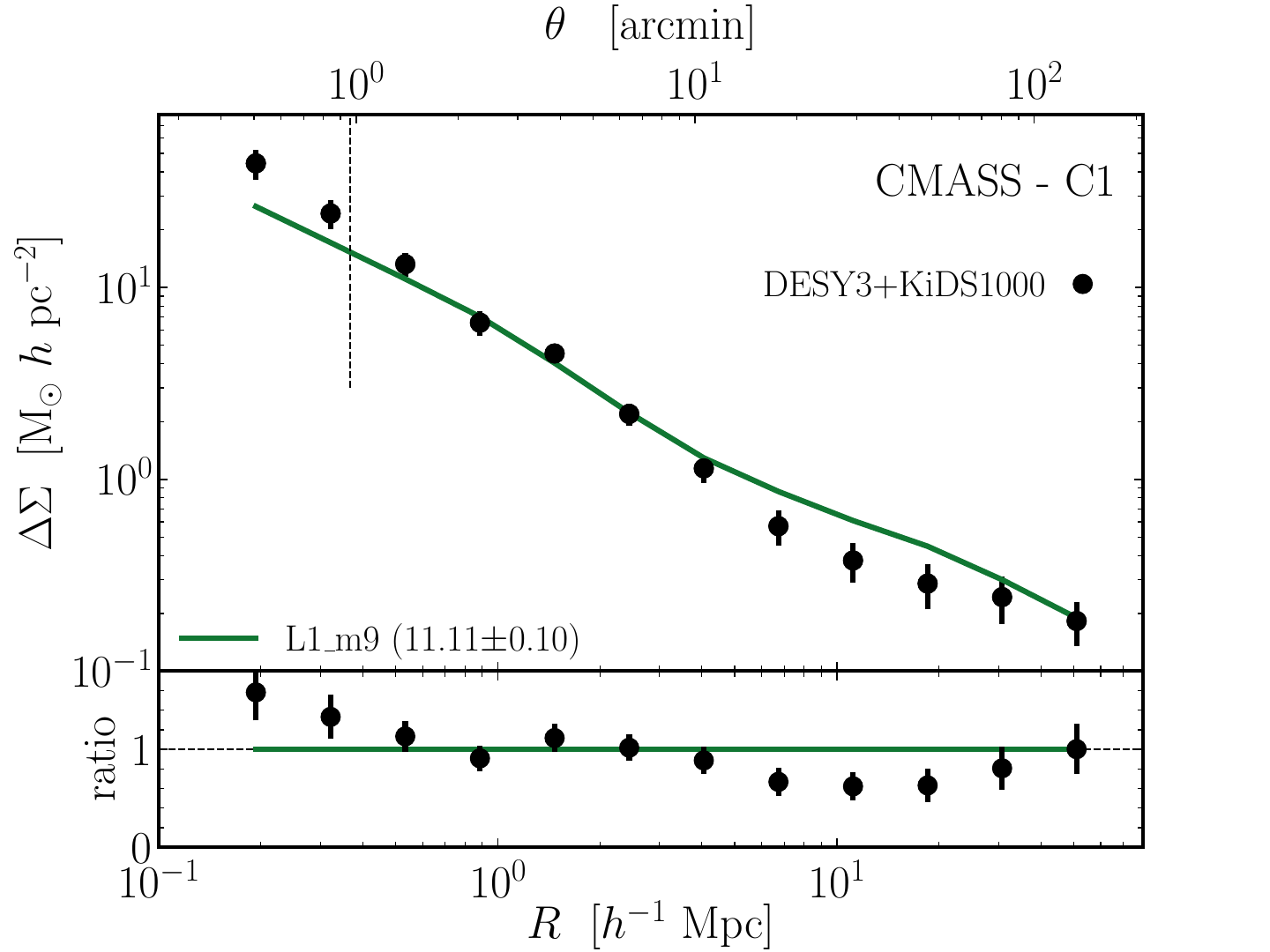}
    \includegraphics[width=0.85\columnwidth]{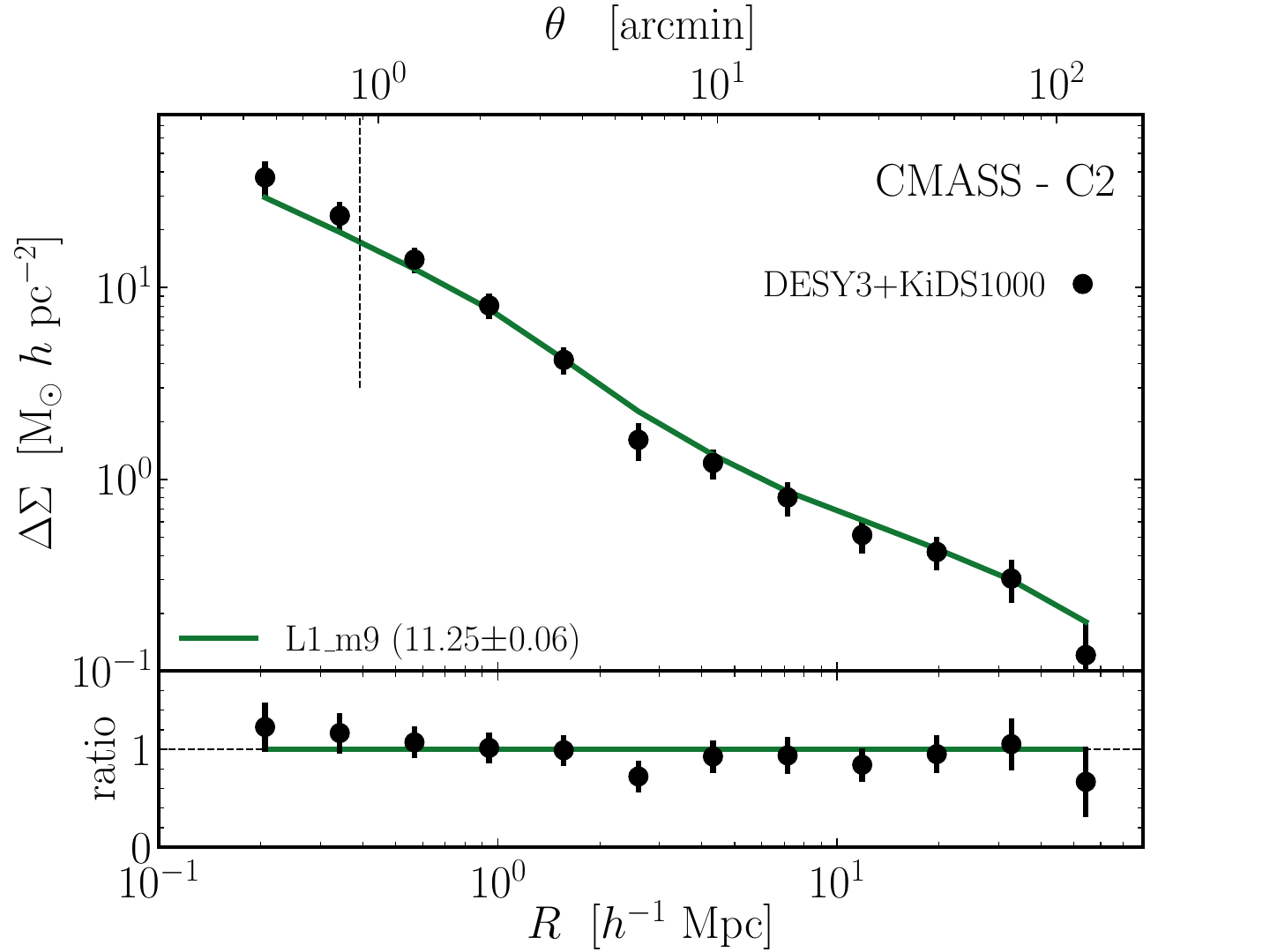}\\    
    \caption{As Fig.~\ref{fig:gg_lensing_fgas}, but showing the impact of selecting only simulated galaxies without significant ongoing star formation for the fiducial \flamingo\ model.  The best-fitting minimum stellar mass consistent with that for our default stellar mass-limited selection in Fig.~\ref{fig:gg_lensing_fgas}.}
    \label{fig:gg_lensing_sSFR_cut}
\end{figure*}

\begin{figure*}
    \includegraphics[width=0.85\columnwidth]{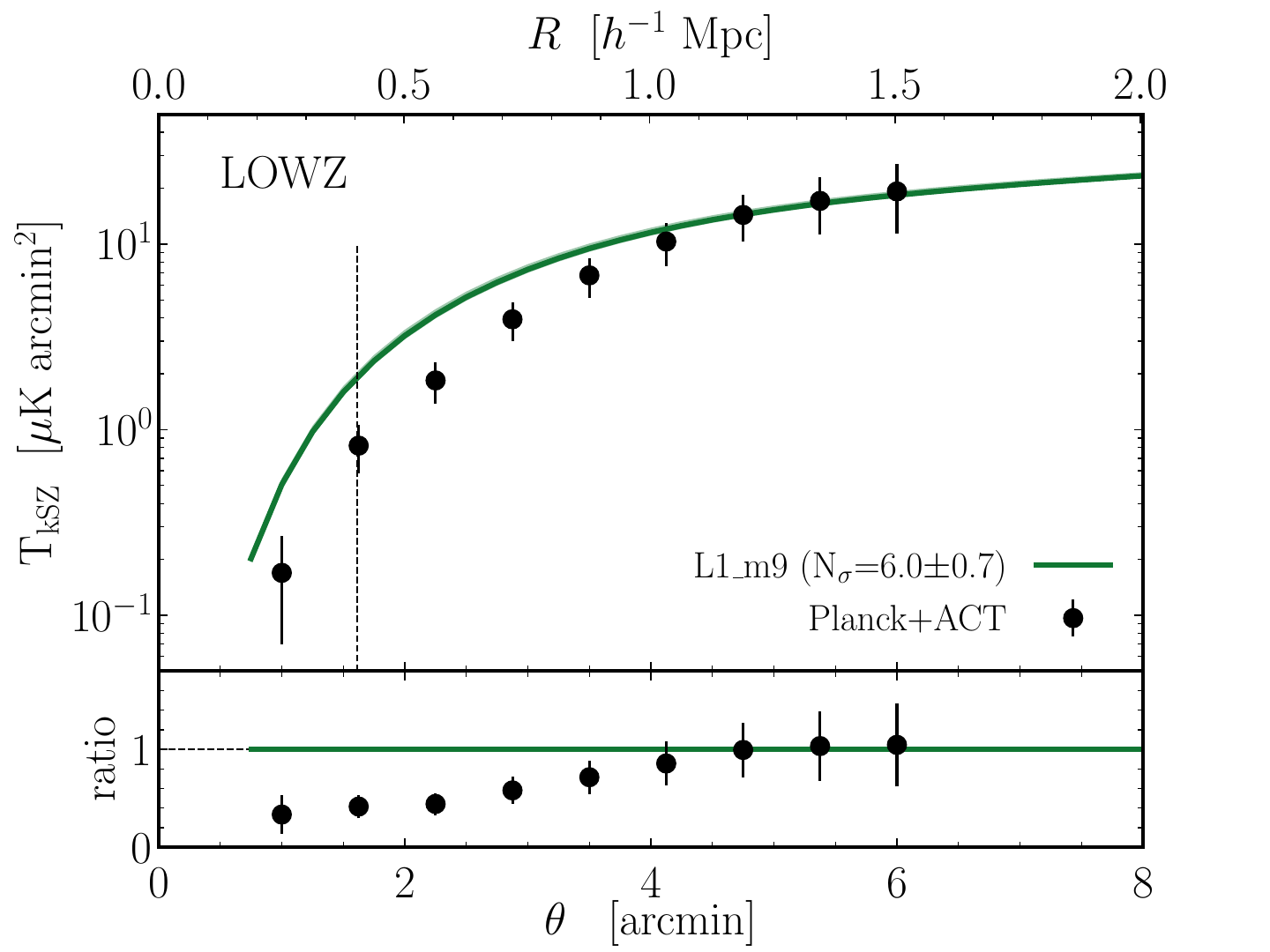}
    \includegraphics[width=0.85\columnwidth]{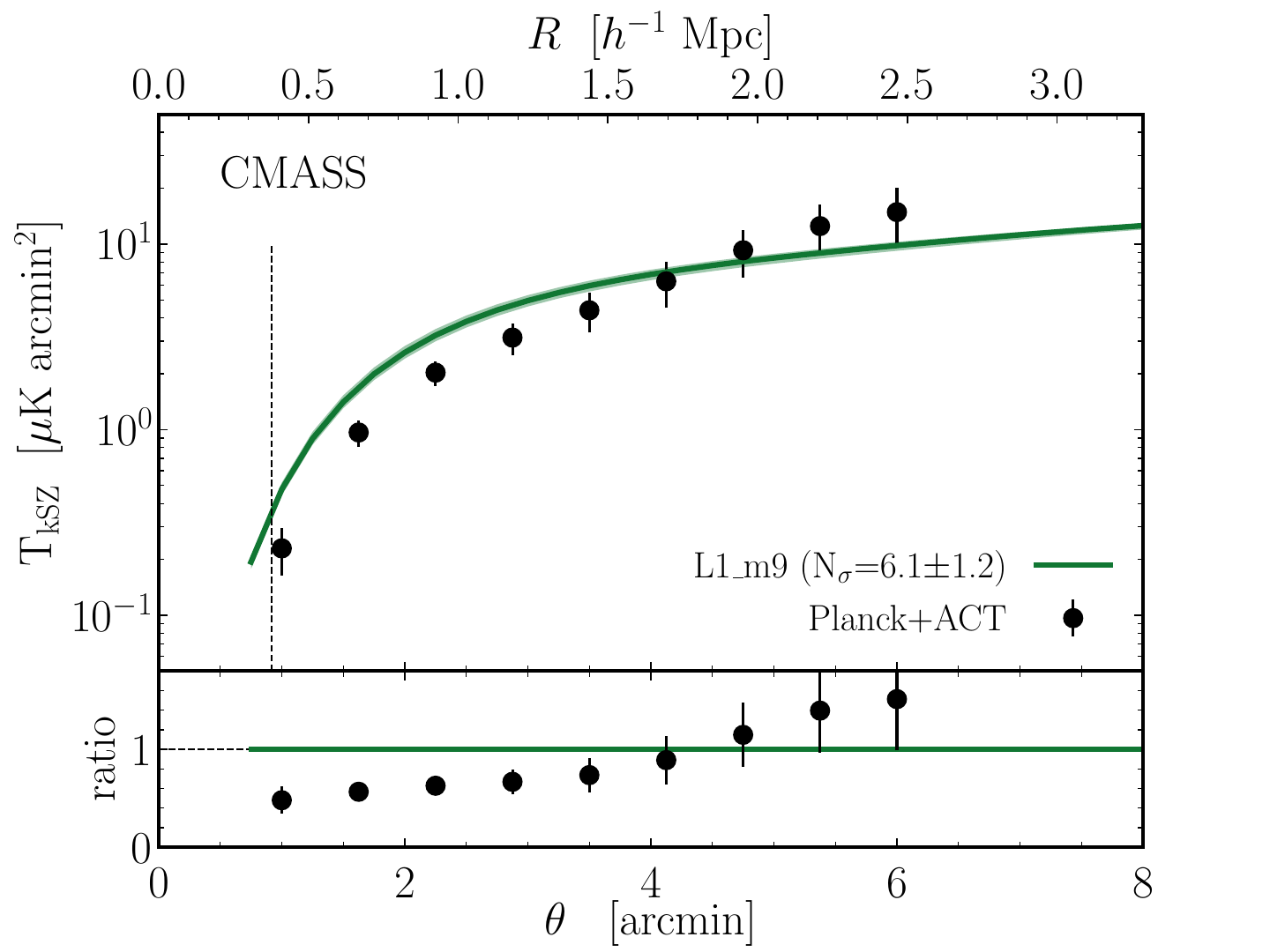}
    \caption{As Fig.~\ref{fig:kSZ_fgas}, but showing the impact of selecting only simulated galaxies without significant ongoing star formation for the fiducial \flamingo\ model.  The solid curves represent the predictions for the best-fitting minimum stellar mass (fitted to the lensing, see Fig.~\ref{fig:gg_lensing_sSFR_cut}).  Selection of quenched galaxies only slightly improves the fit to the observational measurements, but the fiducial model is still strongly ruled out by the data.}
    \label{fig:kSZ_sSFR_cut}
\end{figure*}

We first impose a stellar mass cut as per the default analysis.  From the selected galaxies we compute a histogram of sSFRs.  The peak corresponds to the star-forming main sequence.  Following \citet{Kukstas2023}, we use the galaxies with sSFRs above the star-forming main sequence to determine the standard deviation (width) of this sequence.  We designate quenched galaxies as those with sSFRs that are at least 3 sigma below the sSFR of the star-forming main sequence.  We select these galaxies and compute their galaxy-galaxy lensing signal.  We repeat this process for different choices of the minimum stellar mass cut and determine the value which best matches the galaxy-galaxy lensing measurements.  

We show the results for the fiducial feedback model in Fig.~\ref{fig:gg_lensing_sSFR_cut}.  The quality of the fit to the measurements is similar to that for our default stellar mass-limited analysis (Fig.~\ref{fig:gg_lensing_fgas}) and the best-fitting minimum stellar mass for the quenched sample is consistent with that for the default selection.

We show the resulting kSZ effect predictions for this quenched selection in Fig.~\ref{fig:kSZ_sSFR_cut}.  In agreement with our default stellar mass cut-only analysis (in Fig.~\ref{fig:kSZ_fgas}), we find the fiducial \flamingo\ feedback model is ruled out at high significance for the quenched selection.  Thus, our main conclusion, that the kSZ effect measurements of \citet{Schaan2021} prefer stronger feedback than in the fiducial \flamingo\ model, for which the feedback was calibrated using X-ray cluster observations, is insensitive to the details the simulated galaxy selection so long as the selection matches the galaxy-galaxy lensing measurements.

\section{Resolution dependence}
\label{sec:res_dependence}
\begin{figure*}
    \includegraphics[width=\columnwidth]{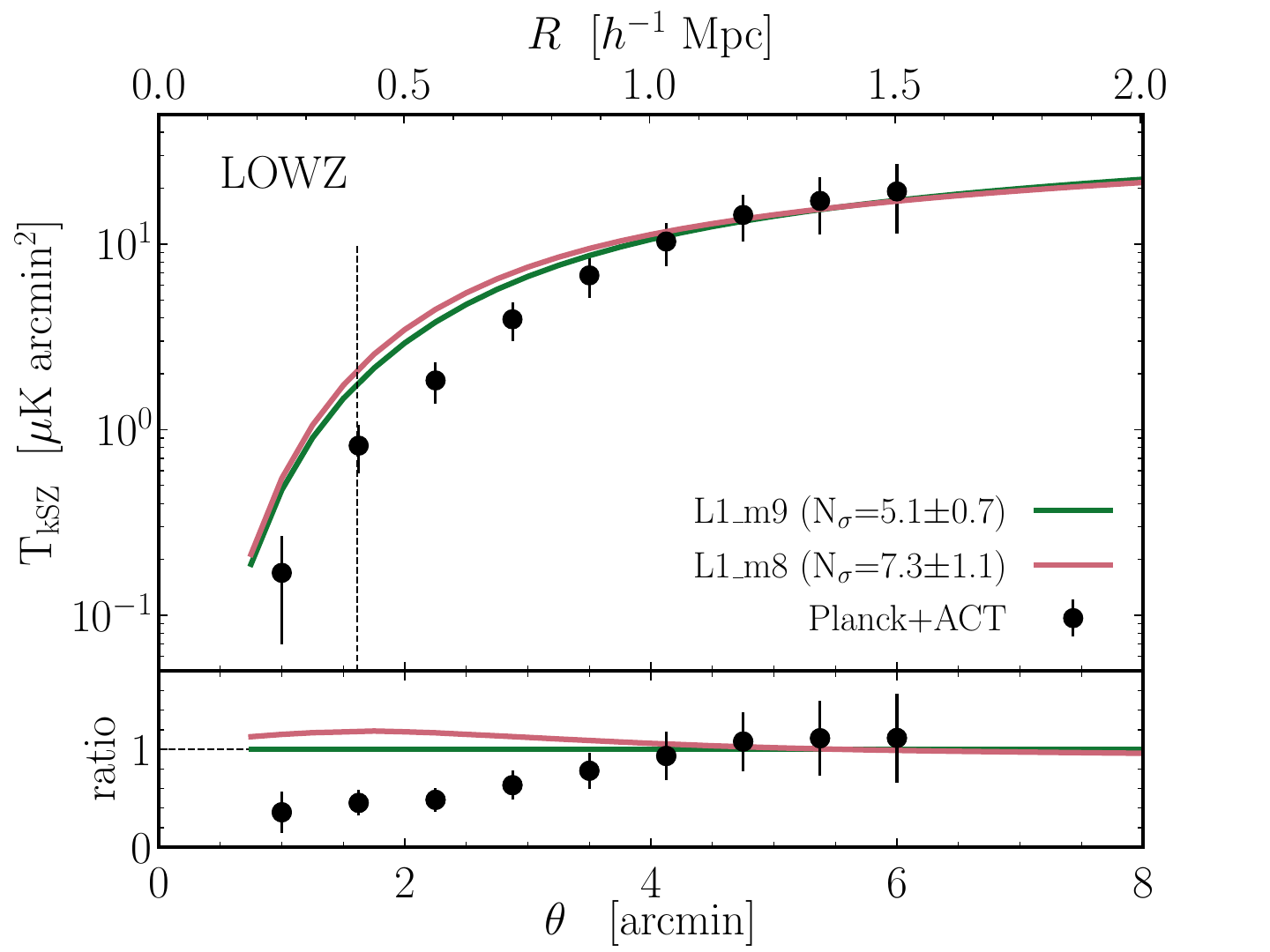}
    \includegraphics[width=\columnwidth]{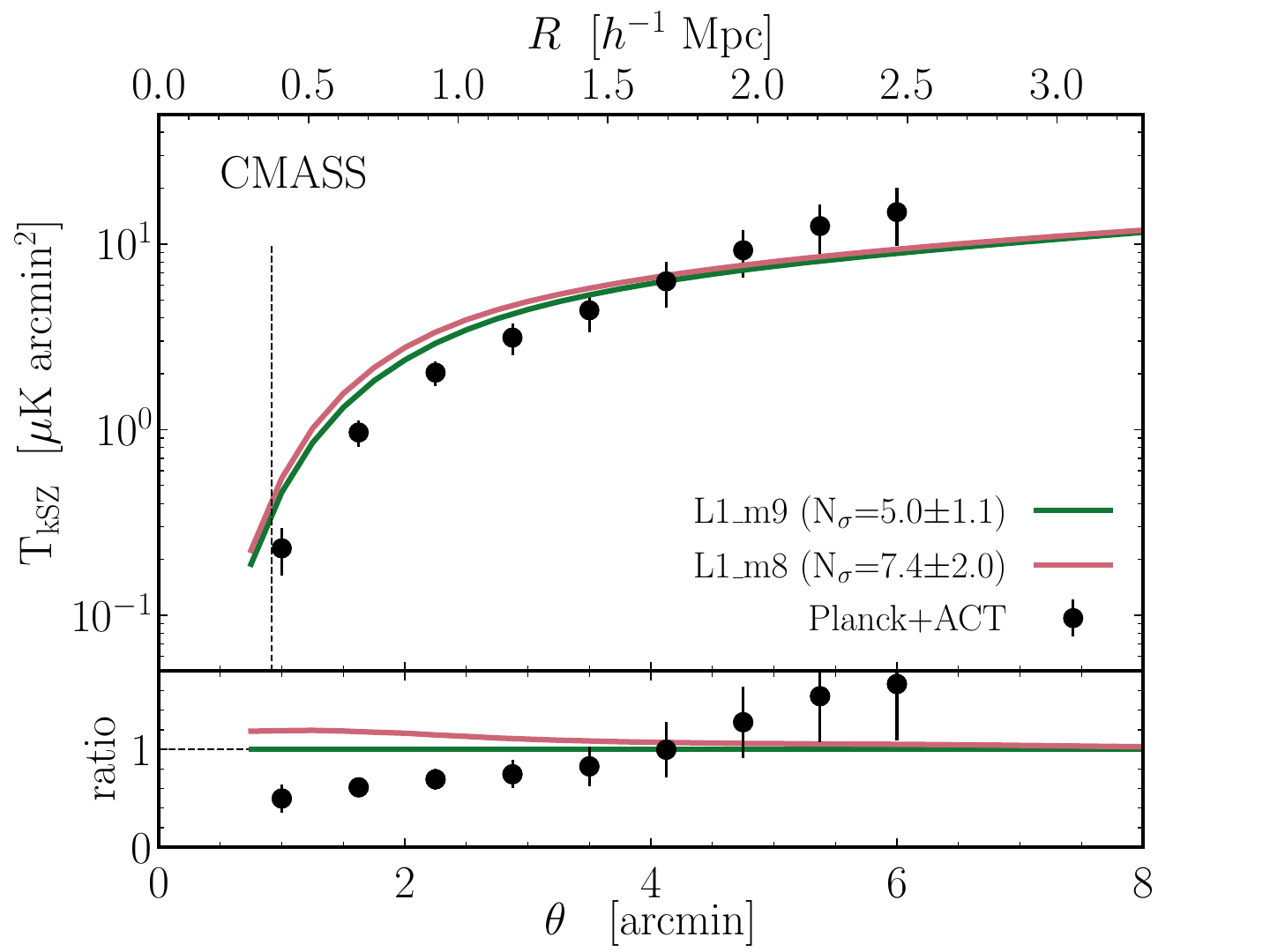}
    \caption{As Fig.~\ref{fig:kSZ_fgas} but comparing the fiducial resolution calibrated model (L1\_m9) with the high-resolution calibrated model (L1\_m8).  Overall the profiles are similar (typically deviating from each by less than the observational measurement errors) and both are in strong tension with the measurements of \citet{Schaan2021}.}
    \label{fig:kSZ_res}
\end{figure*}

In Fig.~\ref{fig:kSZ_res} we compare the predictions of the high-resolution calibrated \flamingo\ model (L1\_m8) with the fidicual resolution calibrated model (L1\_m9).  As per our default analysis, to predict the kSZ effect we first determine the minimum galaxy stellar mass limit that best matches the galaxy-galaxy lensing measurements of \citet{Amon2023} independently for the two models.  For the high-resolution model, the best-fit masses are $\log_{10}[M_\text{star}/\textrm{M}_\odot] = 11.35 \pm 0.04$ and $11.26 \pm 0.06$ for the LOWZ and CMASS samples, respectively. These are the best-fit masses from fitting the two redshift bins in each sample (e.g., LOWZ-L1 and LOWZ-L2) jointly.  These values are somewhat higher than the best-fit joint masses for the fiducial resolution run, which are $\log_{10}[M_\text{star}/\textrm{M}_\odot] = 11.26 \pm 0.02$ and $11.18 \pm 0.04$ for LOWZ and CMASS, respectively, (though only the LOWZ difference is statistically significant).   This difference is expected since, although both models were independently calibrated to the observed galaxy stellar mass function, the calibration to the data is not perfect. Furthermore, the simulations were only calibrated up to $\log_{10}[M_\text{star}/\textrm{M}_\odot] = 11.5$ and higher mass do affect the sample.  We note that in the galaxy stellar mass function the high-resolution run is offset from the fiducial resolution run (and the observational measurements) by $\approx0.1-0.2$ dex at a stellar mass scale of $\sim10^{11} \textrm{M}_\odot$, resulting in larger stellar masses compared to the fiducial resolution run at a halo mass scale of $\sim10^{13} \textrm{M}_\odot$ (see fig.~9 of \citealt{Schaye2023}).  Thus, to match the galaxy-galaxy lensing signal, a higher stellar mass cut is required for the high-resolution run.

When the galaxy selection is fixed by the lensing, the predicted kSZ effect profiles are similar (see Fig.~\ref{fig:kSZ_res}), typically deviating from each by less than the observational measurement errors, and both are in strong tension with the measurements of \citet{Schaan2021}.

\section{Galaxy-galaxy lensing of additional \flamingo\ variations}
\label{sec:galgal_other}

\begin{figure*}
    \includegraphics[width=0.85\columnwidth]{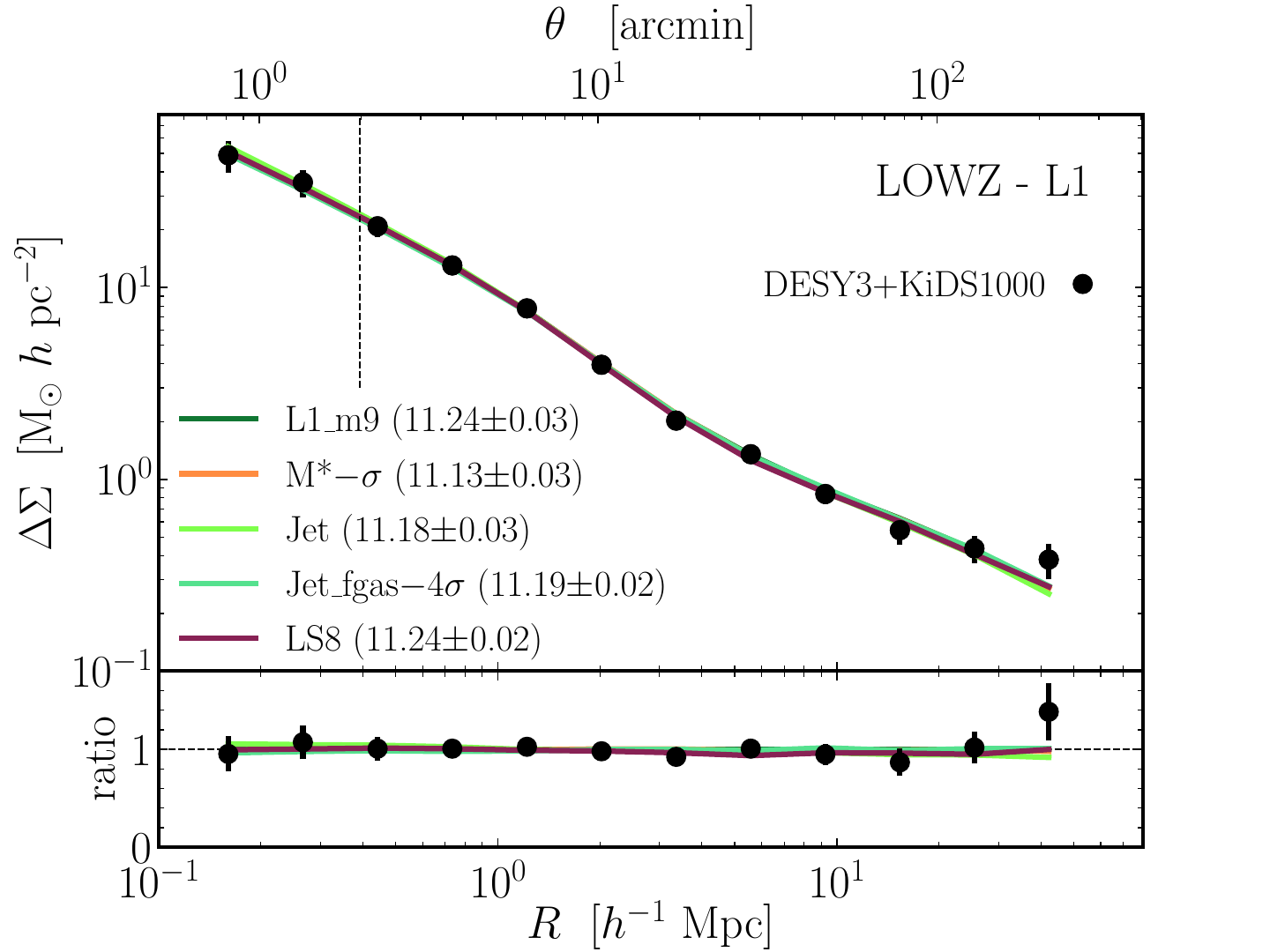}
    \includegraphics[width=0.85\columnwidth]{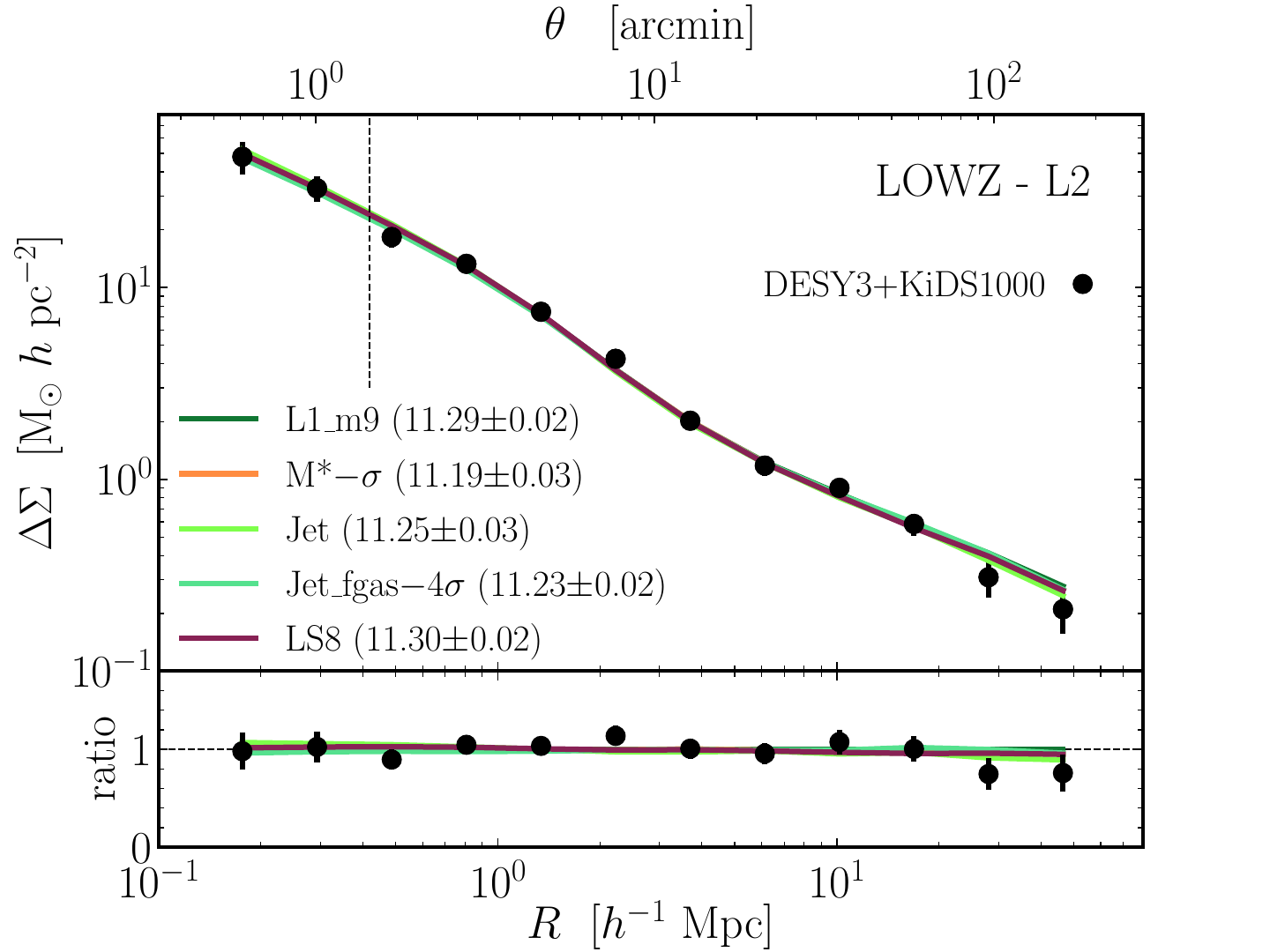}\\
    \includegraphics[width=0.85\columnwidth]{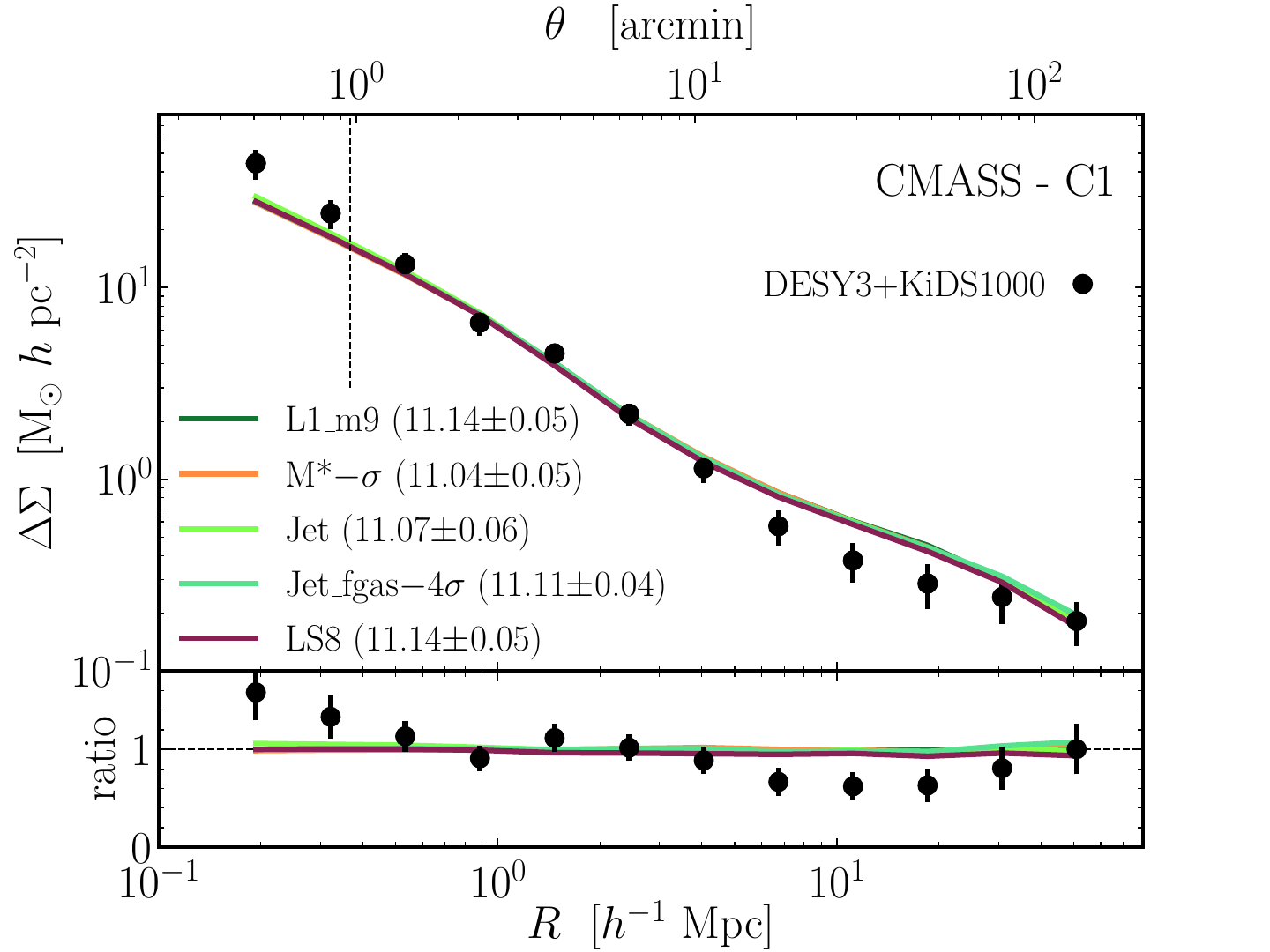}
    \includegraphics[width=0.85\columnwidth]{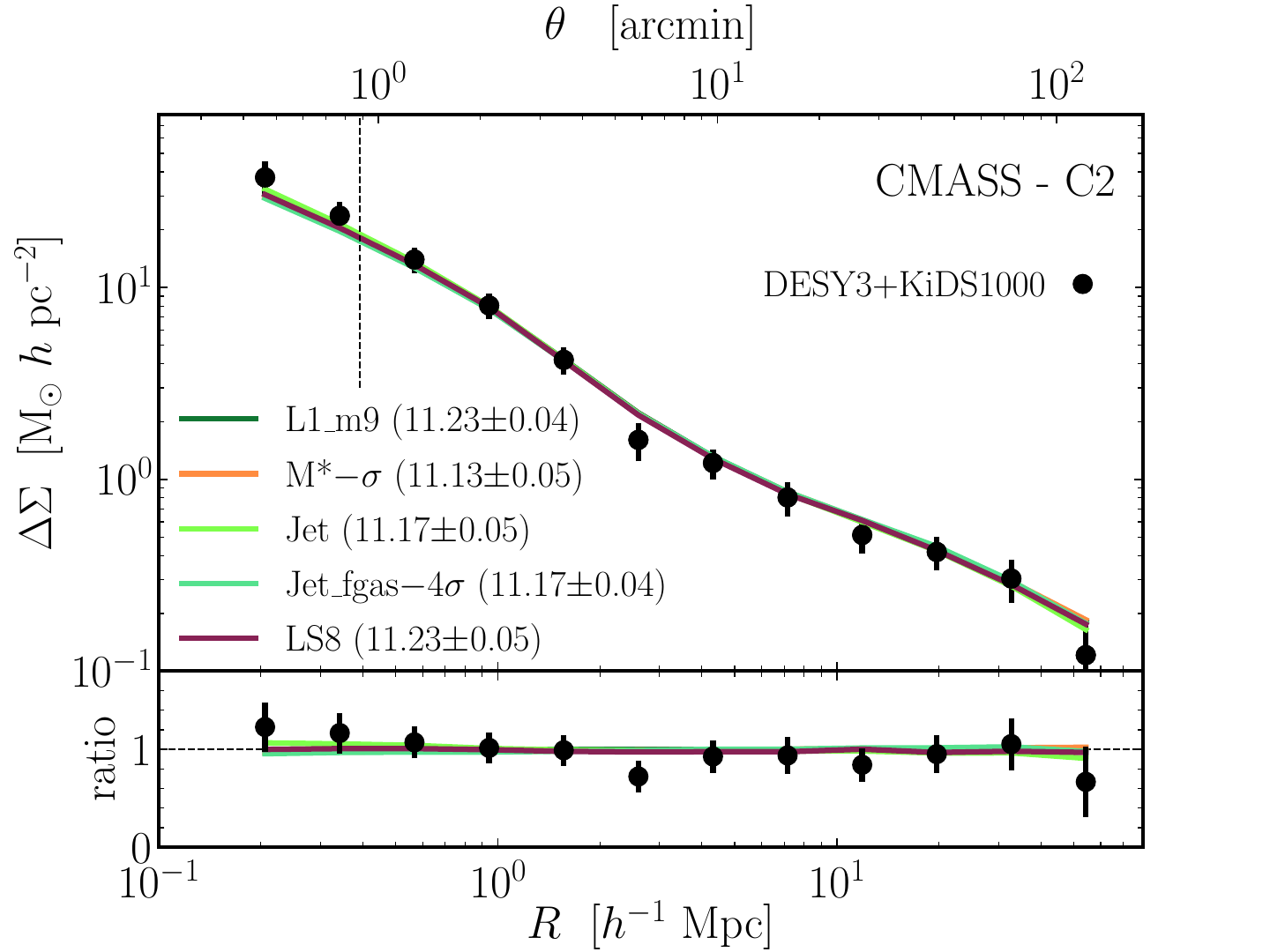}\\    
    \caption{As Fig.~\ref{fig:gg_lensing_fgas}, but showing the dependence of the galaxy-galaxy lensing profiles on other feedback variations, namely variations in the stellar mass function (both the fiducial and reduced cluster gas fractions) and the fiducial and strong jet models of AGN feedback.  The solid coloured curves correspond to the \flamingo\ simulations as baryon models are varied, with the best-fitting stellar masses provided in the legend.}
    \label{fig:gg_lensing_other_var}
\end{figure*}

In Fig.~\ref{fig:gg_lensing_other_var} we show the best-fitting galaxy-galaxy lensing profiles for the \flamingo\ variations shown in Fig.~\ref{fig:kSZ_other_var}.  Consistent with the analyses of the other runs, we find that all models yield an acceptable fit to the DES Y3 + KiDS 1000 measurements given an appropriate choice of the minimum stellar mass.  Note that, as expected, the minimum stellar mass required to match the lensing measurements differs significantly for the run that varied the stellar mass function (M*$-\sigma$) by approximately 0.1 dex.  While the minimum stellar mass is lower, we highlight that the derived mean halo mass of the selected sample agrees remarkably well with that of the other runs.


\bsp	
\label{lastpage}
\end{document}